\documentclass[a4paper,11pt]{article}
\usepackage{a4wide}
\usepackage{xcolor}
\usepackage{graphicx}  
\usepackage{amsmath}   
\usepackage[compress]{cite}
\usepackage{amssymb}   
\usepackage{bm} 
\usepackage{dcolumn}
\usepackage{color}
\usepackage{mathrsfs}
\usepackage{amsfonts}
\usepackage{varioref}
\numberwithin{equation}{section}
\usepackage{tabularx}

\usepackage[caption=false]{subfig}
\usepackage[export]{adjustbox}
\usepackage{multirow}
\usepackage{graphicx}
\usepackage{float}
\RequirePackage[colorlinks,citecolor=blue,urlcolor=magenta,linkcolor=blue]{hyperref}
\input epsf

\usepackage[bottom=0.9in,top=0.5in,right=0.5in,left=0.5in]{geometry}

\labelformat{subfigure}{Fig.~\thefigure#1}
\hypersetup{
    colorlinks,
    linkcolor={green!50!black},
    citecolor={magenta},
    urlcolor={blue!80!black}
}

	\title{\bf Probing hairy Kerr black holes through quasi-periodic oscillations II: a study based on the resonance models}
	\author{ \textbf{Anirban Dasgupta$^{a}$\thanks{522ph1007@nitrkl.ac.in ; These authors contributed equally to this work}, \textbf{Supragyan Priyadarshinee$^{a, b}$}\thanks{supragyanpriyadarshinee52@gmail.com ; These authors contributed equally to this work}, Indrani Banerjee$^{a}$}\thanks{banerjeein@nitrkl.ac.in}, \textbf{Subhash Mahapatra$^{a}$}\thanks{mahapatrasub@nitrkl.ac.in}
			\\\\\textit{{\small $^a$ Department of Physics and Astronomy, National Institute of Technology Rourkela, }}\\
			\textit{{\small Rourkela - 769008, India}}\\
			\textit{{\small $^b$ Department of Physics, IISER Mohali,
Sector 81, }}\\
			\textit{{\small Knowledge City, Mohali, Punjab, 140306, India}}
			}
	
	\date{}

\begin{document}
\maketitle
\begin{abstract}

Hairy black holes (BHs), characterized by nontrivial scalar or matter fields, introduce additional degrees of freedom beyond mass, charge, and spin, providing a framework for testing deviations from the Kerr paradigm. We investigate a class of rotating hairy black holes constructed within the gravitational decoupling framework, whose spacetime satisfies the Einstein field equations with a conserved energy-momentum tensor consistent with the strong energy conditions. In our previous work (Paper I) \cite{Dasgupta:2025blbe}, we studied their horizon structure and test-particle dynamics and confronted two kinematic high-frequency quasi-periodic oscillation (HFQPO) models with observations of six black hole sources. Building on that analysis, we here test nine resonance models against the same HFQPO data, enabling a systematic comparison of eleven models (two kinematic and nine resonance) within the hairy Kerr framework. 
A comprehensive analysis based on the two kinematic models (Paper I) and the nine resonance models reveals that the HFQPO data with the current level of precision cannot rule out the hairy Kerr scenario. By implementing the Akaike Information Criterion and by examining the consistency of the inferred spins with previous independent spin estimates, our analysis provides a systematic framework for assessing the relative performance of different HFQPO models for each source. 
These results highlight the potential of HFQPOs to probe black hole hair and the need for more precise observations and a better understanding of the physical origin of QPOs.


\end{abstract}

\section{Introduction}

Black holes (BHs), among the most fascinating predictions of general relativity (GR), provide a unique arena for exploring gravity in the strong-field regime. Over the past decade, rapid progress in observational astrophysics has transformed black holes from purely theoretical objects into observationally accessible systems. Electromagnetic observations from missions such as RXTE, Chandra, and NuSTAR, together with the direct detection of gravitational waves by the LIGO--VIRGO collaboration \cite{LIGOScientific:2016aoc,LIGOScientific:2018mvr,LIGOScientific:2017vwq}, and the horizon-scale imaging performed by the Event Horizon Telescope (EHT) \cite{EHT:2019ds,EventHorizonTelescope:2022wkp}, have opened unprecedented opportunities to test gravity near the event horizon. So far, these observations remain broadly consistent with the predictions of GR and the Kerr paradigm. Nevertheless, the possibility of deviations from the Kerr geometry continues to motivate extensive theoretical and observational investigations.

In this context, hairy black holes have emerged as important alternatives to the standard Kerr solution. Such spacetimes typically arise in modified gravity theories or in Einstein gravity coupled to additional matter sectors and scalar fields. Unlike the classical no-hair conjecture \cite{ruffini1971introducing,Bekenstein:1971hc,Bekenstein:1995un,Sudarsky:1995zg}, hairy black holes can possess additional parameters that encode deviations from the vacuum Kerr geometry. Over the years, numerous counterexamples to the no-hair conjecture have been constructed \cite{bocharova1970exact,Bekenstein:1974sf,Bekenstein:1975ts,Bronnikov:1978mx,torii1999toward,zloshchastiev2005coexistence,berti2013numerical,garfinkle1991charged,brodbeck1996instability,volkov1995number,bizon1991n,bizon1991stability,zhou1991nonlinear,bizon1990colored,Mahapatra:2020wym,Priyadarshinee:2021rch,Priyadarshinee:2023cmi, Herdeiro:2018wub}, demonstrating that black holes with nontrivial matter configurations are physically viable in a variety of gravitational settings.

One particularly successful framework for constructing hairy compact objects is gravitational decoupling through the minimal geometric deformation (MGD) approach \cite{Ovalle:2017fgl,Ovalle:2018gic,Ovalle:2020kpd,Ovalle:2018umz,Ovalle:2017wqi}. In this method, a known solution of Einstein's equations is deformed by coupling the geometry to an additional matter sector governed by a decoupling parameter. This approach has been applied extensively to anisotropic compact stars \cite{Ovalle:2017fgl,Torres-Sanchez:2019wjv}, wormholes \cite{Panyasiripan:2024kyu}, regular black holes \cite{Casadio:2015gea,Misyura:2024fho,Hua:2025qwu}, and several classes of hairy black hole solutions \cite{Ovalle:2020kpd,Mahapatra:2022xea}. A notable advantage of gravitational decoupling is that new solutions can be generated without specifying an explicit matter Lagrangian for the additional sector.

The rotating hairy Kerr black hole obtained through the extended gravitational decoupling framework \cite{Sharif:2020arn,ovalle2018black,Ovalle:2020kpd,Contreras:2021yxe} is characterized by the deformation parameter $\alpha$ and the hair length scale $l$. The combination $l_0=\alpha l$ plays a crucial role in determining the thermodynamic and observational properties of the spacetime \cite{Mahapatra:2022xea}. Physical viability further requires the spacetime to remain asymptotically flat, free of pathological singularities outside the horizon, and compatible with appropriate energy conditions. These hairy Kerr geometries have attracted considerable interest in recent years because of their potential astrophysical relevance and improved dynamical stability properties \cite{Islam:2021dyk}. They have been investigated in the contexts of black hole shadows and gravitational lensing \cite{Islam:2021dyk,Afrin:2021imp}, accretion disk physics \cite{Meng:2023htc,Meng:2025ivb,Ali:2025dje}, quasi-normal modes and gravitational-wave phenomenology \cite{Zi:2023omh}, black hole spectroscopy \cite{Guimaraes:2025jsh}, and EHT-based parameter estimation \cite{Afrin:2021imp,Vagnozzi:2022moj}. For additional developments concerning hairy black holes and their observational implications, see~\cite{wang2025exploring,Li:2023htz,Li:2022hkq,Heydarzade:2023dof,Al-Badawi:2024iax}.

Among the various observational probes of strong gravity, quasi-periodic oscillations (QPOs) observed in the X-ray fluxes of accreting compact objects play a particularly important role. These oscillations appear as narrow peaks in the power-density spectra of black hole and neutron-star systems and are broadly classified into low-frequency QPOs (LFQPOs) and high-frequency QPOs (HFQPOs) according to their characteristic frequencies \cite{vanderKlis:2000ca,Maselli:2014fca,Torok:2004xs,Abramowicz:2011xu,Aschenbach:2004kj,Torok:2011qy}. In stellar-mass black hole binaries, HFQPOs typically occur in the frequency range of a few hundred Hertz, corresponding to the dynamical timescales of matter orbiting within the innermost regions of the accretion disk \cite{ingram2019review,remillard2006x,PhysRevLett8217}. Since these regions lie deep within the strong gravitational field near the event horizon, HFQPOs are believed to carry valuable information regarding the geometry of spacetime around black holes \cite{PhysRevLett8217}.

One of the most intriguing observational features of HFQPOs is the appearance of frequency pairs in simple rational ratios, particularly the celebrated $3:2$ ratio observed in several black hole X-ray binaries. This feature has motivated the development of resonance-based models, in which the observed oscillations arise from nonlinear resonances between orbital and epicyclic frequencies of matter in the accretion disk. Different resonance mechanisms have been proposed in the literature, including parametric resonances, forced resonances, Keplerian resonances, warped-disk oscillation models \cite{
kluzniak2002parametric,Abramowicz:2003xy,rebusco2004twin,Nowak:1996hg,kato2001basic}. Numerical simulations of accretion flows also support the emergence of preferred rational frequency ratios in the inner disk dynamics \cite{Smith:2020npb,Remillard:2006fc,refId0,Abramowicz:2004rr,
Stuchlik:2013esa}. Moreover, sources such as GRS~1915+105 exhibit multiple HFQPO frequencies, suggesting that the underlying phenomenology may involve several coupled oscillation modes and more sophisticated resonance mechanisms.

In a recent work (henceforth Paper I) \cite{Dasgupta:2025blbe}, the hairy Kerr spacetime was investigated using the kinematic models of HFQPOs, where the observed frequencies were directly associated with the fundamental orbital and epicyclic frequencies of geodesic motion. The analysis demonstrated that while several sources allow a broad region of the parameter space, certain black hole systems exhibiting QPO triplets can impose significantly stronger constraints on the hairy Kerr parameters. However, resonance models provide a fundamentally different physical interpretation of the HFQPOs by introducing nonlinear couplings between the characteristic frequencies of accretion flow oscillations. Since the resonance conditions depend sensitively on the background geometry, they offer an independent and potentially stronger probe of deviations from the Kerr spacetime compared to the continuum spectra or the Fe-line.

Motivated by these considerations, in the present work we investigate the phenomenological viability of the rotating hairy Kerr black hole using the nine available resonance models of HFQPOs. These include several classes of parametric, forced, Keplerian, and warped-disk resonance mechanisms that have been extensively studied in the literature. We systematically analyze the parameter space of the hairy Kerr geometry using the available HFQPO observations of six astrophysical black hole sources and determine the $l-\alpha$ parameter space consistent with the observed frequencies. In particular, we explore how the resonance conditions modify the allowed ranges of the deformation parameters $(l,\alpha)$ and compare the resulting constraints across different resonance scenarios. Our analysis therefore extends previous studies of the hairy Kerr spacetimes \cite{Liu:2023ggz, Dasgupta:2025blbe}, and aims to provide a comprehensive understanding of the role of resonance phenomena in testing strong-field deviations from the Kerr paradigm.

The paper is structured as follows. In \autoref{s2}, we provide a brief overview of the rotating hairy Kerr black-hole spacetime and discuss the fundamental frequencies. In \autoref{sec3}, we provide a summary of the different resonance models that are studied in the literature.
Next we explore the ability of the hairy Kerr spacetime to account for the observed HFQPOs of the six black holes within the framework of the various resonance models in \autoref{Sec4}. Our parameter constraints are verified using the MCMC approach and the robustness of our analysis is discussed in \autoref{MCMC}.
We then assess the relative performance of these models for the six BH sources considered and summarize the constraints on the deformation, the length parameters and the spin parameters of these BHs in \autoref{sec6}.
Finally, in \autoref{sec7}, we summarize the main findings of our work, discuss their physical implications, and highlight possible directions for future investigation. Throughout, we adopt the mostly-plus metric signature and use geometrized units, $G=c=1$.

\section{Fundamental frequencies in rotating hairy Kerr spacetime}
\label{s2}

In this section, we describe the essential properties of a rotating hairy black hole (BH) spacetime obtained as a solution of the Einstein field equations with a conserved energy--momentum tensor obeying the strong energy conditions (SECs). The EM tensor effectively represents additional fundamental fields of unknown origin, potentially related to dark matter and/or dark energy. Instead of specifying the detailed nature of these fields, the framework relies on a few general physical requirements:  
(i) the EM tensor must remain regular everywhere outside the event horizon,  
(ii) it must satisfy the SECs in the exterior spacetime, and  
(iii) the resulting geometry must admit a well-defined event horizon \cite{Ovalle:2020kpd,Ovalle:2017fgl}.
The rotating hairy BH solution is obtained through a two-step procedure. First, a static and spherically symmetric hairy black hole background is generated using the extended gravitational decoupling approach. This static seed metric is then transformed into a stationary and axisymmetric spacetime by means of the gravitational decoupling method or the Newman--Janis algorithm \cite{Ovalle:2020kpd,Ovalle:2018gic, Ovalle:2017fgl,Mahapatra:2022xea,Contreras:2021yxe}.

It is important to note that, at present, a Lagrangian formulation of the hairy gravitational theory considered here is not known. Within the gravitational decoupling framework, one instead works directly at the level of the Einstein field equations by introducing an additional effective energy-momentum tensor, $T_{\mu\nu}$, which deforms a known seed solution in a controlled manner \cite{Ovalle:2018gic, Ovalle:2017fgl, Ovalle:2020kpd,Contreras:2021yxe}. This approach provides a systematic procedure for generating new gravitational solutions through the decoupling of distinct gravitational sources, without requiring an underlying fundamental Lagrangian. Such a construction is intrinsic to the gravitational decoupling framework and has been widely employed to investigate phenomenological extensions of General Relativity \cite{Vagnozzi:2022moj,Afrin:2021imp,Islam:2021dyk,Yang:2025hqk,Meng:2023htc, Meng:2025ivb, Ali:2025dje,Zi:2023omh,Guimaraes:2025jsh}.

The absence of a well-defined matter Lagrangian underlying the hairy black hole solution is primarily associated with the presently unknown microphysical nature of the matter sector that the hair is intended to represent, such as the dark sector. Consequently, an action principle from which the hairy black hole solution can be derived has not yet been established. Instead, the matter sector is subject only to a set of minimal physical requirements, as discussed above. These conditions are deliberately general, allowing the framework to accommodate a broad class of possible matter fields while remaining consistent with the phenomenology that the hairy black hole is intended to capture.

 The stationary and axisymmetric hairy Kerr BH written in Boyer--Lindquist coordinates, takes the form
\begin{align}
ds^2 = g_{tt}(r,\theta)\,dt^2 
+ 2 g_{t\phi}(r,\theta)\,dt\,d\phi
+ g_{\phi\phi}(r,\theta)\,d\phi^2
+ g_{rr}(r,\theta)\,dr^2
+ g_{\theta\theta}(r,\theta)\,d\theta^2 ,
\label{hairyrotmetric}
\end{align}
where the non-vanishing metric functions are given by
\begin{align}
g_{tt}(r,\theta) &=
-\frac{r^2\!\left(1-\frac{2\mathcal{M}}{r}
+ \alpha e^{-\frac{r}{\mathcal{M}-\alpha l/2}}\right)
+ a^2 \cos^2\theta}{\Sigma}, \nonumber \\
g_{t\phi}(r,\theta) &=
-\frac{a r^2 \sin^2\theta}{\Sigma}
\left(\frac{2\mathcal{M}}{r}
- \alpha e^{-\frac{r}{\mathcal{M}-\alpha l/2}}\right), \nonumber \\
g_{\phi\phi}(r,\theta) &=
\Sigma \sin^2\theta
\left[
1 + a^2 \sin^2\theta
\frac{r^2\!\left(1+\frac{2\mathcal{M}}{r}
- \alpha e^{-\frac{r}{\mathcal{M}-\alpha l/2}}\right)
+ a^2 \cos^2\theta}{\Sigma^2}
\right], \nonumber \\
g_{rr}(r,\theta) &=
\frac{\Sigma}{
a^2 + r^2\!\left(1-\frac{2\mathcal{M}}{r}
+ \alpha e^{-\frac{r}{\mathcal{M}-\alpha l/2}}\right)}, \nonumber \\
g_{\theta\theta}(r,\theta) &= \Sigma ,
\label{rotatinghairmet}
\end{align}
with
\begin{align}
\Sigma = r^2 + a^2 \cos^2\theta .
\end{align}
Here, $\mathcal{M}$ is the ADM mass of the BH. All metric components, except for $g_{\theta\theta}$, acquire corrections due to the presence of the deviation parameter $\alpha$ and the hair parameter $l$. In the limit $\alpha \rightarrow 0$, the above geometry reduces to the Kerr spacetime, while setting both $\alpha \rightarrow 0$ and $a \rightarrow 0$ yields the Schwarzschild solution with mass $\mathcal{M}$. Asymptotic flatness requires $\alpha l \leq 2$, and the Kerr solution is exactly recovered when $\alpha l = 2$. This spacetime therefore represents a family of non-Kerr black holes characterized by the additional parameters $\{\alpha, l\}$. 
It may be noted that the present framework does not specify the underlying microscopic origin of the hair. Nevertheless, the parameters $\alpha$ and $l$ may be regarded as effective quantities that capture the collective influence of additional matter sectors or possible modifications of the gravitational interaction, without requiring a commitment to any particular fundamental theory. Such effective parametrizations are widely adopted in phenomenological tests of gravity, where the primary objective is to use astrophysical observations to constrain potential deviations from General Relativity.
It is important to note that when $\alpha \neq 0$ but $l = 0$, the effective EM tensor violates the strong energy conditions; hence, this branch is excluded from the phenomenological analysis.

A discussion of the effective energy--momentum tensor sourcing the rotating hairy black hole can be found  in \cite{Contreras:2021yxe, Mahapatra:2022xea,Dasgupta:2025blbe}
The strong energy conditions impose the inequalities
\begin{align}
\tilde{\rho} + \tilde{p}_r + 2\tilde{p}_\theta &\geq 0, \nonumber \\
\tilde{\rho} + \tilde{p}_r &\geq 0, \nonumber \\
\tilde{\rho} + \tilde{p}_\theta &\geq 0 .
\end{align}
These conditions are satisfied provided that the energy density and the angular pressure are non-negative, i.e., $\tilde{\rho} \geq 0$ and $\tilde{p}_\theta \geq 0$. Restricting attention to motion in the equatorial plane, these requirements translate into the constraints (for details see \cite{Dasgupta:2025blbe})
\begin{align}
r &> \mathcal{M}-\frac{\alpha l}{2}, \nonumber \\
r &> 2\left(\mathcal{M}-\frac{\alpha l}{2}\right).
\label{SEC}
\end{align}
In the remainder of this work, the following quantities are expressed in units of the ADM mass $\mathcal{M}$, such that
\begin{align}
l \equiv \frac{l}{\mathcal{M}}, \qquad
a \equiv \frac{a}{\mathcal{M}}, \qquad
r \equiv \frac{r}{\mathcal{M}} .
\end{align}
Requiring the strong energy conditions to hold everywhere outside the event horizon then leads to the condition
\begin{align}
r \geq r_{\mathrm{SEC}}, \qquad
r_{\mathrm{SEC}} = 2\left(1-\frac{\alpha l}{2}\right),
\end{align}
which is independent of the black hole spin parameter $a$. Depending on the values of the deformation length parameter $l$, we get three distinct regimes, which have been discussed in detail in \cite{Dasgupta:2025blbe}. Below we briefly mention the main results: \\
\\
\textbf{Regime I: $l > e^{-2}$}\\
\\
In this regime, we get
\begin{align}
\alpha_{\rm crit} = \frac{2}{l} - e^{2}\,,
\end{align}
below which the SEC is violated. Within the interval $\alpha_{\rm crit} \le \alpha \le \alpha_{\rm coin} = 1/l$, the SEC is satisfied only up to a maximum spin $a_{\rm SEC} < a_{\max}$, where
\begin{align}
a_{\rm SEC}=\sqrt{2r_{\rm SEC}-r^2_{\rm SEC}(1+\alpha e^{-2}) }\,,
\end{align}
which is obtained by solving $\Delta(r_{\rm SEC})=0$  for fixed $(l,\alpha)$ and $a_{\max}$ corresponds to the spin of the extremal BH.
If we increase the spin beyond $a_{\rm SEC}$,  the outer event horizon (EH) becomes less than $r_{\rm SEC}$, such that SEC does not hold everywhere outside the outer EH. So, the BHs with extremal spin value are excluded in this region.  For $ \alpha_{\rm coin} \lesssim \alpha \lesssim \alpha_{\max}=2/l$, the BH can explore  its full spin range from $-a_{\max}\leq a \leq a_{\max}$.\\
\\
\textbf{Regime II: $l = e^{-2}$}\\
\\
The case $l = e^{-2}$ represents a transition between spin-restricted and spin-unrestricted configurations. At this value of $l$, the two  deformation scales coincide,
\begin{equation}
\alpha_{\rm crit} = \alpha_{\rm coin} = e^{2}.
\end{equation}
At $\alpha = e^{2}$, the extremal black hole corresponds to a non-rotating black hole with spin $0$, and its event horizon coincides with $r_{\rm SEC}$.  Rotating black hole solutions are not permitted at this point.
For $\alpha > e^{2}$, non-zero values of spin are allowed, and the SEC is satisfied for the entire allowed range of the spin parameter. \\
\\
\textbf{Regime III: $l < e^{-2}$}\\
\\
In this regime, a critical deformation parameter $\alpha_{\rm crit}$ continues to exist, but once $\alpha \ge \alpha_{\rm crit}$, the SEC is satisfied for all allowed values of the spin parameter. In contrast to the previous regime, no upper bound on the spin is imposed by the SEC. The outer horizon is always outside $r_{\rm SEC}$.\\

In what follows, we will be interested in exploring the role of the hair parameter and the length parameter of the hairy Kerr BH in explaining the high-frequency quasiperiodic oscillations (HFQPOs) observed in certain BH sources. This requires one to investigate the motion of timelike test particles in the hairy Kerr spacetime. This is because the theoretical models proposed to explain the observed HFQPOs are based on the three fundamental frequencies associated with the motion of massive test particles. 
This has been discussed in detail in Paper I \cite{Dasgupta:2025blbe}, where we consider the motion of massive test particles rotating around the stationary, axisymmetric, hairy Kerr spacetime in a stable, circular, equatorial orbit. The angular velocity $\Omega$ of the test particle is given by,
\begin{eqnarray}
\Omega=\frac{c^3}{GM}\frac{-\partial_{r}g_{t\phi}\pm\sqrt{(\partial_{r}g_{t\phi})^{2}-(\partial _{r}g_{\phi\phi})(\partial_{r}g_{tt})}}{\partial _{r}g_{\phi\phi}}\,,
\end{eqnarray}
where $\Omega=2\pi f_\phi$, with $f_\phi$ being the orbital frequency at the circular radius $r$ while $\frac{c^3}{GM}$ takes care of the dimension.\\

Now, we apply a small perturbation to the circular equatorial orbit both in the radial and vertical directions. Then, we get the radial and vertical epicyclic frequencies as follows,
\begin{eqnarray}
f_{r}^2=\frac{c^6}{G^2M^2}\Big[\frac{(g_{tt}+g_{t\phi}\Omega)^2}{2(2\pi))^{2}g_{rr}}\Big(\frac{\partial^2{U}}{\partial r^2}\Big)_{r_c,\frac{\pi}{2}}\Big]\,,
\label{3.13}
\end{eqnarray}
\begin{eqnarray}
f_{\theta}^2=\frac{c^6}{G^2M^2}\Big[\frac{(g_{tt}+g_{t\phi}\Omega)^2}{2(2\pi)^{2}g_{\theta\theta}}\Big(\frac{\partial^2{U}}{\partial\theta^2}\Big)_{r_c,\frac{\pi}{2}}\Big]\,,
\label{3.14}
\end{eqnarray}
where the factor $\frac{c^6}{G^2M^2}$ is used in order to match the dimension of frequency squared. $U(r,\theta)$ is the effective potential in which the particle moves, which is given by
\begin{eqnarray}
U(r,\theta)=g^{tt}-2\Big(\frac{L}{E}\Big)g^{t\phi}+\Big(\frac{L}{E}\Big)^{2}g^{\phi\phi}\,,
\end{eqnarray}
where $L$ and $E$ correspond to the specific angular momentum and the specific energy of the massive test particle. For more details, one is referred to Paper I \cite{Dasgupta:2025blbe}.

\section{Resonance models explaining the HFQPOs}
\label{sec3}
In this paper, we will consider resonance models to describe the origin of the HFQPOs. In Paper I \cite{Dasgupta:2025blbe}, the role of the kinematic models in explaining the HFQPOs has been addressed.
In \autoref{t1}, we have listed six black holes (five stellar mass and one supermassive) which exhibit HFQPOs in their power spectrum, and these HFQPOs can be described by resonance models. Most importantly, these resonance models can well explain the observed $3:2$ ratio of the twin peaks of the observed HFQPOs. The basic foundation of these models relies on resonance between various oscillation modes in the accretion disk \cite{2001A&A...374L..19A,kluzniak2002parametricsss,2001AcPPB..32.3605K}.
The model-dependent HFQPOs are described as the linear combinations of the three fundamental frequencies, $f_\phi$, $f_\theta$, and $f_r$, introduced in the previous section.  In what follows, we briefly discuss the different resonance models.
\begin{table}[htpb]
    \centering
\begin{adjustbox}{width=0.9\textwidth}
    \begin{tabularx}{\textwidth}{|X|X|X|X|X|}
    \hline
    \textbf{Source} & Mass \boldmath$(M_{\boldsymbol{\odot}})$ & $f_{up1} \pm \Delta f_{up1}$ \textbf{(Hz)} & $f_{up2} \pm \Delta f_{up2}$ \textbf{(Hz)} & $f_{up3} \pm \Delta f_{up3}$ \textbf{(Hz)}\\
    \hline
    \hline
    GRO J1655-40 & $5.4 \pm 0.3$ \cite{beer2002quiescent} & $441 \pm 2$ \cite{motta2014precise} & $298 \pm 4$ \cite{motta2014precise} & $17.3 \pm 0.1$ \cite{motta2014precise} \\
    \hline
    XTE J1550-564 & $9.1 \pm 0.61$\cite{Orosz:2011ki} & $276 \pm 3$ \cite{Torok:2004xs} & $184 \pm 5$ \cite{Torok:2004xs} & - \\
    \hline
    GRS 1915+105 & $12.4^{+2.0}_{-1.8}$ \cite{Reid:2014ywa} & $168 \pm 3$ \cite{Torok:2004xs} & $113 \pm 5$ \cite{Torok:2004xs} & - \\
    \hline
    H 1743+322 & $8.0 - 14.07$ \cite{Pei:2016kka,Bhattacharjee:2017rbl,Petri:2008jc} & $242 \pm 3$ \cite{Remillard:2004sp} & $166 \pm 5$ \cite{Remillard:2004sp}& - \\
    \hline
    Sgr A* & $(3.5-4.9)\times 10^6$\cite{Ghez:2008ms,Gillessen:2008qv} & $(1.445 \pm 0.16)\times10^{-3}$\cite{Torok:2004xs,Stuchlik:2008fy} & $(0.886 \pm 0.04)\times10^{-3}$\cite{Torok:2004xs,Stuchlik:2008fy} & - \\
    \hline
    XTE J1859+226 & $7.85 \pm 0.46$\cite{motta2022black} & $227.5^{+2.1}_{-2.4}$\cite{motta2022black} & $128.6^{+1.6}_{-1.8}$\cite{motta2022black} & $3.65 \pm 0.01$ \\
    \hline
    \end{tabularx}
\end{adjustbox}
\caption{BH Sources where HFQPOs are observed.}
\label{t1}
\end{table}

\begin{enumerate}
\item {\bf Parametric Resonance Model}

\vspace{0.5cm}

The Parametric Resonance Model (PRM) is obtained by assuming that radial perturbations denoted by $\delta r$ dominate over vertical ones $\delta \theta$, such that $\delta r \gg \delta\theta$\cite{kluzniak2002parametricsss,Abramowicz:2003xy,Rebusco:2004ba,2005A&A...436....1T}. Under this assumption, the radial motion remains unaffected by non-linear forcing, while the vertical motion is modulated by the radial perturbation. The equations reduce to
\begin{align}
\delta \ddot{r} + \omega_r^2 \, \delta r &= 0, \\
\delta \ddot{\theta} + \omega_\theta^2 \, \delta \theta &=
- \omega_\theta^2 \, \delta r \, \delta \theta .
\label{Eq3.2}
\end{align}
Here, the radial perturbation has a sinusoidal behavior which, when implemented in \autoref{Eq3.2}, gives rise to parametric resonance when the epicyclic frequencies satisfy
\begin{align}
\frac{f_r}{f_\theta} = \frac{2}{n},
\end{align}
where $n$ is a positive integer. The strongest response is obtained for the smallest allowed values of $n$. In black hole spacetimes where the vertical epicyclic frequency exceeds the radial one (as in the present case, see Paper I \cite{Dasgupta:2025blbe}), the lowest admissible choice is $n=3$. This immediately produces a $3\!:\!2$ ratio between the two oscillation frequencies, which are identified as
\begin{align}
f_1 = f_\theta, \qquad f_2 = f_r .
\end{align}
In this model, the observed HFQPOs are therefore interpreted as a direct manifestation of parametric resonance between the fundamental epicyclic modes.

\vspace{0.5cm}

\item {\bf Forced Resonance Model}
\vspace{0.5cm}

If the accretion disk closely follows Keplerian motion, parametric resonance alone may be sufficient to explain the emergence of HFQPOs. In realistic accretion environments, however, additional physical effects such as pressure forces, viscous dissipation, and magnetic stresses are expected to be operational. These effects introduce non-linear interactions between the radial and vertical perturbations, $\delta r$ and $\delta\theta$, which must be taken into account together with parametric resonance \cite{2001A&A...374L..19A,kluzniak2002parametricsss,2001AcPPB..32.3605K}.

Numerical studies indicate that pressure-related forces can efficiently excite vertical oscillations through resonant driving by radial motion. In this situation, the equations governing the perturbations can be written as
\begin{align}
\delta \ddot{r} &= -\omega_r^2 \, \delta r, \\
\delta \ddot{\theta} + \omega_\theta^2 \, \delta \theta &=
- \omega_\theta^2 \, \delta r \, \delta \theta + F_\theta(\delta \theta),
\end{align}
where $\omega_\theta = 2\pi f_\theta$ is the vertical epicyclic frequency and $F_\theta$ denotes a forcing term associated with non-linear processes in the accretion disk.
Solving this system leads to a resonance condition expressed as
\begin{align}
\frac{f_\theta}{f_r} = \frac{m}{n},
\end{align}
with $m$ and $n$ being positive integers. In the first forced resonance scenario (Forced Resonance Model 1, FRM1), one adopts $m=3$ and $n=1$, corresponding to a $3\!:\!1$ relation between the vertical and radial epicyclic frequencies. The model dependent HFQPO frequencies are then identified as
\begin{align}
f_1 &= f_\theta = 3 f_r, \\
f_2 &= f_\theta - f_r = 2 f_r .
\end{align}
With this choice, the commonly observed $3\!:\!2$ HFQPO ratio arises naturally from the forced resonance mechanism driven by non-linear couplings in the disk.

A second forced resonance scenario (Forced Resonance Model 2, FRM2) can be constructed by assuming a $2\!:\!1$ ratio between the vertical and radial epicyclic frequencies \cite{2005A&A...436....1T,lee2004resonance}. As in the previous case, the vertical motion is governed by a non-linearly forced equation of motion, where the forcing term again reflects the influence of non-linear interactions within the accretion flow.
In this model, the resonance condition is specified by
\begin{align}
\frac{f_\theta}{f_r} = \frac{m}{n} = 2 .
\end{align}
The model dependent upper and lower HFQPO frequencies are identified through linear combinations of the epicyclic frequencies as
\begin{align}
f_1 &= f_\theta + f_r = 3 f_r, \\
f_2 &= f_\theta = 2 f_r .
\end{align}
Hence, FRM2 also reproduces the observed $3\!:\!2$ frequency ratio. In both the forced resonance models, non-linear forcing terms are essential, as they enable frequency combinations that account for the resonant features required to explain the observed HFQPO data.

\vspace{0.5cm}
\item {\bf Keplerian Resonance Model (KRM)}
\vspace{0.5cm}

The Keplerian resonance model interprets HFQPOs as manifestations of a non-linear resonance between the azimuthal (orbital) motion and the radial epicyclic oscillations of matter in the innermost regions of an accretion disk \cite{2005A&A...436....1T,kato2001trapping,2001A&A...374L..19A,Nowak:1996hg}. Since both the orbital and epicyclic frequencies are determined by the underlying spacetime geometry, the model provides a direct connection between observed QPO frequencies and the strong-gravity environment surrounding compact objects.

Several physical mechanisms have been proposed to generate such resonances. One possibility involves the excitation and trapping of non-axisymmetric diskoseismic g-modes through a corotation resonance operating in relativistic thin accretion disks \cite{kato2001trapping}. In this picture, wave modes can become confined within specific regions of the disk, allowing resonant amplification when their characteristic frequencies satisfy suitable commensurability conditions. A second mechanism invokes the interaction between the radially varying orbital angular velocity and a pair of coherent vortices with opposite vorticities that oscillate at the radial epicyclic frequency \cite{2005A&A...436....1T,2010tbha.book.....A}. The resulting non-linear coupling can transfer energy between different oscillatory modes and give rise to enhanced variability at preferred frequencies.

The appeal of the Keplerian resonance scenario lies in its ability to naturally produce frequency ratios close to small integers, a characteristic commonly observed in HFQPO pairs from several microquasars. Nevertheless, the physical realization of the model remains a matter of debate. In particular, later investigations demonstrated that non-axisymmetric g-modes are subject to significant damping due to the corotation resonance \cite{Li:2002yi,kato2003damping2}. This damping suppresses the growth of the oscillations and weakens the resonance mechanism, thereby challenging the applicability of g-mode-based Keplerian resonance models to observed HFQPO phenomena. Despite these difficulties, the model continues to serve as a useful phenomenological framework for exploring the relationship between disk oscillations and the characteristic frequencies predicted by relativistic accretion theory.

Depending on the specific resonance condition imposed between the orbital and radial epicyclic motions, the Keplerian resonance framework can be divided into three variants. The first variant, referred to as Keplerian Resonance Model 1 (KRM1), assumes a direct resonance between the orbital and radial epicyclic frequencies,
\begin{align}
   f_1=f_\phi,\qquad f_2=f_r\,. 
\end{align}
The second variant, Keplerian Resonance Model 2 (KRM2), considers a resonance between the orbital frequency and twice the radial epicyclic frequency,
\begin{align}
f_1=f_\phi,\qquad f_2=2f_r\,.
\end{align}
The third variant, Keplerian Resonance Model 3 (KRM3), assumes a resonance between the orbital frequency and three times the radial epicyclic frequency,
\begin{align}
f_1=3f_r,\qquad f_2=f_\phi\,.
\end{align}
For each of these variants, the model dependent upper and lower HFQPO frequencies are identified with particular combinations of the fundamental geodesic frequencies. Consequently, measurements of twin-peak HFQPOs can be used to constrain the properties of the central compact object, including its mass, spin, and, in modified-gravity scenarios, possible deviations from the Kerr spacetime. 

\vspace{0.5cm}
\item {\bf Warped Disk Oscillation Model (WDOM)}
\vspace{0.5cm}

The warped-disk resonance model attributes HFQPOs to nonlinear interactions between oscillation modes in a relativistic accretion disk and a global disk deformation in the form of a warp \cite{10.1093/pasj/53.1.1,kato2004wave2,kato2004resonant2,kato2008resonant2,kato2005vertical2}. Through this coupling, energy can be transferred between different modes, leading to resonant excitation at preferred locations in the disk.
The model predicts three types of resonances. Horizontal resonances can excite both p-mode and g-mode oscillations, whereas vertical resonances excite only g-mode oscillations \cite{kato2004wave2}. These resonances occur when the oscillation frequencies satisfy specific relationships involving the orbital and epicyclic frequencies of the disk. The model dependent HFQPOs are identified as $f_1 = 2 f_{\phi} - f_r$ and $f_2 = 2 ( f_{\phi} - f_r )$.
A key ingredient of the model is the relativistic behavior of the radial epicyclic frequency, which varies non-monotonically with radius. This feature creates distinct resonant radii where nonlinear mode coupling is enhanced, providing a natural mechanism for generating the characteristic frequencies and frequency ratios observed in HFQPOs \cite{kato2004wave2}. 

Despite its theoretical appeal, the model relies on the presence of a long-lived large-scale warp in the inner accretion disk. Since the formation and persistence of such a warped geometry remain uncertain in realistic accretion flows, the astrophysical applicability of the model is limited \cite{Torok:2011qy,Yagi:2016jml}. Nevertheless, it remains an important framework for studying nonlinear disk oscillations in strong gravitational fields.

\newpage

\item {\bf Non-axisymmetric Disc Oscillation Model (NADO)}
\vspace{0.5cm}

Non-axisymmetric disk-oscillation models explain HFQPOs through resonances involving non-axisymmetric oscillation modes in the accretion flow \cite{2004ApJ...617L..45B,2005AN....326..849B,2005ragt.meet...39B,Torok:2011qy,Kotrlova:2020pqy}. These models can be viewed as extensions of the relativistic precession model that incorporate non-geodesic effects arising from pressure forces in the accretion flow.
To account for these effects, the accretion flow is modeled as a slightly non-slender, pressure-supported perfect-fluid torus rather than a collection of test particles moving on geodesics \cite{Torok:2015tpu,Sramkova:2015bha,Kotrlova:2020pqy}. The resulting oscillation spectrum includes both axisymmetric and non-axisymmetric modes, characterized by the azimuthal wave number ($m$).

Two variants of the model have been proposed. The first, denoted NADO1, associates the observed frequencies with a resonance between the vertical epicyclic frequency, ($f_1=f_\theta$), and the ($m=-1$) non-axisymmetric radial epicyclic frequency, ($f_2=f_\phi-f_r$) \cite{2004ApJ...617L..45B,2005AN....326..849B,2005ragt.meet...39B}. The second, NADO2, invokes a resonance between the ($m=-1$) radial mode, ($f_2=f_\phi-f_r$), and the ($m=-2$) vertical mode, ($f_1=2 f_\phi-f_\theta$) \cite{Torok:2010rk,Torok:2011qy,Kotrlova:2020pqy}.
Although these models provide a natural framework for incorporating pressure effects and non-axisymmetric oscillations, their physical basis remains uncertain. In particular, the mechanisms responsible for coupling an axisymmetric mode to a non-axisymmetric mode, or coupling pairs of non-axisymmetric modes, are not yet well understood, limiting the predictive power of the model \cite{Horak:2008zg,Torok:2011qy}.

\end{enumerate}

\section{Evaluating model-predicted HFQPOs against observations}
\label{Sec4}
In this section, we test the HFQPO frequencies given by the aforesaid nine resonance models and compare them with the observational data shown in \autoref{t1}. We employ the $\chi^2$ minimization method to measure the deviation between the theoretically predicted and observed QPO frequencies. We define the $\chi^2$ function as follows,
\begin{eqnarray}
\label{e1}
\chi_i^{2}(l,\alpha) =\frac{[f_{up1,i} - f_{1}(l,\alpha, a_{\rm min}, M_{\rm min}, r_{\rm em,min})]^2}{\sigma_{f_{up1},i}^2} 
+  \frac{[f_{up2,i} - f_{2}(l,\alpha, a_{\rm min}, M_{\rm min}, r_{\rm em,min})]^2}{\sigma_{f_{up2},i}^2}\,.
\label{chi}
\end{eqnarray}
This expression measures the deviation between the model-predicted and observed HFQPO frequencies. \autoref{t1} lists the black hole sources that exhibit HFQPOs in their power spectra. In this framework,  $f_{up1,i}$ and $f_{up2,i}$ denote the upper and lower HFQPO frequencies of the 
i-th source, as given in \autoref{t1}. The associated uncertainties, $\sigma_{f_{\mathrm{up1},i}}$ and $\sigma_{f_{\mathrm{up2},i}}$ correspond to the observational errors of these frequencies and are also reported in \autoref{t1}. $f_1$ and $f_2$ are the theoretical, model-dependent HFQPO frequencies which can be determined from $f_\phi$, $f_\theta$ and $f_r$ as discussed in \autoref{sec3}. We outline the stages of our analysis as follows:
\begin{itemize}
    \item At first, we select any one of the nine aforesaid models, in which $f_1$ and $f_2$ are functions of $l$, $\alpha$, $M$, $a$ and the emission radius $r_{\rm em}$ (the circular radius at which the HFQPO frequencies are generated), as discussed in the previous sections. 
    \item Now, from \autoref{t1} we select a black hole source whose mass has been estimated from independent observation.
    \item We fix the parameters $l$ and $\alpha$, subject to the constraint $\alpha l \leq 2$ and $\alpha>\alpha_{\rm crit}$ (for a given $l$) to ensure that the strong energy condition (SEC) is satisfied. For these fixed values of $l$ and $\alpha$, the spin parameter $a$ is allowed to vary within the ranges $-a_{\rm \max} \leq a \leq a_{\rm \max}$ or $-a_{\rm SEC} \leq a \leq a_{\rm SEC}$ (see \autoref{s2}), such that the event horizon remains real and positive and SEC holds everywhere outside the event horizon.
    \item The black hole mass $M$ is allowed to vary around the measured value $M_0$, within the range $(M_0 - \Delta M) \leq M \leq (M_0 + \Delta M)$, where $\Delta M$ represents the uncertainty in the mass measurement given in \autoref{t1}. 
    \item Our goal is to compute the frequencies $f_1$ and $f_2$. For each chosen value of the black hole mass $M$, the emission radius $r_{em}$ is allowed to vary from the radius of the marginally stable orbit, $r_{\rm ms}(a,l,\alpha)$, up to $r_{\rm ms}(a,l,\alpha) + 20\,R_g$. Here, $R_g = GM/c^2$ denotes the gravitational radius.
   \item For a fixed choice of $l$ and $\alpha$, we evaluate $\chi_i^2$ for each possible combination of $M$, $a$, and $r_{\rm em}$ corresponding to the chosen source. The parameter set $(M,a,r_{\rm em})$ that produces the smallest value of $\chi_i^2$ is taken as the best fit, giving the preferred values of the mass, spin, and emission radius for the given source for that particular choice of $l$ and $\alpha$. These best-fit parameters are denoted by $M_{\rm \min}$, $a_{\rm \min}$, and $r_{\rm em, \min}$ in \autoref{e1}.
   \item We repeat the same steps for different choices of $l$ and $\alpha$. This allows us to track how the value of $\chi^2$ changes with $l$ and $\alpha$ for the chosen black hole source.
   \item We carry out the same analysis for all the black hole sources listed in \autoref{t1} within the chosen model.
   \item Finally, we run the same analysis for the other models as well. The smallest value of $\chi^2$ tells us which values of $l$ and $\alpha$ work best, and it also gives the corresponding best-fit mass $M_{\rm \min}$ and spin $a_{\rm \min}$ for each black hole source within the domain of the particular model.
   \end{itemize}

Now, we will discuss our results for each black hole listed in \autoref{t1} where we have tried to constrain $l$ and $\alpha$ using the aforesaid nine resonance models.

\subsection{GRO J1655-40}   
Here we consider the source GRO J1655-40 and try to constrain ($l-\alpha$) parameter space based on the nine resonance models.
\subsubsection{Parametric Resonance Model (PRM)}

\begin{figure}[hbt!]
	\centering
	\subfloat[]
    {\includegraphics[width=0.34\linewidth]{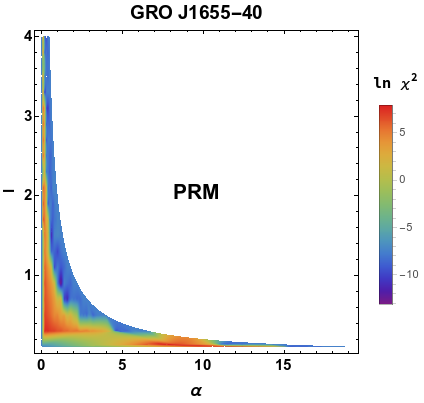}\label{a11}}
	\hfill
	\subfloat[]
    {\includegraphics[width=0.32\linewidth]{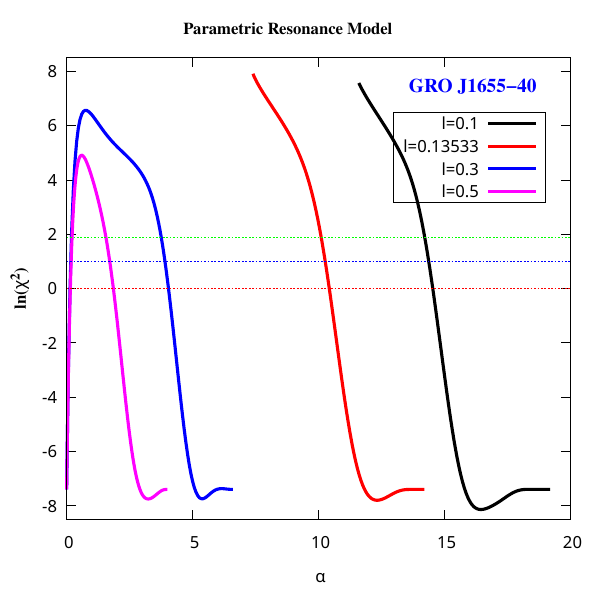}\label{a12}}
	\hfill
	\subfloat[]
    {\includegraphics[width=0.32\linewidth]{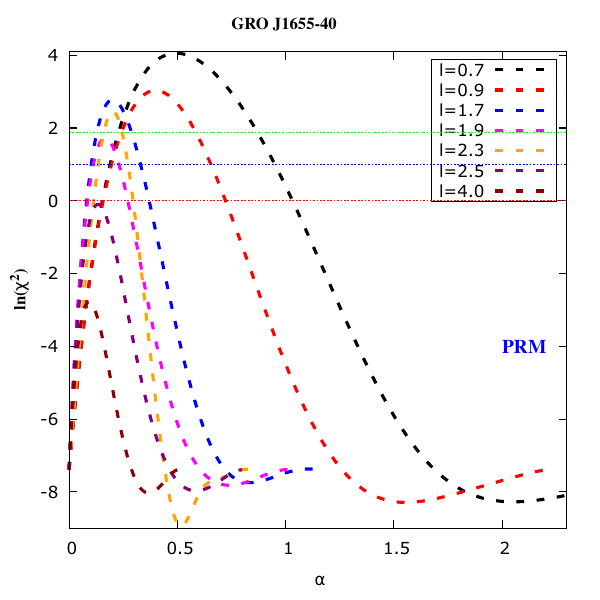}\label{a13}}
	\caption{(a) Density plot showing the variation of $\ln{\chi^{2}}$ with the $(\alpha,l)$ parameter space for the X-ray binary GRO~J1655-40 using the Parametric Resonance Model (PRM), 
    (b) Variation of $\ln \chi^{2}$ with the hair parameter $\alpha$ for selected lower values of $l = 0.1, 0.13533, 0.3,$ and $0.5$. The red, blue, and green horizontal dotted lines represent the $1\sigma$, $2\sigma$, and $3\sigma$ confidence limits, respectively. For  $l=0.1$ and $l=e^{-2}$, values of $\alpha$ near the critical value $\alpha_{\rm crit}$ are excluded outside the $3\sigma$ confidence level, while for $l=0.3$ and $l=0.5$, certain intermediate ranges of $\alpha$ are disfavored . (c) Variation of $\ln \chi^{2}$ with $\alpha$ for comparatively higher values of $l = 0.7,0.9,1.7,1.9,2.3, 2.5,$ and $4.0$. For $l>2.5$, all combinations of $l$ and $\alpha$ are allowed within the 1$\sigma$ limit. }
    \label{a1}
\end{figure}

In \autoref{a11}, we have shown $\ln\chi^2$ as a function of $l$ and $\alpha$ considering the PRM model for GRO J1655-40. Regions with lower values of $\ln\chi^2$, shown by darker blue shades in the color bar, indicate a better agreement between the model predictions and the observed HFQPO frequencies, whereas regions with higher $\ln\chi^2$, shown by yellow–red shades, correspond to poorer fits. In our analysis, we consider the parameter $l$ in the range $0.1 \leq l \leq 4$, sampled in steps of $0.2$. For each chosen value of $l$, the parameter $\alpha$ is varied from a lower limit $\alpha_0$ up to an upper limit $\alpha_{\rm \max} = 2/l$. We use a step size of $0.1$ for $\alpha$ when $l \geq 1.7$, and a step size of $0.2$ when  $0.1 \leq l < 1.7$. This setup is applied to all the black hole sources considered in this work and for all models. Note that for very small values of $l$ ($l < 0.3$), the lower bound $\alpha_0$ is taken to be the critical value $\alpha_{\rm crit}$, while for larger $l$ it is set to zero (as $\alpha_{\rm crit}$ becomes negative for $l>2/e^2$). In \autoref{a12} and \autoref{a13}, we plot how $\ln\chi^2$ changes with $\alpha$ for different values of $l$. For $l = 0.1$ and $l = e^{-2} \simeq 0.13533$, the parameter $\alpha$ is not varied starting from $\alpha = 0$, as shown in \autoref{a12}. This is because the strong energy condition is satisfied only for $\alpha \geq \alpha_{\rm crit}$, where $\alpha_{\rm crit} \simeq 11.4$ for $l = 0.1$ and $\alpha_{\rm crit} \simeq 7.4$ for $l = e^{-2}$. Accordingly, $\alpha$ is scanned from $\alpha_{\rm crit}$ up to the upper limit $\alpha_{\rm \max} \simeq 2/l$. SEC is not satisfied for $l=0$, for any choice of $\alpha$, hence we abstain ourselves to include this in our work. The red, blue, and green dotted lines indicate the $1\sigma$, $2\sigma$, and $3\sigma$ confidence levels, respectively. These levels correspond to $\chi^2_{\rm \min} + 1$, $\chi^2_{\rm \min} + 2.71$, and $\chi^2_{\rm \min} + 6.63$ \cite{avni1976energy}. In \autoref{a12}, for $l \lesssim e^{-2}$, values of $\alpha$ close to $\alpha_{\rm crit}$ are ruled out at more than the $3\sigma$ level, as indicated by the green dotted curve in \autoref{a12}. 
When $l \sim e^{-2}$ and $l \sim 0.1$, the range of $\alpha$ allowed within the $1\sigma$ confidence level lies roughly in $10 \lesssim \alpha \lesssim \alpha_{\rm \max} = 2/l$ and $15 \lesssim \alpha \lesssim \alpha_{\rm \max} = 2/l$, respectively. 
For intermediate values of $l$ in the range $0.3 \lesssim l \lesssim 2.3$ (see \autoref{a12} and \autoref{a13}), mid-range values of $\alpha$ are less favored by the data, with the fits improving toward either small $\alpha$ or values close to the upper limit $\alpha_{\rm \max}$. For $l \gtrsim 2.5$, the entire range of $\alpha$ from $0 \lesssim \alpha \lesssim \alpha_{\rm \max}$ is allowed within the $1\sigma$ confidence level. Hence, when we consider the PRM model, we can describe the BH GRO J1655-40 either by the Kerr model or by the hairy Kerr model with $\alpha\to 0$ or $l\alpha$ close to 2.

\subsubsection{Forced Resonance Model 1 (FRM1)}
\begin{figure}[hbt!]
		\centering
		\subfloat[]
       {\includegraphics[width=0.48\linewidth, height=7cm]{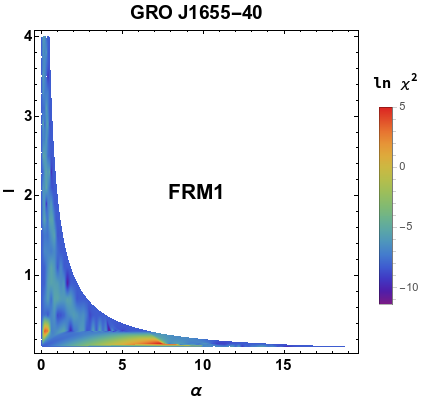}\label{a21}}
		\hfill
		\subfloat[]
        {\includegraphics[width=0.48\linewidth, height=7cm]{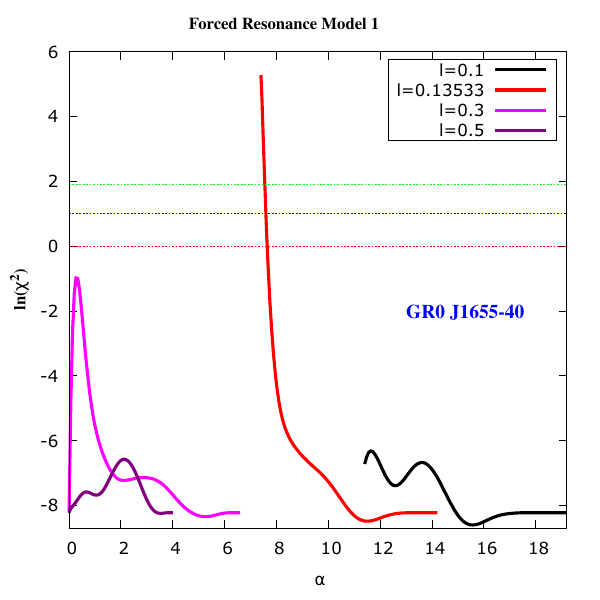}\label{a22}}
		\caption{(a) Density plot showing the variation of $\ln \chi^{2}$ with the $(\alpha,l)$ parameter space for the X-ray binary GRO~J1655-40 using the Forced Resonance Model 1 (FRM1) (b) Variation of $\ln \chi^{2}$ with the hair parameter $\alpha$ for selected values of the parameter $l = 0.1,0.13533,0.3,$ and $0.5$. The red, blue, and green horizontal dotted lines represent the $1\sigma$, $2\sigma$, and $3\sigma$ confidence limits, respectively. For  $l=0.13533$, values of $\alpha$ close to the critical value $\alpha_{\rm crit}$ are excluded outside the $3\sigma$ confidence level. This model cannot strongly constrain the $l-\alpha$ parameter space for the source GRO J1655-40. }
        \label{a2}
	\end{figure}
    
In \autoref{a21}, we have shown $\ln\chi^2$ as a function of $l$ and $\alpha$ considering FRM1 model for the BH GRO J1655-40.
As the blue shades in the color bar represent the best fit, it is clear from \autoref{a21} that, for almost all combinations of $l$ and $\alpha$, the theoretical value is very close to the observational data. In \autoref{a22}, we show changes of  $\ln\chi^2$ with $\alpha$ for different values of $l$. 3$\sigma$ interval can exclude only $l=0.13533=e^{-2}, \alpha=\alpha_{\rm crit}$. All other values of $l$ and $\alpha$ are allowed within the 1$\sigma$ interval. Hence, this model cannot constrain the $l-\alpha$ parameter space using the HFQPO data of GRO J1655-40.

\subsubsection{Forced Resonance Model 2 (FRM2), Warped Disc Oscillation Model (WDOM)}
\begin{figure}[hbt!]
		\centering
		\subfloat[]
        {\includegraphics[width=0.48\linewidth, height=7cm]{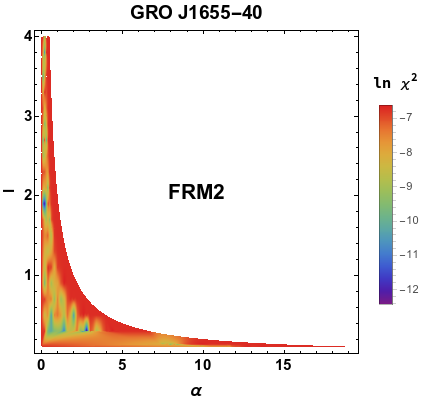}\label{a31}}
		\hfill
		\subfloat[]
        {\includegraphics[width=0.48\linewidth, height=7cm]{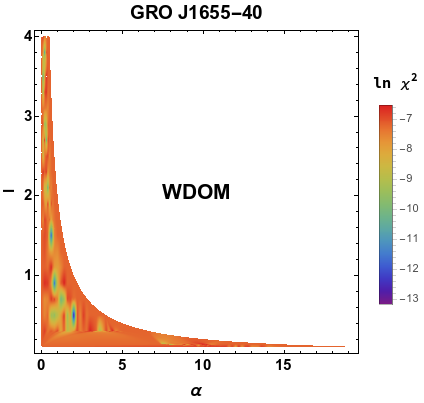}\label{a32}}
		\caption{Density plot showing the variation of $\ln \chi^{2}$ with the $(\alpha,l)$ parameter space for the black hole source GRO J1655-40 using the (a) Forced Resonance Model 2 (FRM2), and the (b) Warped Disc Oscillation Model (WDOM). It is evident that this deviation remains very small and shows only weak variation across the parameter space, leading to a broad nearly degenerate allowed region. Consequently, no constraints on the model parameters $\alpha$ and $l$ can be established for the source GRO J1655-40 within the $1\sigma$/$3\sigma$ confidence level using these models.}
        \label{a3}
	\end{figure}
    We next test the models FRM2 and WDOM with the HFQPO data of GRO J1655-40.
From \autoref{a31} and \autoref{a32} it is clear that the variation of $l$ and $\alpha$ have very less effect on the $\chi^2$ value which is also very small. Therefore, based on these models both the Kerr and the hairy Kerr scenario seem to be equally favored for the source GRO J1655-40.

\subsubsection{Keplerian Resonance Model 1 (KRM1)}
\begin{figure}[hbt!]
	\centering
	\subfloat[]
    {\includegraphics[width=0.34\linewidth]{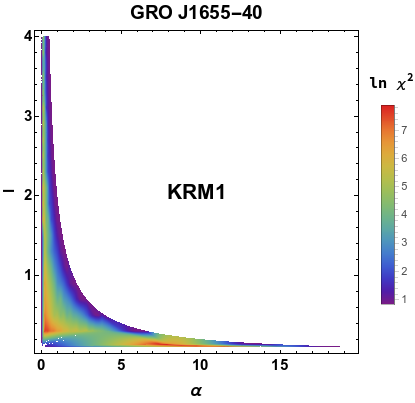}\label{a41}}
	\hfill
	\subfloat[]
    {\includegraphics[width=0.32\linewidth]{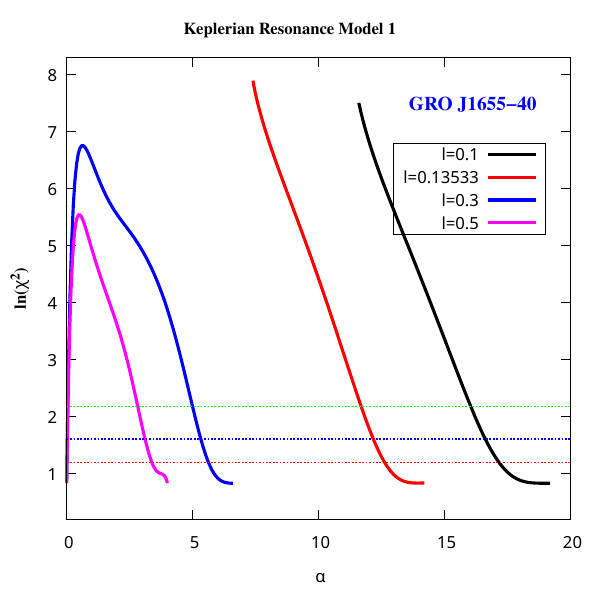}\label{a42}}
	\hfill
	\subfloat[]
    {\includegraphics[width=0.32\linewidth]{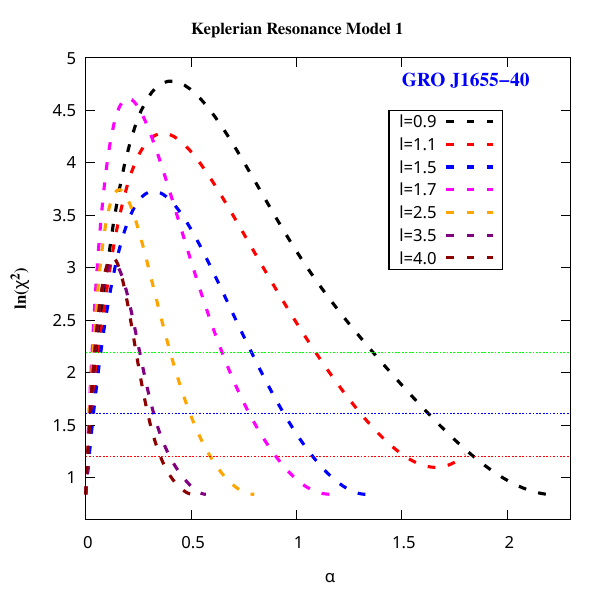}\label{a43}}
	\caption{(a) Density plot showing the variation of $\ln \chi^{2}$ with the $(\alpha,l)$ parameter space for the black hole binary GRO~J1655-40 using the Keplerian Resonance Model 1 (KRM1) (b) Variation of $\ln \chi^{2}$ with the hair parameter $\alpha$ for selected lower values of $l = 0.1,0.13533,0.3,$ and $0.5$. The red, blue, and green horizontal dotted lines represent the $1\sigma$, $2\sigma$, and $3\sigma$ confidence limits, respectively. For $l=0.1$ and $l=0.13533$, values of $\alpha$ near the critical value $\alpha_{\rm crit}$ are excluded outside the $3\sigma$ confidence level, while for $l=0.3$ and $l=0.5$ certain intermediate ranges are disfavored. (c) Variation of $\ln \chi^{2}$ with $\alpha$ for comparatively higher values of $l = 0.9,1.1,1.5,1.7,2.5, 3.5,$ and $4.0$. Intermediate $\alpha$ values are not allowed within $1\sigma$ limit for all these $l$ values.  }
    \label{a4}
\end{figure}

We next test the viability of the KRM1 model using the HFQPO data of GRO J1655-40.
 In \autoref{a41}, we show the variation of $\ln\chi^2$ with $(l, \alpha)$ using the KRM1 model. 
 In \autoref{a42}, for $l \lesssim e^{-2}$, values of $\alpha$ close to $\alpha_{\rm crit}$ are ruled out at more than the $3\sigma$ level, as indicated by the green dotted curve in \autoref{a42}. 
When $l \sim e^{-2}$ and $l \sim 0.1$, the range of $\alpha$ allowed within the $1\sigma$ confidence level lies roughly in $13 \lesssim \alpha \lesssim \alpha_{\rm \max} = 2/l$ and $17 \lesssim \alpha \lesssim \alpha_{\rm \max} = 2/l$, respectively. For higher values of $l$ (see \autoref{a42} and \autoref{a43}), the data prefer either small values of 
$\alpha$ or values close to the upper limit $\alpha_{\rm max}$, while intermediate $\alpha$
 values exhibit considerably higher $\chi^2$. Using this model and the HFQPO data for the source GRO J1655-40, the Kerr scenario or the hairy Kerr scenario with $\alpha l$ close to 2 seem to be equally favored.
 
\subsubsection{Keplerian Resonance Model 2 (KRM2)}
\begin{figure}[hbt!]
		\centering
		\subfloat[]
        {\includegraphics[width=0.48\linewidth, height=7cm]{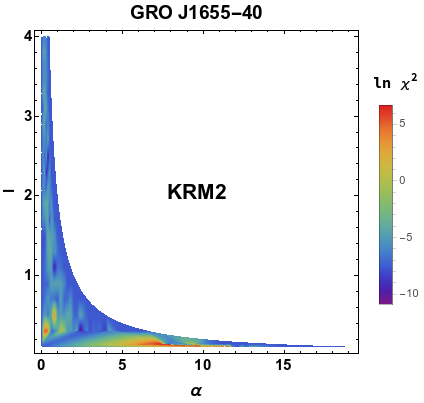}\label{a51}}
		\hfill
		\subfloat[]
        {\includegraphics[width=0.48\linewidth, height=7cm]{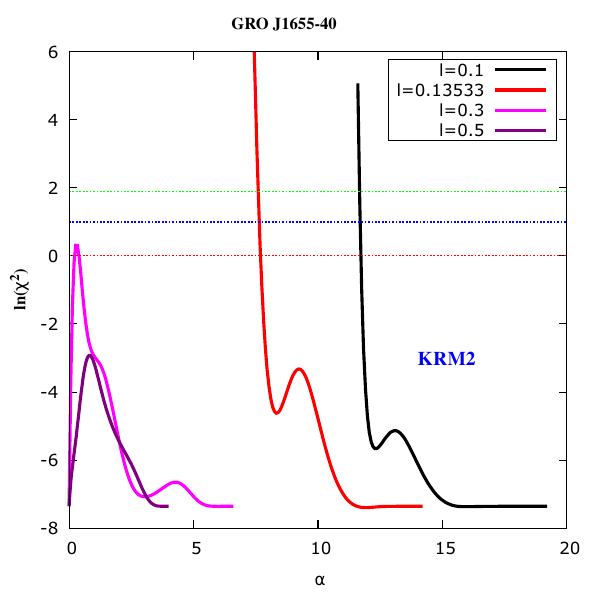}\label{a52}}
		\caption{(a)  Density plot showing the variation of $\ln \chi^{2}$ with the $(\alpha,l)$ parameter space for the black hole binary GRO~J1655-40 using the Keplerian Resonance Model 2 (KRM2) (b) Variation of $\ln \chi^{2}$ with the hair parameter $\alpha$ for selected values of the parameter $l = 0.1,0.13533,0.3,$ and $0.5$. The red, blue, and green horizontal dotted lines represent the $1\sigma$, $2\sigma$, and $3\sigma$ confidence limits, respectively. For lower values of $l$ ($l=0.1$ and $0.13533$), values of $\alpha$ close to the critical value $\alpha_{\rm crit}$ are excluded outside the $3\sigma$ confidence level. For $l=0.3$, few intermediate values of $\alpha$ are excluded outside $1\sigma$. For still larger values of $l$, all allowed values of $\alpha$ are included within 1$\sigma$. }
        \label{a5}
	\end{figure}
    
In \autoref{a51}, we represent the dependence of $\ln\chi^2$ as a function of $l$ and $\alpha$ using the KRM2 model for GRO J1655-40. The best-fitting regions appear as blue shades in the contour plot. Hence, for almost all combinations of $l$ and $\alpha$, the model dependent QPO frequencies are in close agreement with the observational data. In \autoref{a52} , the plot illustrates the change in $\ln\chi^2$ as $\alpha$ varies for different $l$.  Only $l=0.13533,\alpha=\alpha_{\rm crit}$ and $l=0.1, \alpha=\alpha_{\rm crit}$  can be excluded outside the 3$\sigma$ interval. 
For higher $l$ values, all combinations of $l$ and $\alpha$ are allowed within 1$\sigma$ interval. Therefore, this model cannot strongly constrain the $l-\alpha$ parameter space assuming the HFQPO data for GRO J1655-40.

\subsubsection{Keplerian Resonance Model 3 (KRM3)}
\begin{figure}[hbt!]
		\centering
		\subfloat[]
        {\includegraphics[width=0.48\linewidth, height=6cm]{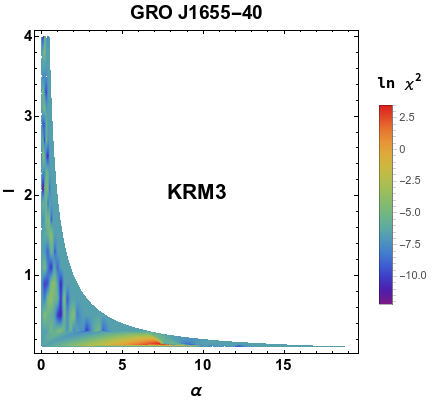}\label{a61}}
		\hfill
		\subfloat[]
        {\includegraphics[width=0.48\linewidth, height=6cm]{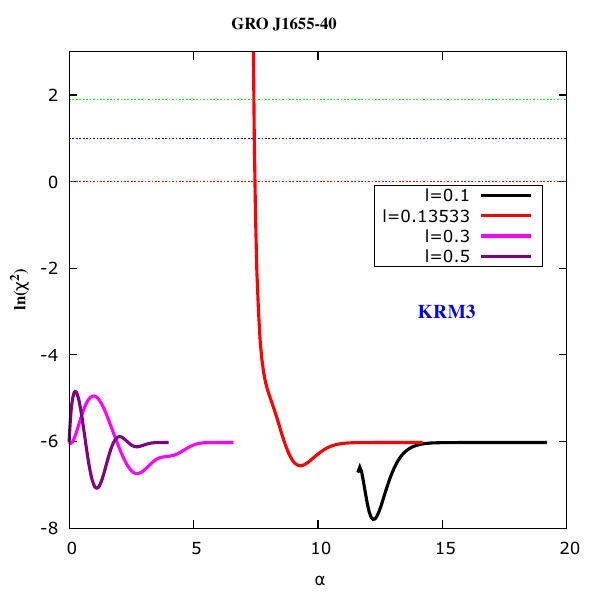}\label{a62}}
		\caption{(a) Density plot showing the variation of $\ln \chi^{2}$ with the $(\alpha,l)$ parameter space for the black hole source GRO~J1655-40 using the Keplerian Resonance Model 3 (KRM3) (b) Variation of $\ln \chi^{2}$ with the hair parameter $\alpha$ for selected values of the parameter $l = 0.1,0.13533,0.3,$ and $0.5$. The red, blue, and green horizontal dotted lines represent the $1\sigma$, $2\sigma$, and $3\sigma$ confidence limits, respectively. For  $l=0.13533$, values of $\alpha$ close to the critical value $\alpha_{\rm crit}$ are excluded outside the $3\sigma$ confidence level. For all other values of $l,\alpha$, the magnitude of $\chi^2$ is very small and the variation of $\ln \chi^{2}$ is also not substantial such that strong constraints on the $l-\alpha$ parameter space can be established.}
        \label{a6}
	\end{figure}
    
In \autoref{a61}, we present the dependence of $\ln\chi^2$ on the parameters $l$ and $\alpha$ taking into account the KRM3 model for GRO J1655-40. The blue shaded region corresponds to the best fit. As a result, the theoretical predictions based on the KRM3 model are mostly consistent with the observational data for almost all combinations of $l$ and $\alpha$, except for $l\simeq e^{-2}\simeq 0.13533$ and $\alpha\simeq \alpha_{\rm crit}$ (see \autoref{a62}). Hence, based on this model, the hairy Kerr BH with $l\simeq e^{-2}\simeq 0.13533$ and $\alpha\simeq \alpha_{\rm crit}$ are ruled out outside 3$\sigma$, however, for other $l,\alpha$ combinations, the data cannot distinguish the Kerr or the hairy Kerr scenario.

\subsubsection{Non-Axisymmetric Disc Oscillation Model 1 (NADO1)}
\begin{figure}[hbt!]
		\centering
		\subfloat[]
        {\includegraphics[width=0.48\linewidth, height=6cm]{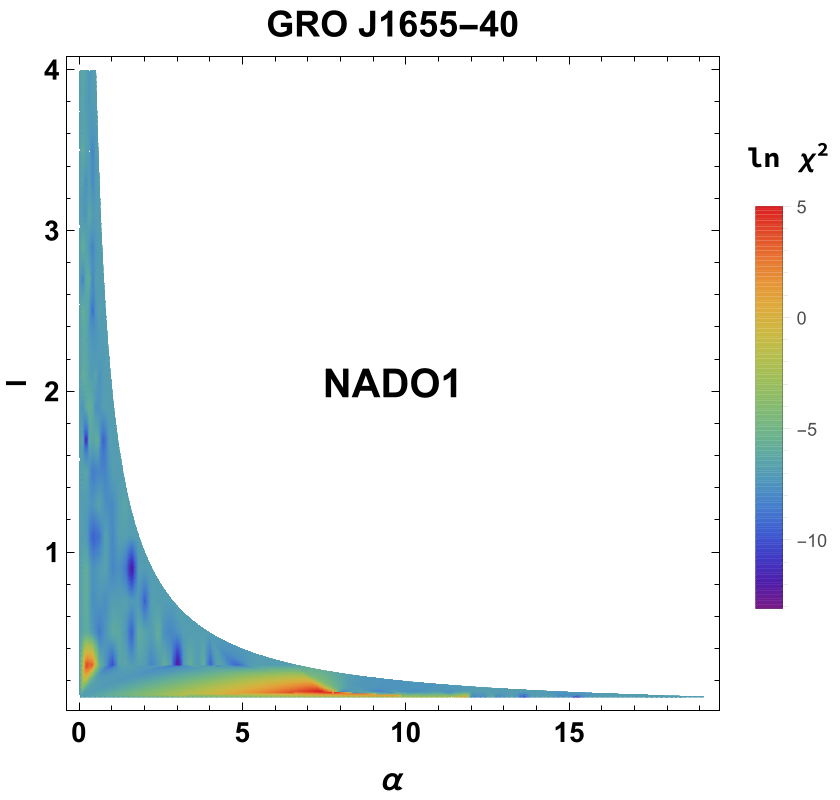}\label{a81}}
		\hfill
		\subfloat[]
        {\includegraphics[width=0.48\linewidth, height=6cm]{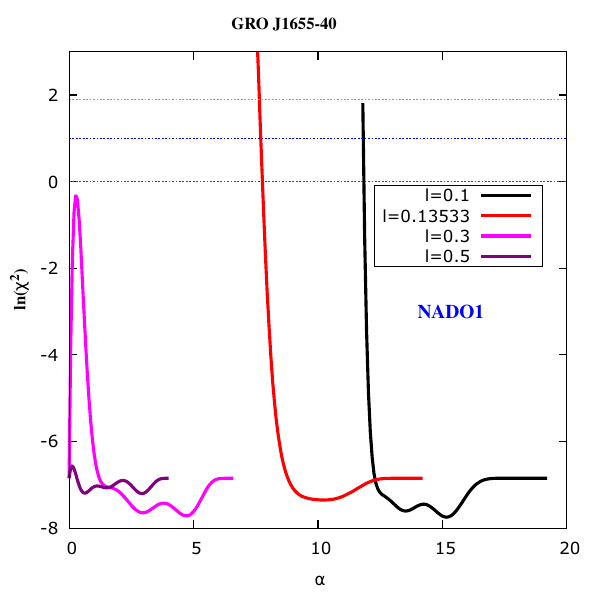}\label{a82}}
		\caption{(a)  Density plot showing the variation of $\ln \chi^{2}$ with the $(\alpha,l)$ parameter space for the black hole source GRO~J1655-40 using the Non Axisymmetric Disc Oscillation Model 1 (NADO1)(b) Variation of $\ln \chi^{2}$ with the hair parameter $\alpha$ for selected values of the parameter $l = 0.1,0.13533,0.3,$ and $0.5$. The red, blue, and green horizontal dotted lines represent the $1\sigma$, $2\sigma$, and $3\sigma$ confidence limits, respectively. For lower values of $l$ ($l=0.1$ and $0.13533$), values of $\alpha$ close to the critical value $\alpha_{\rm crit}$ are excluded outside the $3\sigma$ confidence level. For comparatively larger values (e.g $l\gtrsim0.3$),  the entire $l-\alpha$ parameter space is allowed within 1$\sigma$.}
        \label{a8}
	\end{figure}
    
We next test the viability of the model NADO1 in explaining the HFQPO data of GRO J1655-40.
In \autoref{a81}, we represent the dependence of $\ln\chi^2$ as a function of $l$ and $\alpha$. The best fit regions are denoted by blue shades in the contour plot. Hence, for almost all combinations of $l$ and $\alpha$, the theoretical (model predicted) HFQPO frequencies are nearly in agreement with the observational data. In \autoref{a82} , we show changes of  $\ln\chi^2$ with $\alpha$ for different values of $l$. Only $l=0.13533,\alpha=\alpha_{\rm crit}$ and $l=0.1, \alpha=\alpha_{\rm crit}$  can be excluded outside the 3$\sigma$ interval. All other combinations of $l$ and $\alpha$ are allowed within the 1$\sigma$ interval. Therefore, based on this model, the HFQPO data of GRO J1655-40 cannot distinguish between the Kerr and the hairy Kerr scenario (except for $l\lesssim e^{-2}$ and $\alpha\simeq \alpha_{\rm crit}$ which the data strongly rules out).

\subsubsection{Non-Axisymmetric Disc Oscillation Model 2 (NADO2)}  
\begin{figure}[hbt!]
		\centering
		\subfloat[]
        {\includegraphics[width=0.48\linewidth, height=6cm]{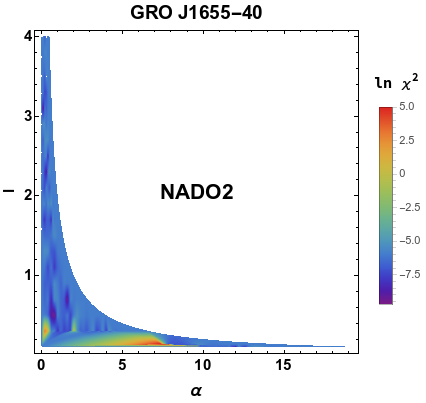}\label{a91}}
		\hfill
		\subfloat[]
        {\includegraphics[width=0.48\linewidth, height=6cm]{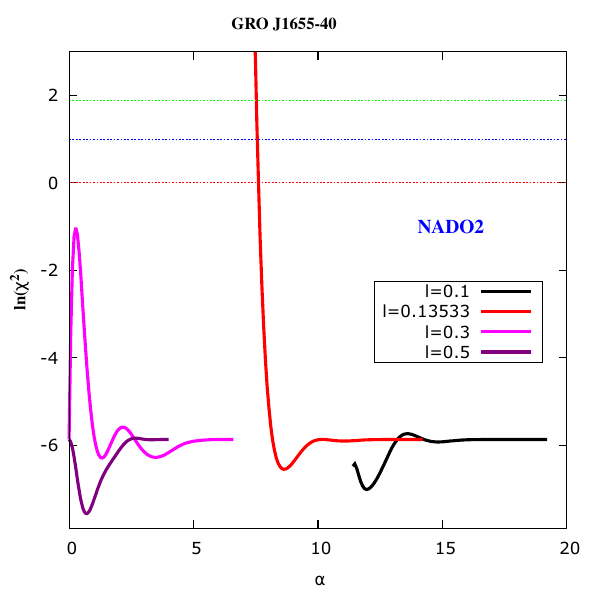}\label{a92}}
		\caption{(a)  Density plot showing the variation of $\ln \chi^{2}$ with the $(\alpha,l)$ parameter space for the black hole source GRO~J1655-40 using the Non Axisymmetric Disc Oscillation Model 2 (NADO2) (b) Variation of $\ln \chi^{2}$ with the hair parameter $\alpha$ for selected values of the parameter $l = 0.1,0.13533,0.3,$ and $0.5$. The red, blue, and green horizontal dotted lines represent the $1\sigma$, $2\sigma$, and $3\sigma$ confidence limits, respectively. For  $l=0.13533$, values of $\alpha$ close to the critical value $\alpha_{\rm crit}$ are excluded outside the $3\sigma$ confidence level. For comparatively larger values (e.g $l\gtrsim0.3$),  the entire $l-\alpha$ parameter space is allowed within 1$\sigma$.}
        \label{a9}
	\end{figure}

In \autoref{a91}, we show $\ln\chi^2$ as a function of $l$ and $\alpha$ considering the NADO2 model for GRO J1655-40. Once again, the blue shades in the color bar represent the best fit region. From the figure it is evident that for almost all combinations of $l$ and $\alpha$, the model predicted HFQPO frequency is mostly in agreement with the observational data. In \autoref{a92} , we show changes of  $\ln\chi^2$ with $\alpha$ for selected values of $l$. 3$\sigma$ interval can exclude only $l=0.13533, \alpha=\alpha_{\rm crit}$. All other values of $l$ and $\alpha$ are allowed within 1$\sigma$ interval. Hence based on this model, the HFQPO data of GRO J1655-40 equally favors the Kerr and the hairy Kerr scenario.

The best-fit model parameters for GRO J1655-40
corresponding to all the nine resonance models are summarized in \autoref{gro}. The results for the kinematic models studied in Paper I \cite{Dasgupta:2025blbe} are also reported in \autoref{gro} for completeness.
We postpone the discussion pertaining to the most suitable model for each source and whether the source exhibits a preference towards the Kerr or the hairy Kerr scenario to the next section. In what follows we present the plots illustrating the variation of $\chi^2$ with $l$ and $\alpha$ for the remaining five sources in \autoref{t1}. Since the discussion pertaining to the most favorable model parameters is similar to the present section, we do not discuss the plots, rather we present a summary of the results in the next section in \autoref{xte1}-\autoref{xte2}.

\newpage

\subsection{XTE J1550-564}
Here we present the variation of $\chi^2$ with $l$ and $\alpha$ for the source XTE J1550-564 considering the nine resonance models.
\begin{enumerate}
\item{ \bf{Parametric Resonance Model (PRM)}}
\begin{figure}[hbt!]
	\centering
	\subfloat[]
    {\includegraphics[width=0.34\linewidth, height=6.2cm]{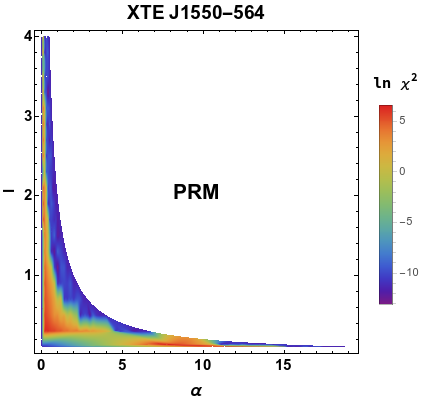}\label{b11}}
	\hfill
	\subfloat[]
    {\includegraphics[width=0.32\linewidth, height=6.2cm]{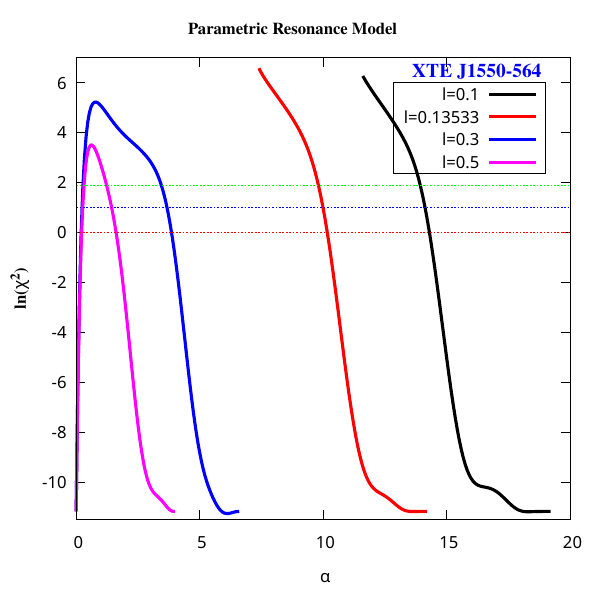}\label{b12}}
	\hfill
	\subfloat[]
    {\includegraphics[width=0.32\linewidth, height=6.2cm]{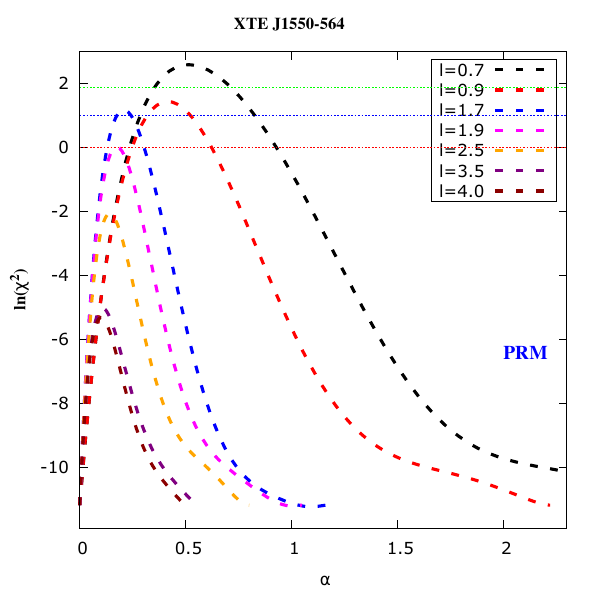}\label{b13}}
	\caption{(a) Density plot showing the variation of $\ln \chi^{2}$ with  $(\alpha,l)$ for the black hole source XTE J1550-564 using the Parametric Resonance Model  (PRM) (b) Variation of $\ln \chi^{2}$ with the hair parameter $\alpha$ for selected lower values of $l = 0.1, 0.13533, 0.3,$ and $0.5$. The red, blue, and green horizontal dotted lines represent the $1\sigma$, $2\sigma$, and $3\sigma$ confidence limits, respectively. For  $l=0.1$ and $l=0.13533$ ($e^{-2}$), values of $\alpha$ near the critical value $\alpha_{\rm crit}$ are excluded outside the $3\sigma$ confidence level, while for $l=0.3$ and $l=0.5$, certain intermediate ranges are disfavored. (c) Variation of $\ln \chi^{2}$ with $\alpha$ for comparatively higher values of $l = 0.7,0.9,1.7,1.9,2.5, 3.5,$ and $4.0$. For $l>1.8$, all combinations of $l$ and $\alpha$ are allowed within the $1\sigma$ limit.}
    \label{b1}
\end{figure}




\item{\bf{Forced Resonance Model 1 (FRM 1)}}
\begin{figure}[hbt!]
		\centering
		\subfloat[]
        {\includegraphics[width=0.48\linewidth, height=6.0cm]{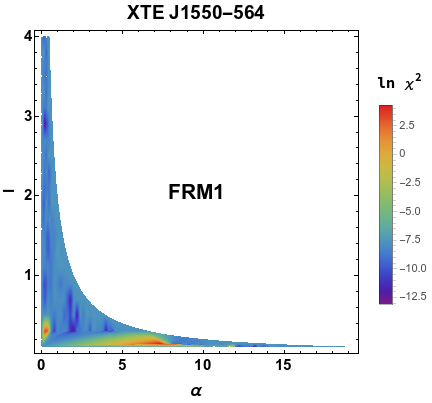}\label{b21}}
		\hfill
		\subfloat[]
        {\includegraphics[width=0.48\linewidth, height=6.0cm]{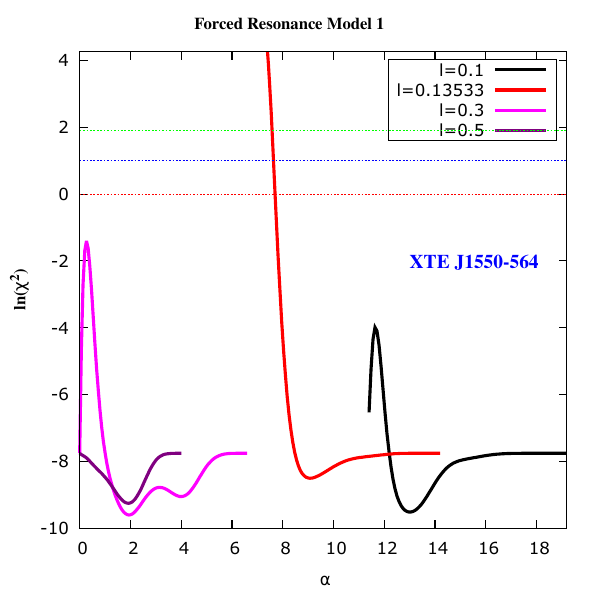}\label{b22}}
		\caption{(a) Density plot showing the variation of $\ln \chi^{2}$ with $(\alpha,l)$ for the black hole source XTE J1550-564 using the Forced Resonance Model 1 (FRM1) (b) Variation of $\ln \chi^{2}$ with the hair parameter $\alpha$ for selected values of the parameter $l = 0.1,0.13533,0.3,$ and $0.5$. The red, blue, and green horizontal dotted lines represent the $1\sigma$, $2\sigma$, and $3\sigma$ confidence limits, respectively. For  $l=0.13533$, values of $\alpha\to\alpha_{\rm crit}$ are excluded outside the $3\sigma$ confidence level. For all other values of $l$, the variation of $\chi^{2}$ with $\alpha$ is not substantial to adequately constrain the $l-\alpha$ parameter space.}
        \label{b2}
	\end{figure}



\item{\bf{Forced Resonance Model 2 (FRM2) and Warped Disc Oscillation Model (WDOM)}}
\begin{figure}[H]
		\centering
		\subfloat[]
        {\includegraphics[width=0.48\linewidth]{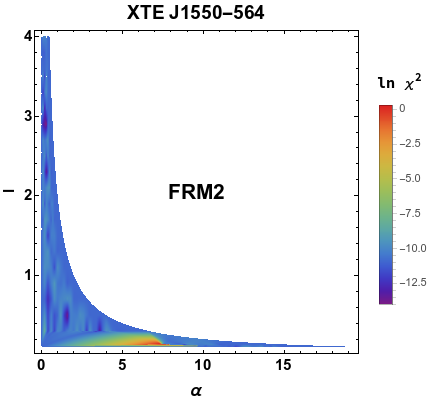}\label{b31}}
		\hfill
		\subfloat[]
        {\includegraphics[width=0.48\linewidth]{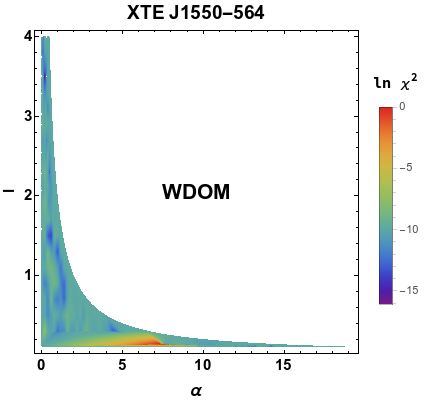}\label{b32}}
		\caption{Density plot showing the variation of $\ln \chi^{2}$ with  $(\alpha,l)$ parameter space for the black hole source XTE J1550-564 using the (a) Forced Resonance Model 2 (FRM2), and the (b) Warped Disc Oscillation Model (WDOM). The $\chi^2$ values are very small and the variation of $\chi^2$ with $l,\alpha$ is also not substantial to impose strong constraints on the $l-\alpha$ parameter space. 
        Consequently, based on these two models, the source cannot distinguish between the Kerr and the hairy Kerr scenario.  
        }
        \label{b3}
	\end{figure}

\item{\bf{Keplerian Resonance Model 1 (KRM1)}}
\begin{figure}[H]
	\centering
	\subfloat[]
    {\includegraphics[width=0.34\linewidth, height=6.5cm]{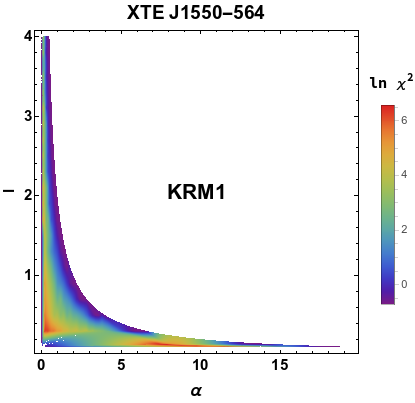}\label{b41}}
	\hfill
	\subfloat[]
    {\includegraphics[width=0.32\linewidth, height=6.5cm]{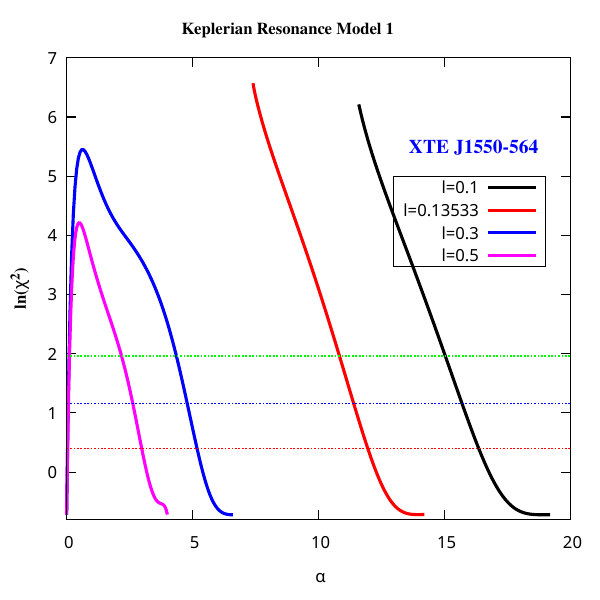}\label{b42}}
	\hfill
	\subfloat[]
    {\includegraphics[width=0.32\linewidth, height=6.5cm]{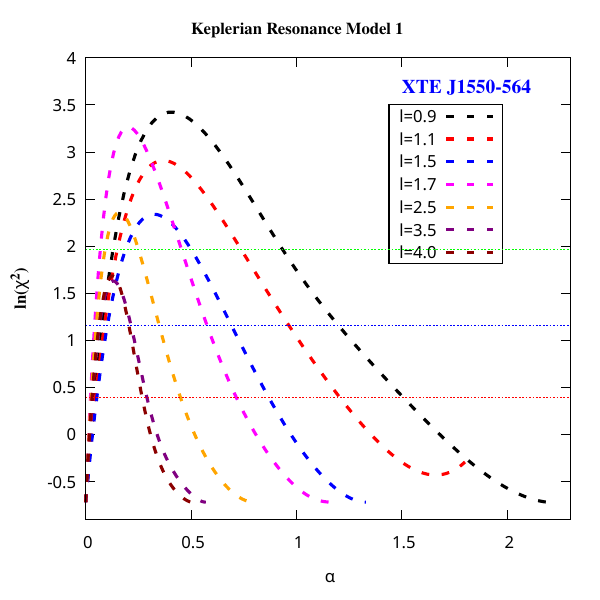}\label{b43}}
	\caption{(a) Density plot showing the variation of $\ln \chi^{2}$ with the $(\alpha,l)$ parameter space for the black hole source XTE J1550-564 using the Keplerian Resonance Model 1 (KRM1) (b) Variation of $\ln \chi^{2}$ with the hair parameter $\alpha$ for selected lower values of $l = 0.1,0.13533,0.3,$ and $0.5$. The red, blue, and green horizontal dotted lines represent the $1\sigma$, $2\sigma$, and $3\sigma$ confidence limits, respectively. For $l=0.1$ and $l=0.13533$, values of $\alpha$ near the critical value $\alpha_{\rm crit}$ are excluded outside the $3\sigma$ confidence level, while for $l=0.3$ and $l=0.5$ certain intermediate ranges are disfavored. (c) Variation of $\ln \chi^{2}$ with $\alpha$ for comparatively higher values of $l = 0.9,1.1,1.5,1.7,2.5, 3.5,$ and $4.0$. Intermediate $\alpha$ values are not allowed within the 1$\sigma$ limit for all $l$ values.}
    \label{b4}
\end{figure}


\item{\bf{Keplerian Resonance Model 2 (KRM2)}}
\begin{figure}[H]
		\centering
		\subfloat[]
        {\includegraphics[width=0.48\linewidth, height=7.5cm]{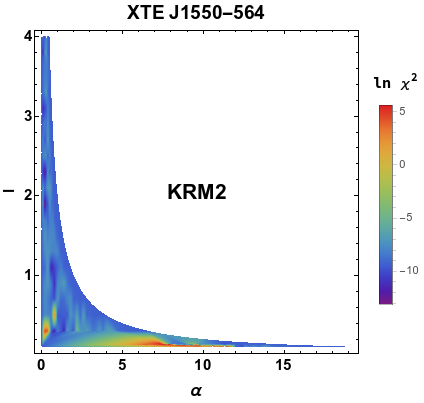}\label{b51}}
		\hfill
		\subfloat[]
        {\includegraphics[width=0.48\linewidth, height=7.5cm]{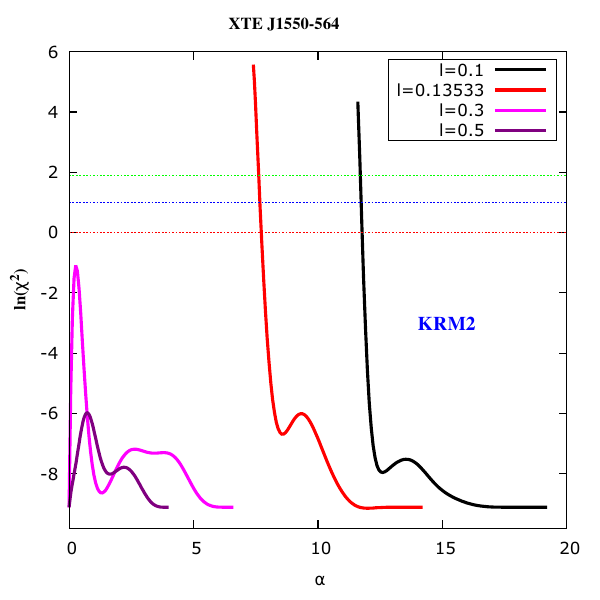}\label{b52}}
		\caption{(a) Density plot showing the variation of $\ln \chi^{2}$ with the $(\alpha,l)$ parameter space for the black hole source XTE J1550-564 using the Keplerian Resonance Model 2 (KRM2) (b) Variation of $\ln \chi^{2}$ with the hair parameter $\alpha$ for selected values of the parameter $l = 0.1, 0.13533, 0.3$ and $0.5$. The red, blue, and green horizontal dotted lines represent the $1\sigma$, $2\sigma$, and $3\sigma$ confidence limits, respectively. For lower values of $l$ ($l\lesssim e^{-2}$), values of $\alpha$ close to the critical value $\alpha_{\rm crit}$ are excluded outside the $3\sigma$ confidence level. For comparatively larger values (e.g $l\gtrsim0.3$),  the magnitude of $\chi^2$ substantially decreases and the variation of $\ln \chi^{2}$ with $l,\alpha$ is also not substantial to establish strong constraints on the $l,\alpha$ parameter space.}
        \label{b5}
	\end{figure}

\item{\bf{Keplerian Resonance Model 3 (KRM3)}}
\begin{figure}[H]
		\centering
		\subfloat[]
        {\includegraphics[width=0.48\linewidth, height=7.5cm]{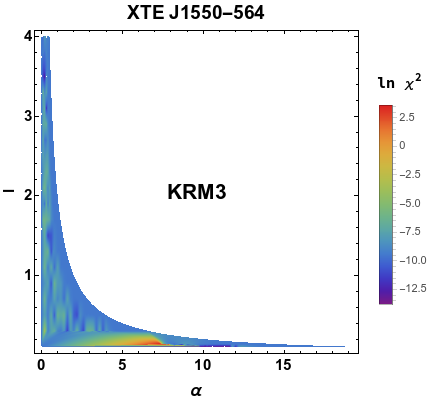}\label{b61}}
		\hfill
		\subfloat[]
        {\includegraphics[width=0.48\linewidth, height=7.5cm]{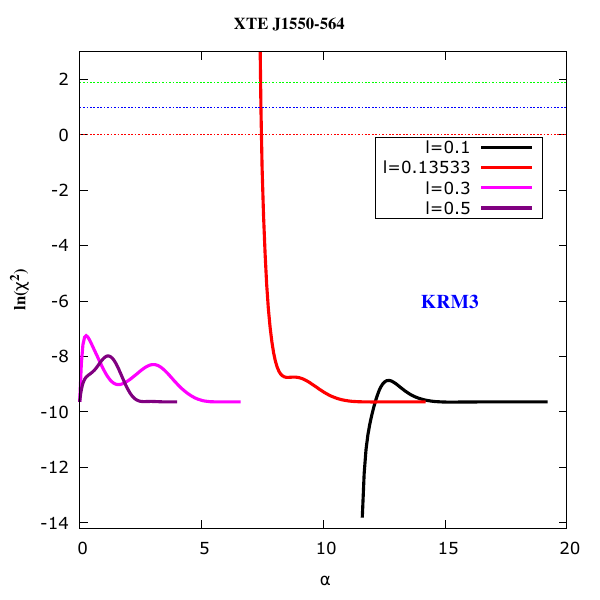}\label{b62}}
		\caption{(a) Density plot showing the variation of $\ln \chi^{2}$ with the $(\alpha,l)$ parameter space for the black hole source XTE J1550-564 using the model Keplerian Resonance Model 3 (KRM3)(b) Variation of $\ln \chi^{2}$ with the hair parameter $\alpha$ for selected values of the parameter $l = 0.1,0.13533,0.3,$ and $0.5$. The red, blue, and green horizontal dotted lines represent the $1\sigma$, $2\sigma$, and $3\sigma$ confidence limits, respectively. For  $l=0.13533$, $\alpha\simeq\alpha_{\rm crit}$ are excluded outside the $3\sigma$ confidence level. However, the data of XTE J1550-564 allow all other $l,\alpha$ and hence cannot generally distinguish between the Kerr and the hairy Kerr scenario.}
        \label{b6}
	\end{figure}

\item{\bf{Non-Axisymmetric Disc Oscillation Model 1 (NADO1)}}
\begin{figure}[H]
		\centering
		\subfloat[]
        {\includegraphics[width=0.48\linewidth, height=7.5cm]{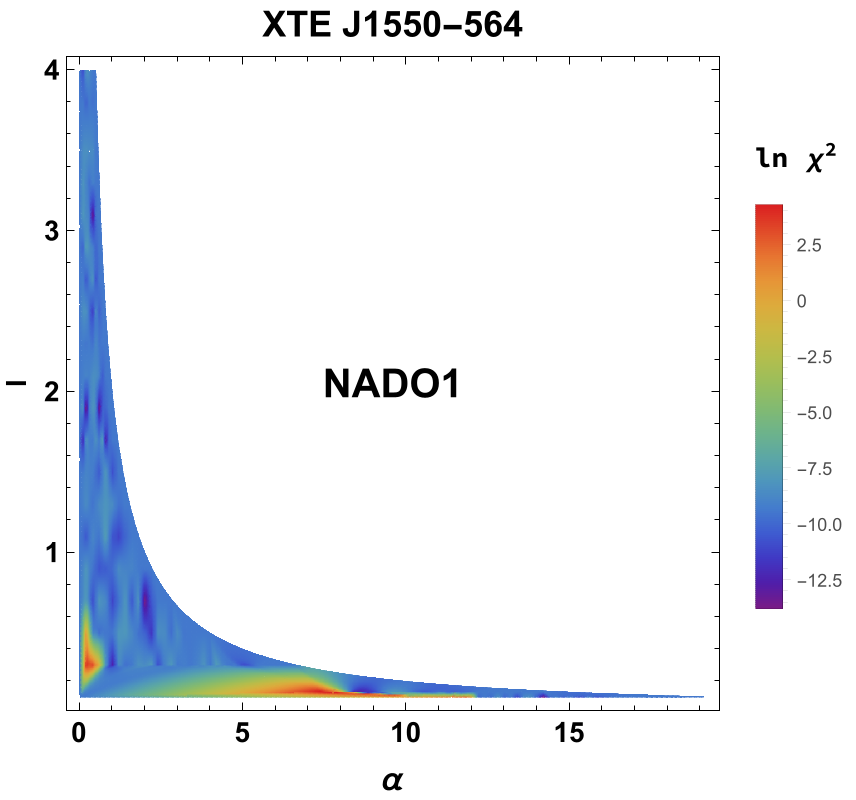}\label{b81}}
		\hfill
		\subfloat[]
        {\includegraphics[width=0.48\linewidth, height=7.5cm]{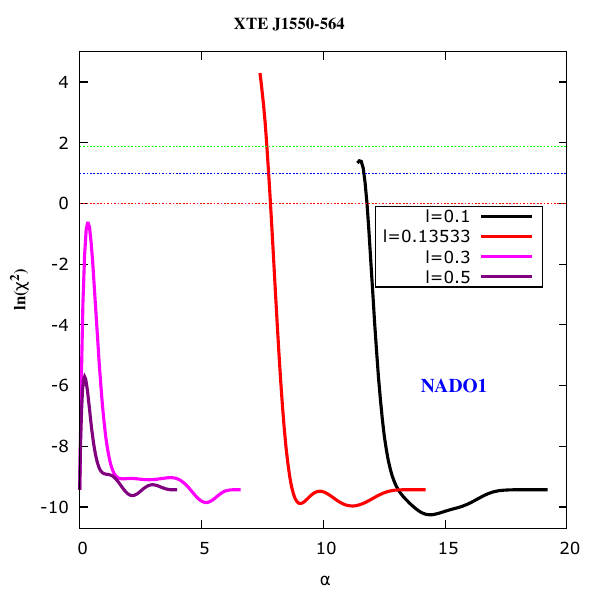}\label{b82}}
		\caption{(a) Variation of $\ln \chi^{2}$ with the $(\alpha,l)$ parameter space for the black hole source XTE J1550-564 using the Non Axisymmetric Disc Oscillation Model 1 (NADO1) (b) Variation of $\ln \chi^{2}$ with the hair parameter $\alpha$ for selected values of the parameter $l = 0.1, 0.13533, 0.3$ and $0.5$. The red, blue, and green horizontal dotted lines represent the $1\sigma$, $2\sigma$, and $3\sigma$ confidence limits, respectively. For lower values of $l$ ($l=0.1$ and $0.13533$), $\alpha\simeq \alpha_{\rm crit}$ are excluded outside the $3\sigma$ confidence level. For comparatively larger values (e.g, $l\gtrsim 0.3$), all values of $l-\alpha$ are within the 1$\sigma$ confidence interval. 
        }
        \label{b8}
	\end{figure}

\item{\bf{Non-Axisymmetric Disc Oscillation Model 2 (NADO2)}}
\begin{figure}[H]
		\centering
		\subfloat[]
        {\includegraphics[width=0.48\linewidth, height=7.5cm]{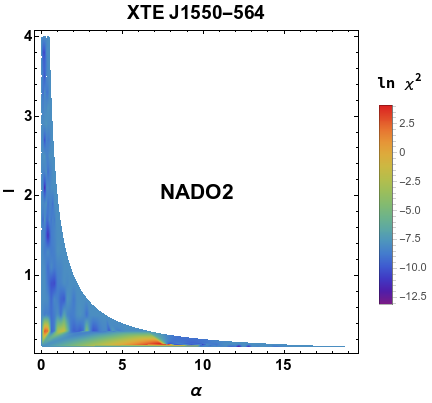}\label{b91}}
		\hfill
		\subfloat[]
        {\includegraphics[width=0.48\linewidth, height=7.5cm]{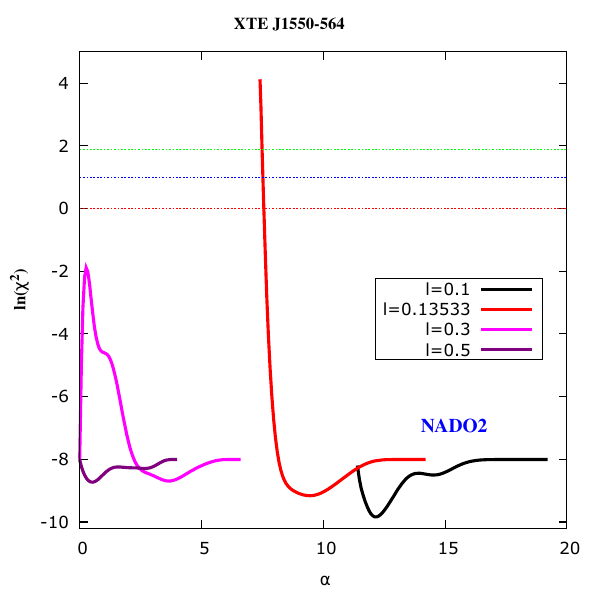}\label{b92}}
		\caption{(a) Variation of $\ln \chi^{2}$ with the $(\alpha,l)$ parameter space for the black hole source XTE J1550-564 using the Non Axisymmetric Disc Oscillation Model 2 (NADO2) (b) Variation of $\ln \chi^{2}$ with the hair parameter $\alpha$ for selected values of the parameter $l = 0.1,0.13533,0.3,$ and $0.5$. The red, blue, and green horizontal dotted lines represent the $1\sigma$, $2\sigma$, and $3\sigma$ confidence limits, respectively. Only for  $l=0.13533$, $\alpha\simeq\alpha_{\rm crit}$ are excluded outside the $3\sigma$ confidence level. All other values of $l,\alpha$, 
        are allowed within the 1$\sigma$ confidence limit.}
        \label{b9}
\end{figure}



\end{enumerate}

\subsection{GRS 1915+105}
Here we present the variation of $\chi^2$ with $l$ and $\alpha$ for the source GRS 1915+105 considering the nine resonance models.

\begin{enumerate}
\item{\bf{Parametric Resonance Model (PRM)}}
\begin{figure}[H]
		\centering
		\subfloat[]
        {\includegraphics[width=0.48\linewidth, height=7.5cm]{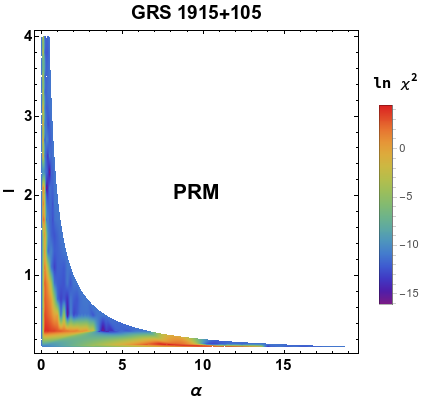}\label{c11}}
		\hfill
		\subfloat[]
        {\includegraphics[width=0.48\linewidth, height=7.5cm]{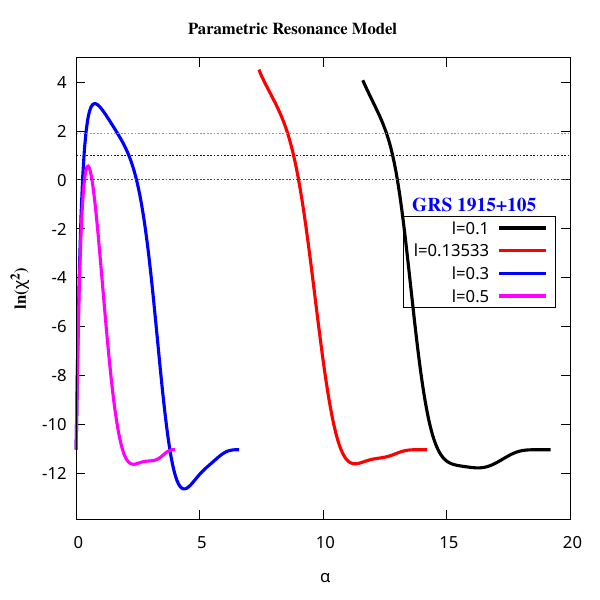}\label{c12}}
		\caption{(a)  Variation of $\ln \chi^{2}$ with $(\alpha,l)$ for the BH GRS 1915+105 using the Parametric Resonance Model  (PRM) (b) Variation of $\ln \chi^{2}$ with the hair parameter $\alpha$ for selected values of the parameter $l = 0.1, 0.13533, 0.3,$ and $0.5$. The red, blue, and green horizontal dotted lines represent the $1\sigma$, $2\sigma$, and $3\sigma$ confidence limits, respectively. For lower values of $l$ ($l=0.1$ and $0.13533$), values of $\alpha$ close to the critical value $\alpha_{\rm crit}$ are excluded outside the $3\sigma$ confidence level while for $l\gtrsim 0.5$, all $(l,\alpha)$ are allowed within 1$\sigma$.}
        \label{c1}
\end{figure}

\item{\bf{Keplerian Resonance Model 1 (KRM1)}}
\begin{figure}[H]
		\centering
		\subfloat[]
        {\includegraphics[width=0.48\linewidth, height=7.5cm]{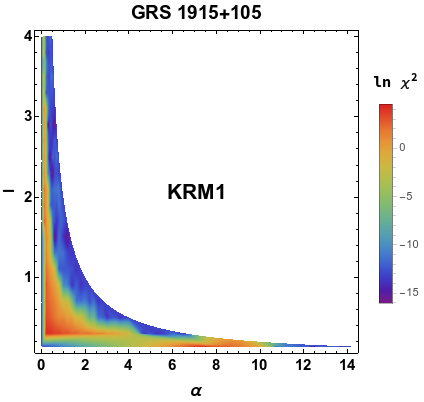}\label{c21}}
		\hfill
		\subfloat[]
        {\includegraphics[width=0.48\linewidth, height=7.5cm]{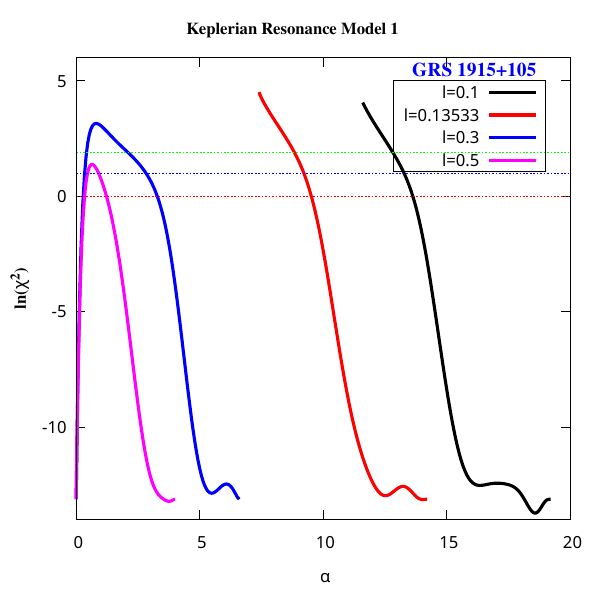}\label{c22}}
		\caption{(a) Variation of $\ln \chi^{2}$ with the $(\alpha,l)$ for the BH GRS 1915+105 using the Keplerian Resonance Model 1 (KRM1) (b) Variation of $\ln \chi^{2}$ with $\alpha$ for selected values of the parameter $l = 0.1, 0.13533, 0.3,$ and $0.5$. The red, blue, and green horizontal dotted lines represent the $1\sigma$, $2\sigma$, and $3\sigma$ confidence limits, respectively. }
        \label{c2}
\end{figure}


\item{\bf{FRM1, FRM2, KRM2, KRM3, WDOM, NADO1, NADO2}}

\begin{figure}[hbt!]
	\centering
	\subfloat[]{
		\includegraphics[width=0.3\linewidth]{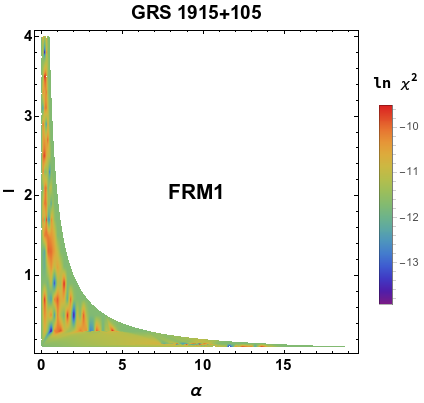}\label{c3}
	}
	\hfill
	\subfloat[]{
		\includegraphics[width=0.3\linewidth]{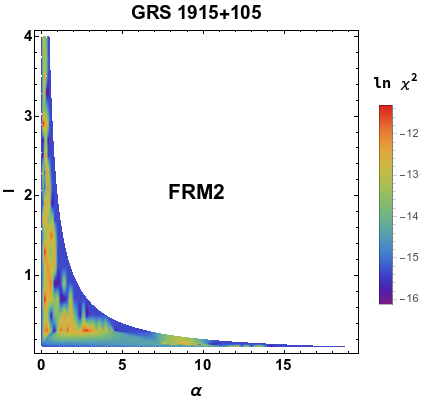}\label{c4}
	}
	\hfill
	\subfloat[]{
		\includegraphics[width=0.3\linewidth]{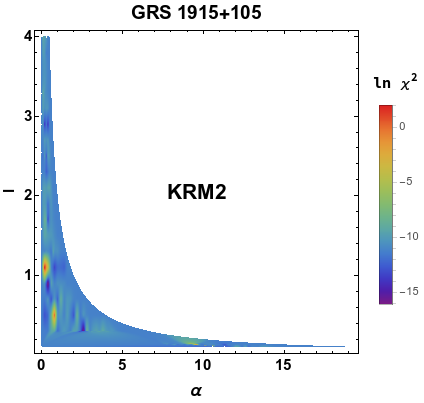}\label{c5}
	}

	\vspace{0.33cm}

	\subfloat[]{
		\includegraphics[width=0.3\linewidth]{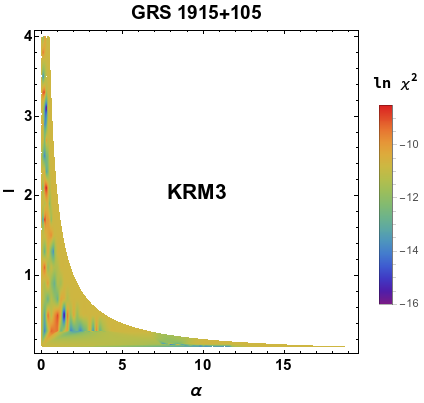}\label{c6}
	}
	\hfill
	\subfloat[]{
		\includegraphics[width=0.3\linewidth]{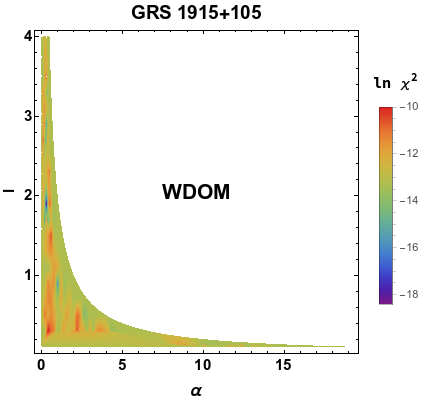}\label{c7}
	}
	\hfill
	\subfloat[]{
		\includegraphics[width=0.3\linewidth]{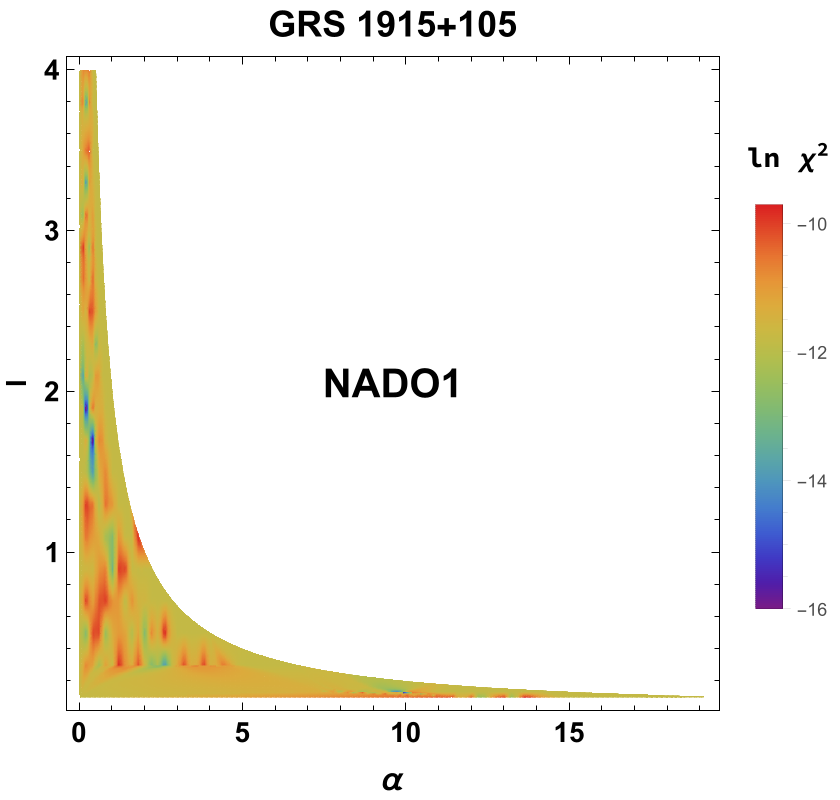}\label{c8}
	}

	\vspace{0.33cm}

	\subfloat[]{
		\includegraphics[width=0.3\linewidth]{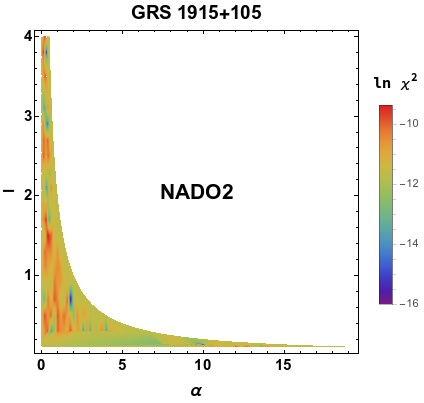}\label{c9}
	}

	\caption{Density plot showing the variation of $\ln \chi^{2}$ with the $(\alpha,l)$ parameter space for the BH source GRS 1915+105 using the  (a) Forced Resonance Model 1 (FRM1), (b) Forced Resonance Model 2 (FRM2), (c) Keplerian Resonance Model 2  (KRM2) , (d) Keplerian Resonance Model 3 (KRM3) , (e) Warped Disc Oscillation Model (WDOM), (f) Non Axisymmetric Disc Oscillation Model 1 (NADO1), and (g) Non Axisymmetric Disc Oscillation Model 2 (NADO2).  It is evident that $\chi^2$ values are very small and shows only weak variation across the parameter space, leading to a broad nearly degenerate allowed region for all models. Consequently, the HFQPO data of GRS 1915+105 cannot distinguish between the Kerr and the hairy Kerr scenario, assuming these models. }
	\label{c10}
\end{figure}

\end{enumerate}

\newpage

\subsection{H 1743+322}
Here we present the variation of $\chi^2$ with $(l,\alpha)$ for the black hole source H 1743+322 considering the nine resonance models.

\begin{enumerate}
    \item {\bf{Parametric Resonance Model (PRM)}}
\begin{figure}[hbt!]
		\centering
		\subfloat[]
        {\includegraphics[width=0.48\linewidth, height=7.0cm]{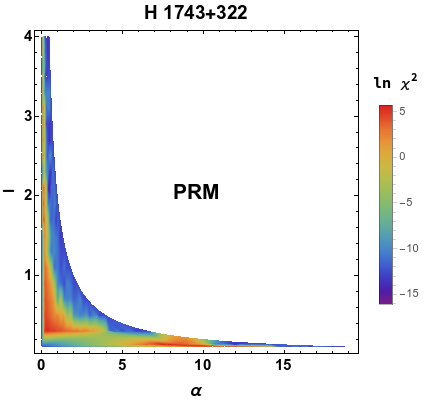}\label{d11}}
		\hfill
		\subfloat[]
        {\includegraphics[width=0.48\linewidth, height=7.05cm]{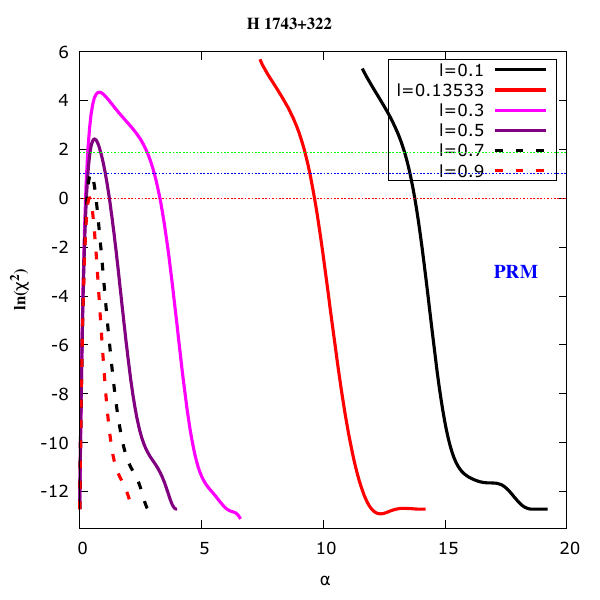}\label{d12}}
		\caption{(a)  Variation of $\ln \chi^{2}$ with $(\alpha,l)$ for the source H1743+322 using the Parametric Resonance Model  (PRM). (b) Variation of $\ln \chi^{2}$ with the hair parameter $\alpha$ for selected values of the parameter $l = 0.1, 0.13533, 0.3, 0.5, 0.7$ and $0.9$. The red, blue, and green horizontal dotted lines represent the $1\sigma$, $2\sigma$, and $3\sigma$ confidence limits, respectively. For lower values of $l$ ($l=0.1$ and $0.13533$), values of $\alpha$ close to $\alpha_{\rm crit}$ are excluded outside the $3\sigma$ confidence level. 
        For $l\gtrsim0.9$, all values of $l,\alpha$ are allowed within 1$\sigma$.}
        \label{d1}
\end{figure}


\item{\bf{Keplerian Resonance Model (KRM1)}}
\begin{figure}[hbt!]
	\centering
	\subfloat[]
    {\includegraphics[width=0.32\linewidth, height=6.5cm]{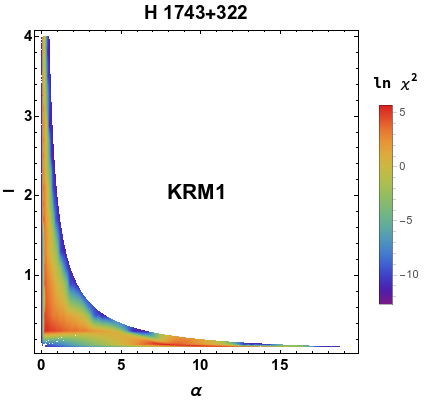}\label{d21}}
	\hfill
	\subfloat[]
    {\includegraphics[width=0.32\linewidth, height=6.5cm]{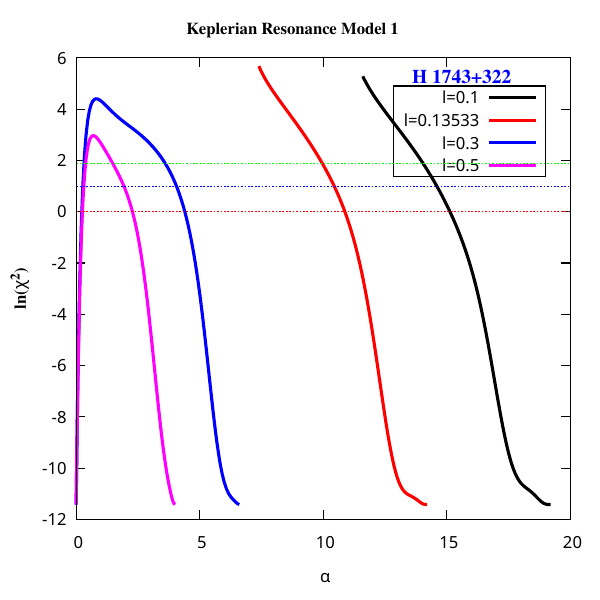}\label{d22}}
	\hfill
	\subfloat[]
    {\includegraphics[width=0.32\linewidth, height=6.5cm]{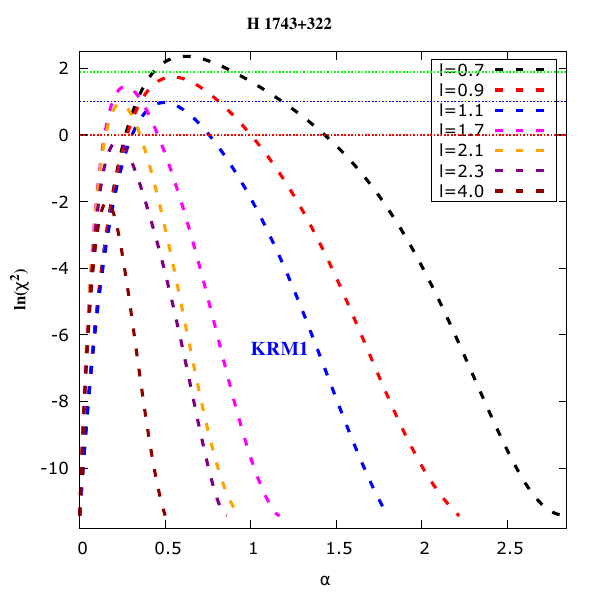}\label{d23}}
	\caption{(a) Variation of $\ln \chi^{2}$ with the $(\alpha,l)$ parameter space for the black hole binary H 1743+322 using the Keplerian Resonance Model 1 (b) Variation of $\ln \chi^{2}$ with $\alpha$ for selected lower values of $l = 0.1, 0.13533, 0.3,$ and $0.5$. The red, blue, and green horizontal dotted lines represent the $1\sigma$, $2\sigma$, and $3\sigma$ confidence limits, respectively. For  $l\lesssim e^{-2}$, $\alpha$ close to $\alpha_{\rm crit}$ are excluded outside the $3\sigma$ confidence level, while for $l=0.3$ and $l=0.5$, certain intermediate ranges are disfavored . (c) Variation of $\ln \chi^{2}$ with $\alpha$ for comparatively higher values of $l = 0.7,0.9,1.1,1.7,2.1, 2.3,$ and $4.0$. For $l>2.1$, all combinations of $l$ and $\alpha$ are allowed within $1\sigma$.}
    \label{d2}
\end{figure}

\item{\bf{FRM1, FRM2, KRM2, KRM3, WDOM, NADO1, NADO2}}

\begin{figure}[hbt!]
	\centering
	\subfloat[]{
		\includegraphics[width=0.3\linewidth]{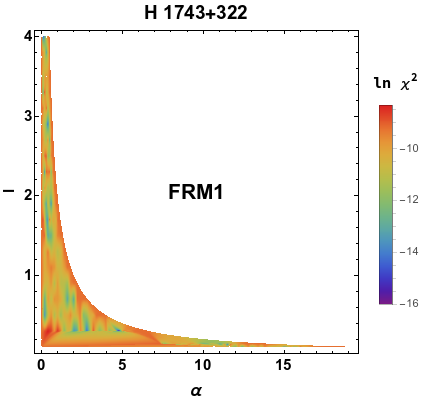}\label{d3}
	}
	\hfill
	\subfloat[]{
		\includegraphics[width=0.3\linewidth]{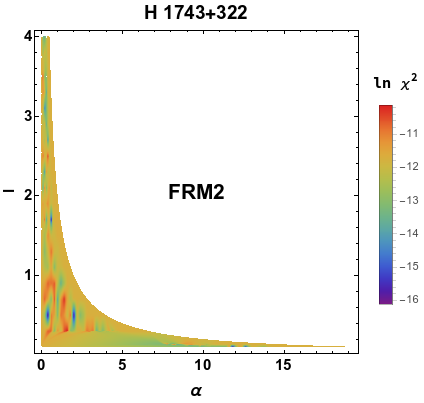}\label{d4}
	}
	\hfill
	\subfloat[]{
		\includegraphics[width=0.3\linewidth]{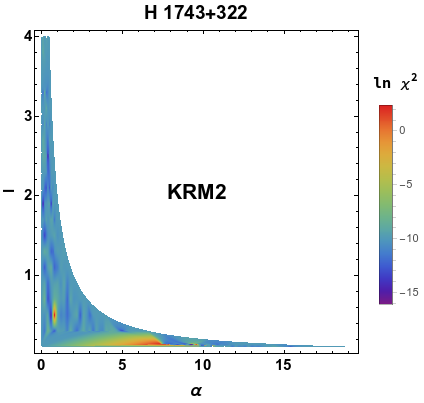}\label{d5}
	}

	\vspace{0.3cm}

	\subfloat[]{
		\includegraphics[width=0.3\linewidth]{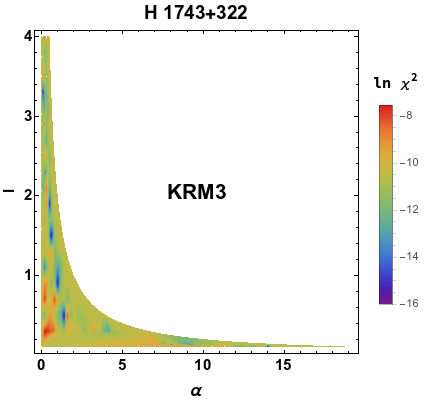}\label{d6}
	}
	\hfill
	\subfloat[]{
		\includegraphics[width=0.3\linewidth]{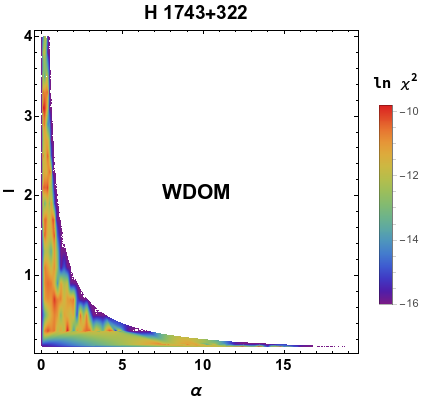}\label{d7}
	}
	\hfill
	\subfloat[]{
		\includegraphics[width=0.3\linewidth]{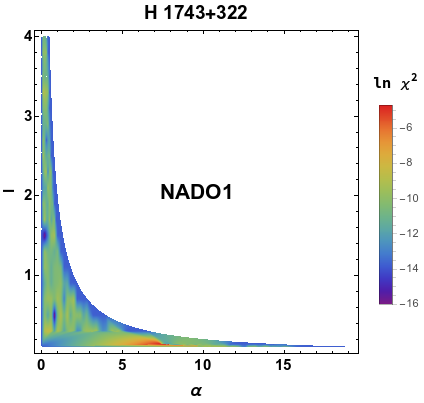}\label{d8}
	}

	\vspace{0.3cm}

	\subfloat[]{
		\includegraphics[width=0.3\linewidth]{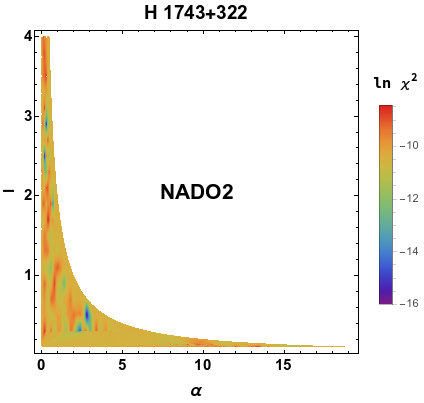}\label{d9}
	}

	\caption{Variation of $\ln \chi^{2}$ with  $(\alpha,l)$ parameter space for the black hole source H 1743+322 using the  (a) Forced Resonance Model 1 (FRM1), (b) Forced Resonance Model 2 (FRM2), (c) Keplerian Resonance Model 2 (KRM2), (d) Keplerian Resonance Model 3 (KRM3), (e) Warped Disc Oscillation Model (WDOM), (f) Non Axisymmetric Disc Oscillation Model 1 (NADO1), and (g) Non Axisymmetric Disc Oscillation Model 2 (NADO2).  It is evident that $\chi^2$ values are very small and shows only weak variation across the parameter space, leading to a broad nearly degenerate allowed region for all models. Consequently, the HFQPO data of H 1743+322 cannot distinguish between the Kerr and the hairy Kerr scenario based on these models. 
    }
	\label{d10}
\end{figure}


\end{enumerate}



\newpage

\subsection{Sgr A*}
Here we present the variation of $\chi^2$ with $(l,\alpha)$ for the blacl hole Sgr A* considering the nine resonance models.

\begin{enumerate}
    \item{\bf{Parametric Resonance Model (PRM)}}

\begin{figure}[hbt!]
	\centering
	\subfloat[]
    {\includegraphics[width=0.34\linewidth, height=6.8cm]{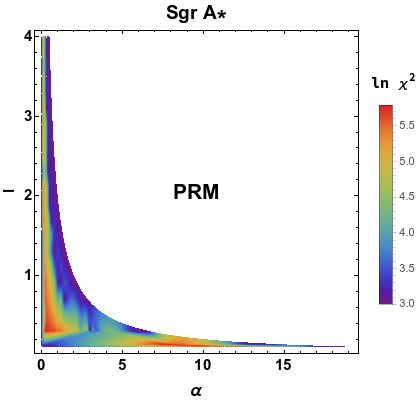}\label{e11}}
	\hfill
	\subfloat[]
    {\includegraphics[width=0.32\linewidth, height=6.8cm]{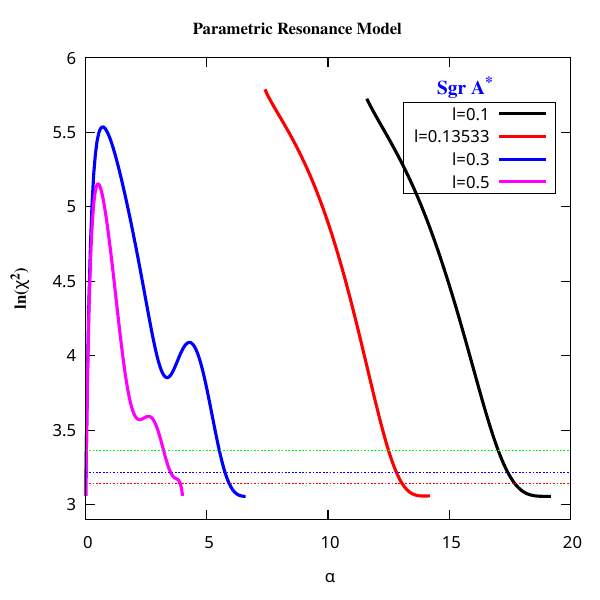}\label{e12}}
	\hfill
	\subfloat[]
    {\includegraphics[width=0.32\linewidth, height=6.8cm]{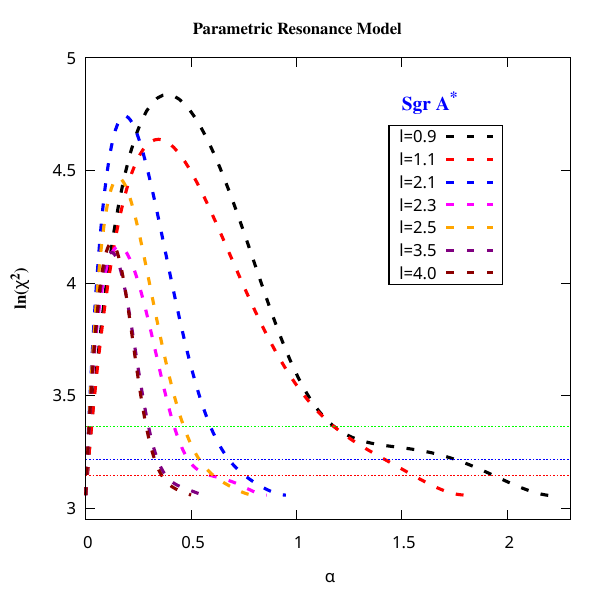}\label{e13}}
	\caption{(a)  Density plot showing the variation of $\ln \chi^{2}$ with $(\alpha,l)$ for the supermassive black hole Sgr A* using the Parametric Resonance Model  (PRM) (b) Variation of $\ln \chi^{2}$ with the hair parameter $\alpha$ for selected lower values of $l = 0.1,0.13533,0.3,$ and $0.5$. (c) Variation of $\ln \chi^{2}$ with $\alpha$ for comparatively higher values of $l = 0.9,1.1,2.1,2.3,2.5, 3.5,$ and $4.0$.
    The red, blue, and green horizontal dotted lines represent the $1\sigma$, $2\sigma$, and $3\sigma$ confidence limits, respectively. For  all $l$, either $\alpha\to0$ or values of $\alpha$ close to $2/l$ are allowed within 1$\sigma$. For a given $l$, intermediate values of $\alpha$ are strongly ruled out.
    }
    \label{f1}
\end{figure}

{\item{\bf{Forced Resonance Model 1}}}
\begin{figure}[hbt!]
	\centering
	\subfloat[]
    {\includegraphics[width=0.34\linewidth, height=6.8cm]{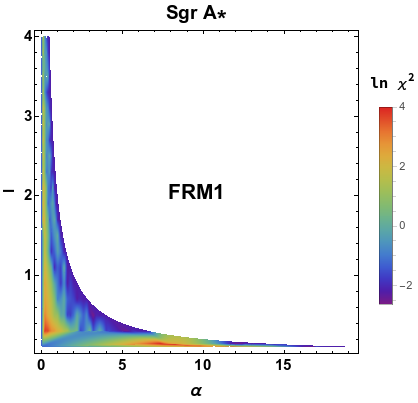}\label{e21}}
	\hfill
	\subfloat[]
    {\includegraphics[width=0.32\linewidth, height=6.8cm]{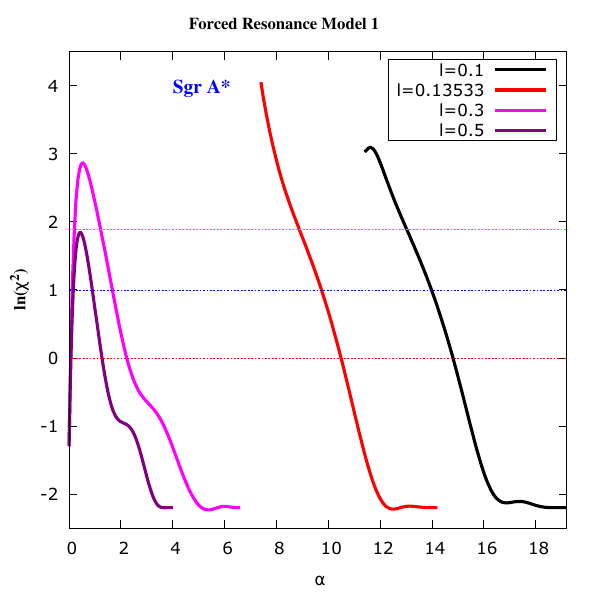}\label{e22}}
	\hfill
	\subfloat[]
    {\includegraphics[width=0.32\linewidth, height=6.8cm]{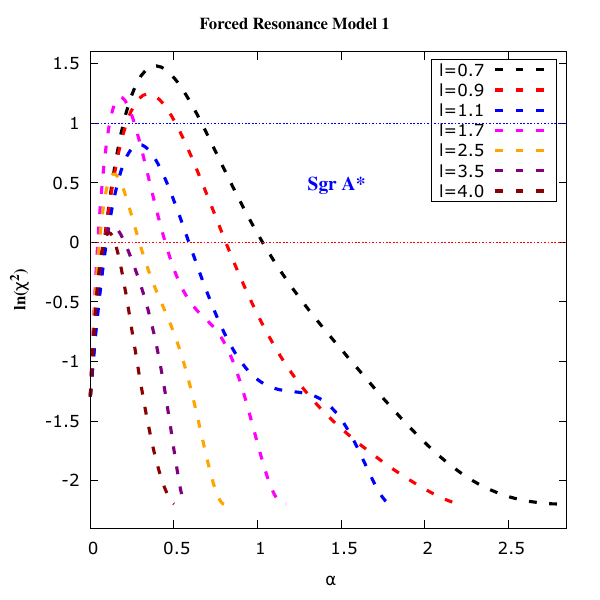}\label{e23}}
	\caption{(a)  Density plot showing the variation of $\ln \chi^{2}$ with $(\alpha,l)$ for Sgr A* using the Forced Resonance Model 1  (FRM1) (b) Variation of $\ln \chi^{2}$ with the hair parameter $\alpha$ for selected lower values of $l = 0.1, 0.13533, 0.3,$ and $0.5$. (c) Variation of $\ln \chi^{2}$ with $\alpha$ for comparatively higher values of $l = 0.7,0.9,1.1,1.7,2.5, 3.5,$ and $4.0$. 
    The red, blue, and green horizontal dotted lines represent the $1\sigma$, $2\sigma$, and $3\sigma$ confidence limits, respectively. For  $l\lesssim e^{-2}$, values of $\alpha$ near the critical value $\alpha_{\rm crit}$ are excluded outside the $3\sigma$ confidence level. For higher $l$, intermediate values of $\alpha$ are generally less favored.
    }
    \label{f2}
\end{figure}

\item{\bf{Forced Resonance Model 2 (FRM 2)}}
\begin{figure}[hbt!]
		\centering
		\subfloat[]
        {\includegraphics[width=0.48\linewidth, height=7.6cm]{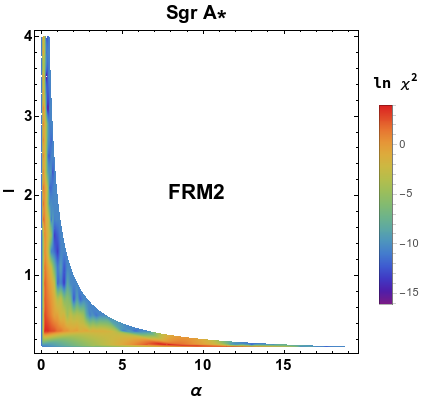}\label{e61}}
		\hfill
		\subfloat[]
        {\includegraphics[width=0.48\linewidth, height=7.6cm]{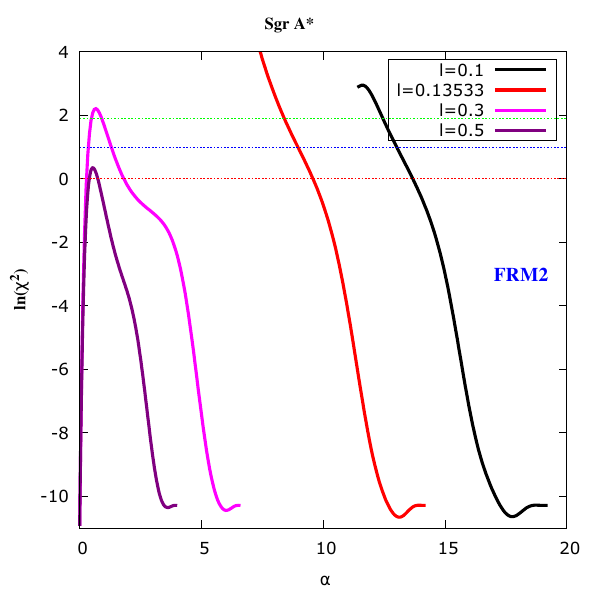}\label{e62}}
		\caption{(a)Variation of $\ln \chi^{2}$ with $(\alpha,l)$ for Sgr A* using the Forced Resonance Model 2, (b) Variation of $\ln \chi^{2}$ with $\alpha$ for selected values of the parameter $l = 0.1,0.13533,0.3,$ and $0.5$. The red, blue, and green horizontal dotted lines represent the $1\sigma$, $2\sigma$, and $3\sigma$ confidence limits, respectively. For lower values of $l$ ($l\lesssim e^{-2}$), values of $\alpha$ close to $\alpha_{\rm crit}$ are excluded outside the $3\sigma$ confidence level. For $l\gtrsim 0.5$, all values of $l,\alpha$ are allowed within 1$\sigma$.
        }
        \label{f6}
\end{figure}

\item{\bf{Keplerian Resonance Model 1}}
\begin{figure}[hbt!]
	\centering
	\subfloat[]
    {\includegraphics[width=0.34\linewidth,  height=7.6cm]{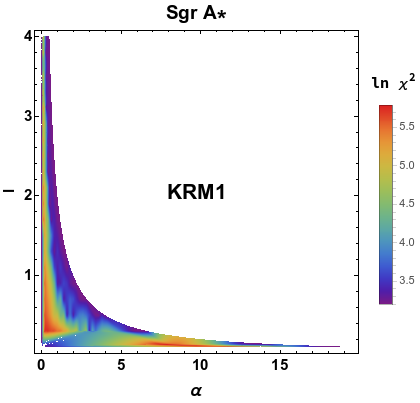}\label{e31}}
	\hfill
	\subfloat[]
    {\includegraphics[width=0.32\linewidth,  height=7.6cm]{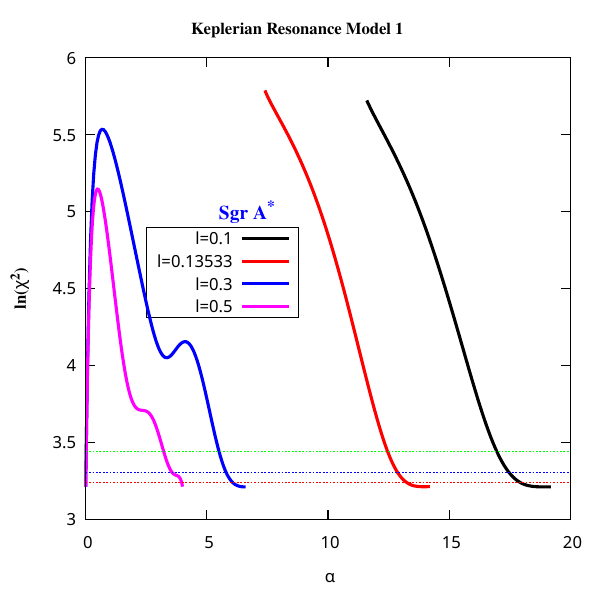}\label{e32}}
	\hfill
	\subfloat[]
    {\includegraphics[width=0.32\linewidth,  height=7.6cm]{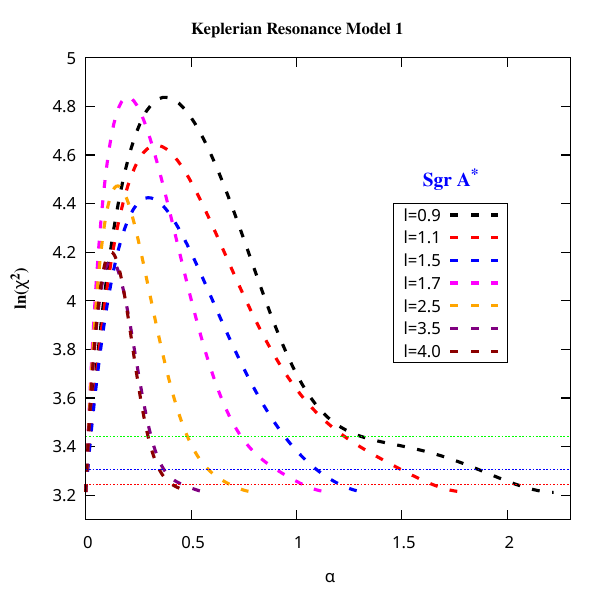}\label{e33}}
	\caption{(a) Density plot showing the variation of $\ln \chi^{2}$ with $(\alpha,l)$ for the supermassive black hole Sgr A* using the Keplerian Resonance Model 1 (KRM1) (b) Variation of $\ln \chi^{2}$ with the hair parameter $\alpha$ for selected lower values of $l = 0.1,0.13533,0.3,$ and $0.5$. (c) Variation of $\ln \chi^{2}$ with $\alpha$ for comparatively higher values of $l = 0.9,1.1,1.5,1.7,2.5, 3.5,$ and $4.0$.
    The red, blue, and green horizontal dotted lines represent the $1\sigma$, $2\sigma$, and $3\sigma$ confidence limits, respectively. For  all $l$, only values of $\alpha$ close to 0 or $\alpha$ close to $2/l$ are allowed within 1$\sigma$. For a given $l$, intermediate values of $\alpha$ are strongly ruled out.
    }
    \label{f3}
\end{figure}

\newpage
\item{\bf{Keplerian Resonance Model 2}}
\begin{figure}[hbt!]
	\centering
	\subfloat[]
    {\includegraphics[width=0.34\linewidth,  height=7.7cm]{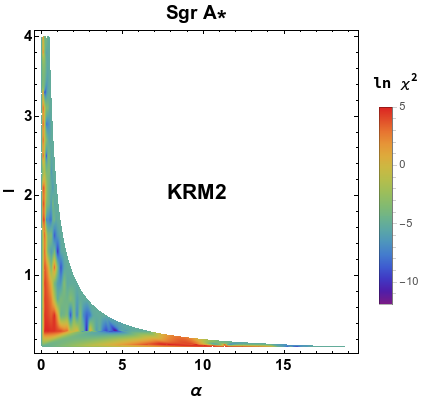}\label{e41}}
	\hfill
	\subfloat[]
    {\includegraphics[width=0.32\linewidth,  height=7.7cm]{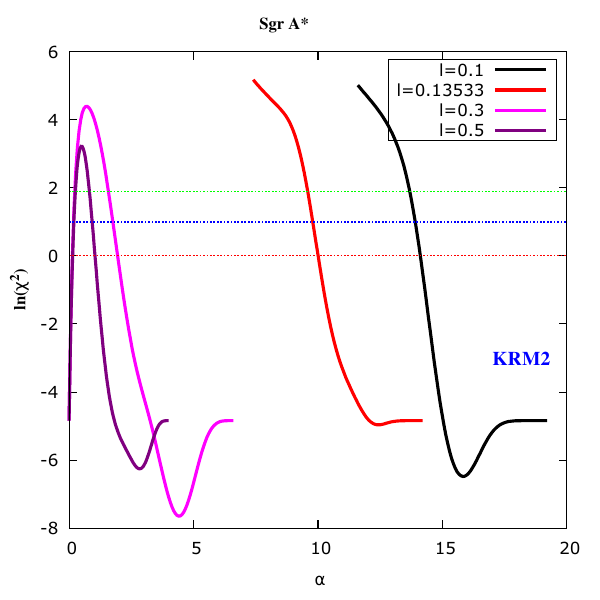}\label{e42}}
	\hfill
	\subfloat[]
    {\includegraphics[width=0.32\linewidth,  height=7.7cm]{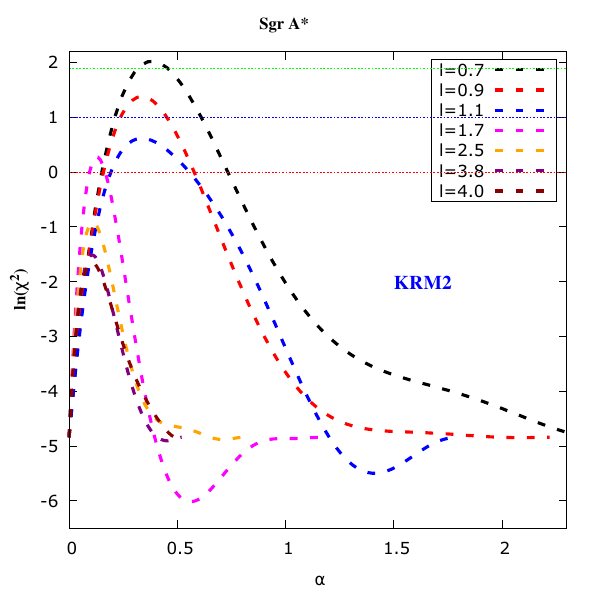}\label{e43}}
	\caption{(a) Density plot showing the variation of $\ln \chi^{2}$ with $(\alpha,l)$ for Sgr A* using the Keplerian Resonance Model 2  (KRM2) (b) Variation of $\ln \chi^{2}$ with the hair parameter $\alpha$ for selected lower values of $l = 0.1, 0.13533, 0.3,$ and $0.5$. (c) Variation of $\ln \chi^{2}$ with $\alpha$ for comparatively higher values of $l = 0.7,0.9,1.1,1.7,2.5, 3.8,$ and $4.0$. 
    The red, blue, and green horizontal dotted lines represent the $1\sigma$, $2\sigma$, and $3\sigma$ confidence limits, respectively. For  $l\lesssim e^{-2}$, values of $\alpha$ near the critical value $\alpha_{\rm crit}$ are excluded outside the $3\sigma$ confidence level. For higher $l$ values intermediate values of $\alpha$ are generally less favored.
    }
    \label{f4}
\end{figure}

\item{\bf{Keplerian Resonance Model 3}}
\begin{figure}[hbt!]
		\centering
		\subfloat[]
        {\includegraphics[width=0.48\linewidth, height=7.7cm]{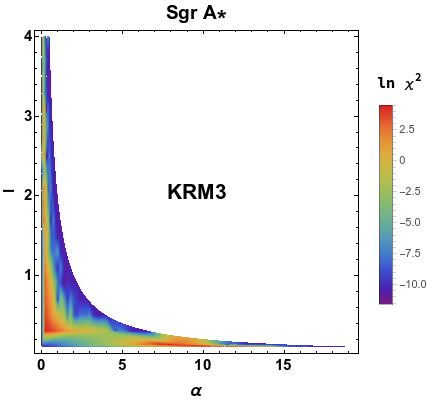}\label{e71}}
		\hfill
		\subfloat[]
        {\includegraphics[width=0.48\linewidth, height=7.7cm]{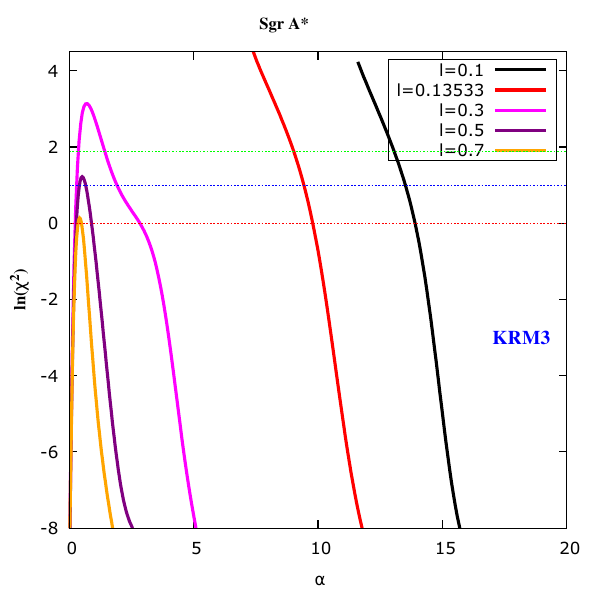}\label{e72}}
		\caption{(a) Variation of $\ln \chi^{2}$ with $(\alpha,l)$ for Sgr A* using the model (KRM3),  (b) Variation of $\ln \chi^{2}$ with the hair parameter $\alpha$ for selected values of the parameter $l = 0.1,0.13533,0.3,0.5$ and $0.7$. The red, blue, and green horizontal dotted lines represent the $1\sigma$, $2\sigma$, and $3\sigma$ confidence limits, respectively. For lower values of $l$ ($l=0.1$ and $0.13533$), values of $\alpha$ close to the critical value $\alpha_{\rm crit}$ are excluded outside the $3\sigma$ confidence level. For $l\gtrsim0.7$ all allowed values of $\alpha$ are included within 1$\sigma$.
        }
        \label{f7}
\end{figure}

\item{\bf{Non-Axisymmetric Disc Oscillation Model 1 (NADO1)}}

\begin{figure}[hbt!]
	\centering
	\subfloat[]
    {\includegraphics[width=0.34\linewidth, height=7.8cm]{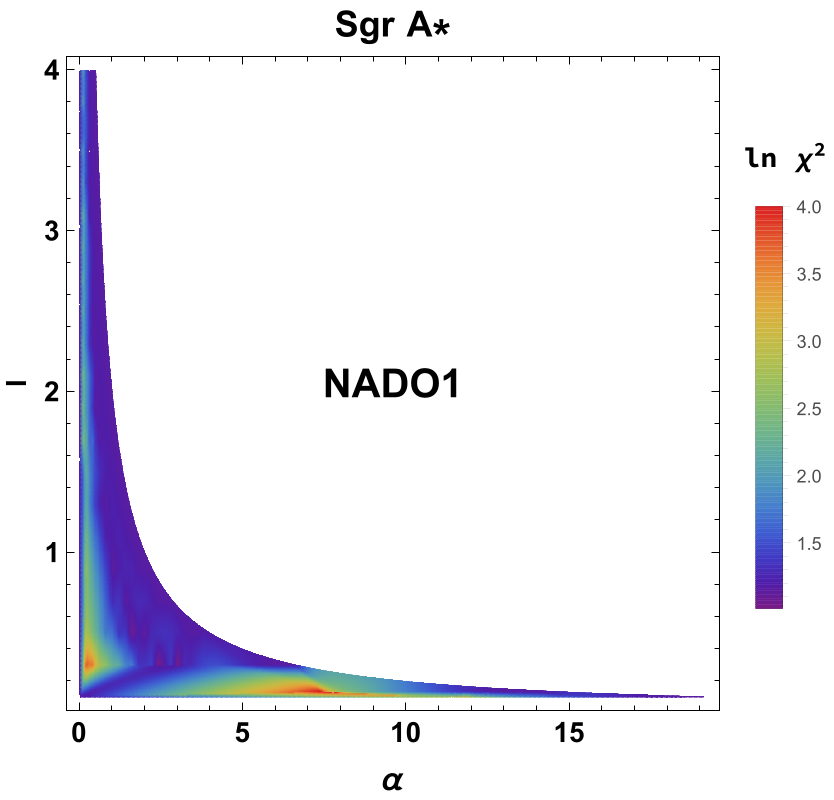}\label{e51}}
	\hfill
	\subfloat[]
    {\includegraphics[width=0.32\linewidth, height=7.8cm]{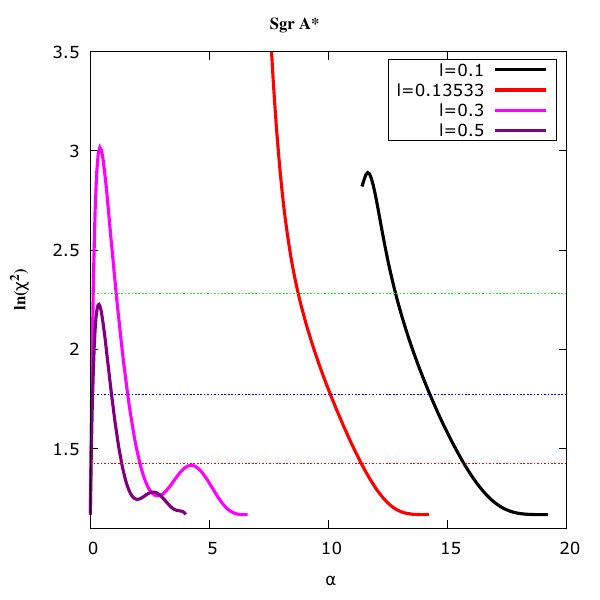}\label{e52}}
	\hfill
	\subfloat[]
    {\includegraphics[width=0.32\linewidth, height=7.8cm]{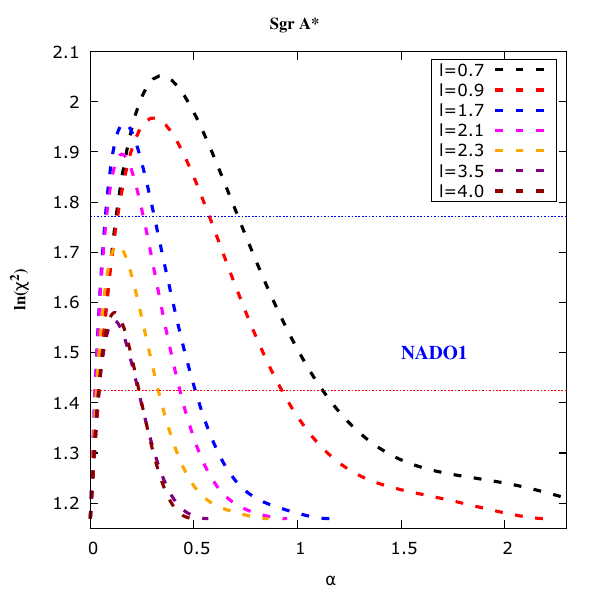}\label{e53}}
	\caption{(a) Density plot showing the variation of $\ln \chi^{2}$ with $(\alpha,l)$ for Sgr A* using the Non-Axisymmetric Disc Oscillation Model 1 (NADO1) (b) Variation of $\ln \chi^{2}$ with the hair parameter $\alpha$ for selected lower values of $l = 0.1, 0.13533, 0.3,$ and $0.5$. (c) Variation of $\ln \chi^{2}$ with $\alpha$ for comparatively higher values of $l = 0.7,0.9,1.7,2.1,2.3, 3.5,$ and $4.0$. 
    The red, blue, and green horizontal dotted lines represent the $1\sigma$, $2\sigma$, and $3\sigma$ confidence limits, respectively. For  $l\lesssim e^{-2}$, values of $\alpha$ near the critical value $\alpha_{\rm crit}$ are excluded outside the $3\sigma$ confidence level. For higher $l$ values intermediate values of $\alpha$ are generally less favored.  
     }
    \label{f5}
\end{figure}


\item{\bf{Non-Axisymmetric Disc Oscillation Model 2 (NADO2)}}
\begin{figure}[hbt!]
		\centering
		\subfloat[]
        {\includegraphics[width=0.48\linewidth,  height=8cm]{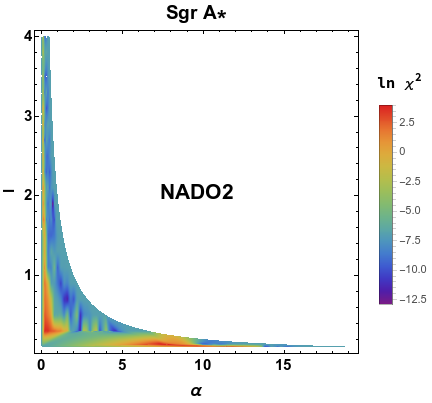}\label{e91}}
		\hfill
		\subfloat[]
        {\includegraphics[width=0.48\linewidth,  height=8cm]{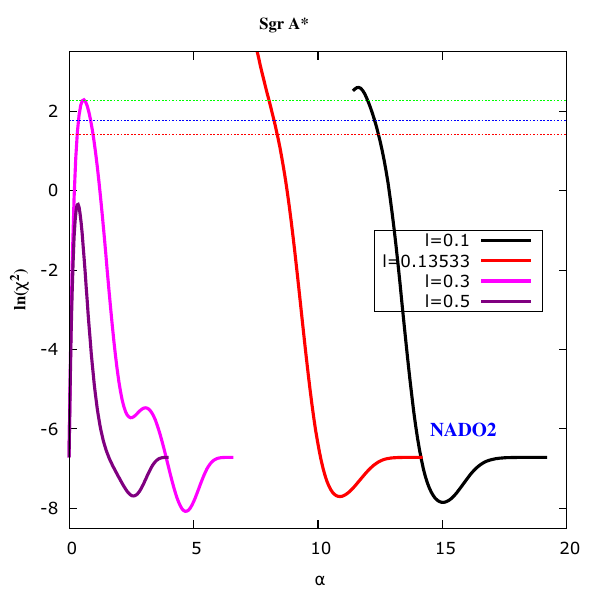}\label{e92}}
		\caption{(a) Variation of $\ln \chi^{2}$ with $(\alpha,l)$ for the BH Sgr A* using the model NADO2, (b) Variation of $\ln \chi^{2}$ with $\alpha$ for selected values of the parameter $l = 0.1,0.13533,0.3,$ and $0.5$. The red, blue, and green horizontal dotted lines represent the $1\sigma$, $2\sigma$, and $3\sigma$ confidence limits, respectively. For $l\lesssim e^{-2}$, values of $\alpha$ close to $\alpha_{\rm crit}$ are excluded outside the $3\sigma$ confidence level. For $l\gtrsim 0.5$, all allowed values of $\alpha$ are included within 1$\sigma$.
        }
        \label{f9}
\end{figure}

\newpage

\item{\bf{Warped Disc Oscillation Model (WDOM)}}
\begin{figure}[hbt!]
		\centering
		\subfloat[]
        {\includegraphics[width=0.48\linewidth]{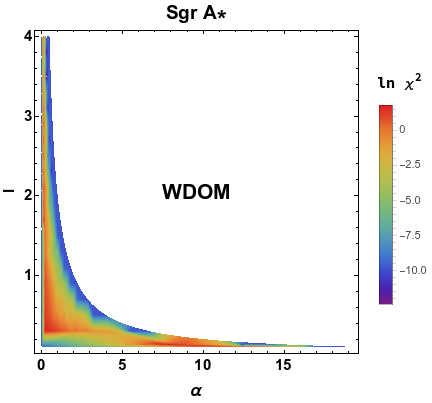}\label{e81}}
		\hfill
		\subfloat[]
        {\includegraphics[width=0.48\linewidth]{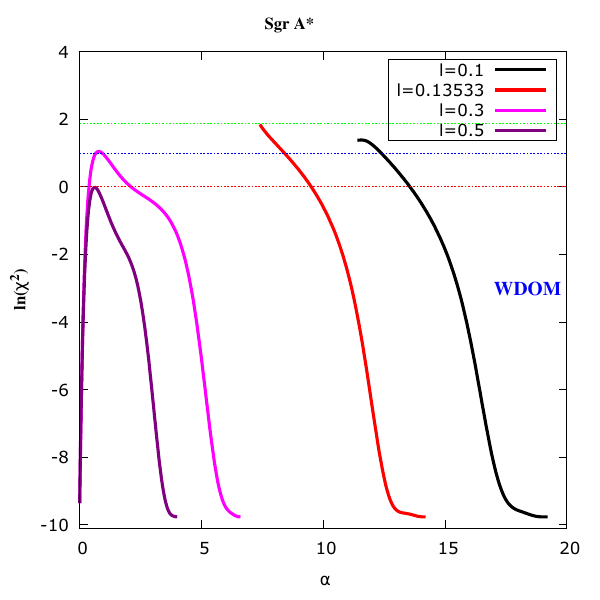}\label{e82}}
		\caption{(a) Variation of $\ln \chi^{2}$ with $(\alpha,l)$ for Sgr A* using the  Warped Disc Oscillation Model,  (b) Variation of $\ln \chi^{2}$ with the hair parameter $\alpha$ for selected values of the parameter $l = 0.1,0.13533,0.3,$ and $0.5$. The red, blue, and green horizontal dotted lines represent the $1\sigma$, $2\sigma$, and $3\sigma$ confidence limits, respectively. For $l\lesssim e^{-2}$, values of $\alpha$ close to $\alpha_{\rm crit}$ are excluded outside the $3\sigma$ confidence level. For  $l\gtrsim 0.5$, all
        allowed values of $\alpha$ are included with 1$\sigma$.}
        \label{f8}
\end{figure}


\end{enumerate}

\newpage

\subsection{XTE J1859+226}
Now we present the variation of $\chi^2$ with $(l,\alpha)$ for the source XTE J1859+226 considering the nine resonance models.

\begin{enumerate}
\item{\bf{Parametric resonance Model (PRM)}}
\begin{figure}[hbt!]
		\centering
		\subfloat[]
        {\includegraphics[width=0.48\linewidth, height=7.5cm]{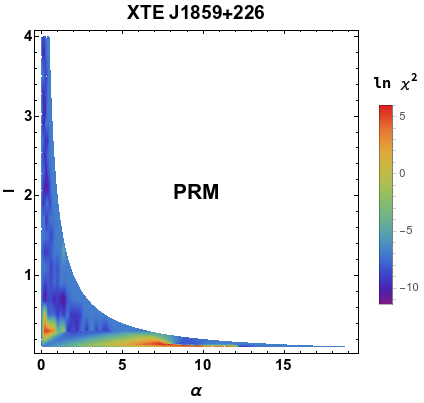}\label{g11}}
		\hfill
		\subfloat[]
        {\includegraphics[width=0.48\linewidth, height=7.5cm]{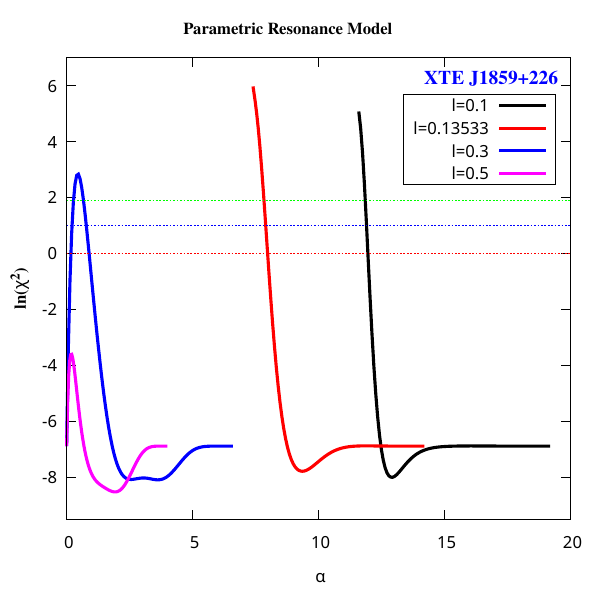}\label{g12}}
		\caption{(a) Variation of $\ln \chi^{2}$ with $(\alpha,l)$  for the  BH XTE J1859+226 using the PRM model, (b) Variation of $\ln \chi^{2}$ with $\alpha$ for  $l = 0.1,0.13533,0.3,$ and $0.5$. The red, blue, and green horizontal dotted lines denote the $1\sigma$, $2\sigma$, and $3\sigma$ confidence limits, respectively. For lower values $l\lesssim e^{-2}$, values of $\alpha\simeq \alpha_{\rm crit}$ are excluded outside the $3\sigma$ confidence level. For $l\gtrsim 0.5$, all allowed values of $\alpha$ are included within 1$\sigma$.
       }
        \label{g1}
\end{figure}

\item{\bf{Forced Resonance Model 2 (FRM2)}}
\begin{figure}[hbt!]
		\centering
		\subfloat[]
        {\includegraphics[width=0.48\linewidth, height=7.5cm]{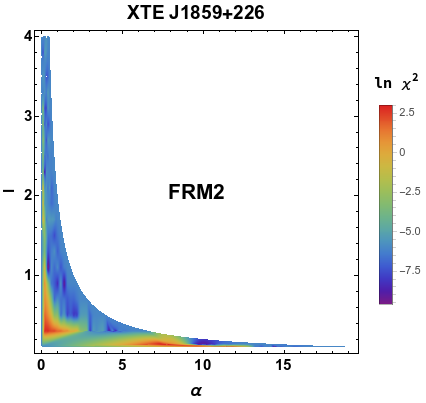}\label{g21}}
		\hfill
		\subfloat[]
        {\includegraphics[width=0.48\linewidth, height=7.5cm]{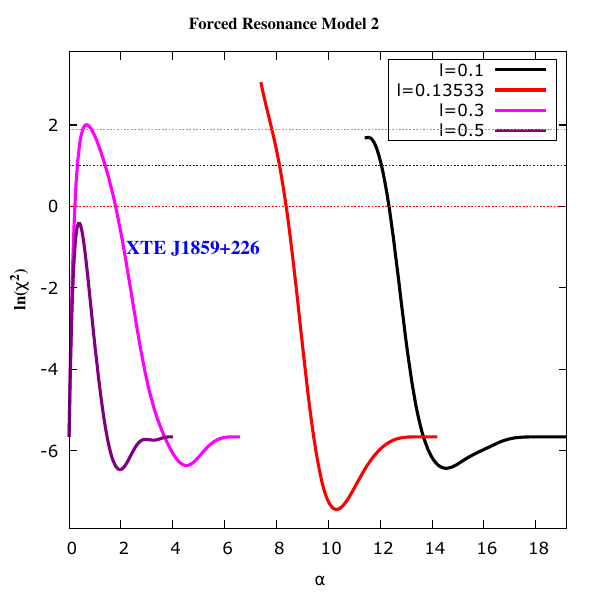}\label{g22}}
		\caption{ (a) Variation of $\ln \chi^{2}$ with $(\alpha,l)$ for XTE J1859+226 using the FRM2, (b) Variation of $\ln \chi^{2}$ with the hair parameter $\alpha$ for $l = 0.1,0.13533,0.3,$ and $0.5$. The red, blue, and green horizontal dotted lines represent the $1\sigma$, $2\sigma$, and $3\sigma$ confidence limits, respectively. For $l\lesssim e^{-2}$, values of $\alpha\to\alpha_{\rm crit}$ are excluded outside the $3\sigma$ confidence level. For $l\gtrsim0.5$, all allowed values of $\alpha$ are included within 1$\sigma$.
        }
        \label{g2}
\end{figure}

\item{\bf{Keplerian Resonance Model 1 (KRM1)}}
\begin{figure}[hbt!]
		\centering
		\subfloat[]
        {\includegraphics[width=0.48\linewidth, height=8.4cm]{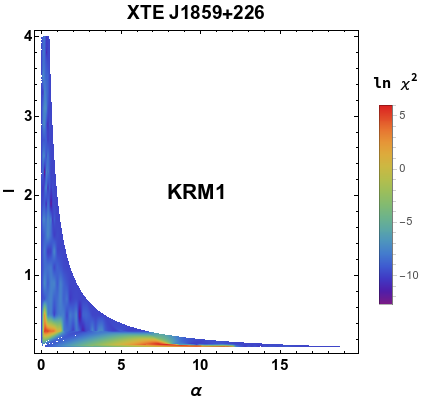}\label{g31}}
		\hfill
		\subfloat[]
        {\includegraphics[width=0.48\linewidth, height=8.4cm]{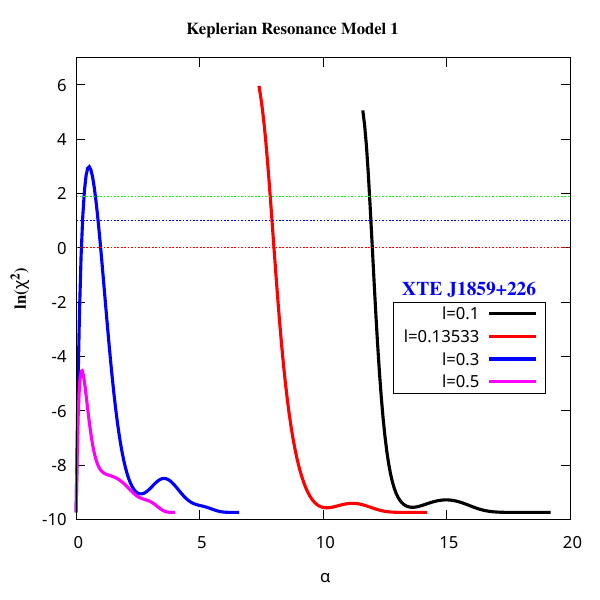}\label{g32}}
		\caption{(a) Variation of $\ln \chi^{2}$ with $(\alpha,l)$ for XTE J1859+226 using the model KRM1,  (b) Variation of $\ln \chi^{2}$ with $\alpha$ for $l = 0.1,0.13533,0.3,$ and $0.5$. The red, blue, and green horizontal dotted lines represent the $1\sigma$, $2\sigma$, and $3\sigma$ confidence limits, respectively. For 
        $l\lesssim e^{-2}$, values of $\alpha\to\alpha_{\rm crit}$ are excluded outside the $3\sigma$ confidence level. For $l>0.3$, all allowed values of $\alpha$ are included within 1$\sigma$.
        }
        \label{g3}
\end{figure}

\item{\bf{Keplerian Resonance Model 3 (KRM3)}}
\begin{figure}[hbt!]
		\centering
		\subfloat[]
        {\includegraphics[width=0.48\linewidth, height=8.5cm]{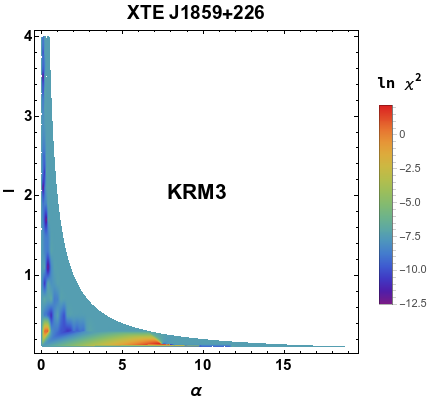}\label{g41}}
		\hfill
		\subfloat[]
        {\includegraphics[width=0.48\linewidth, height=8.5cm]{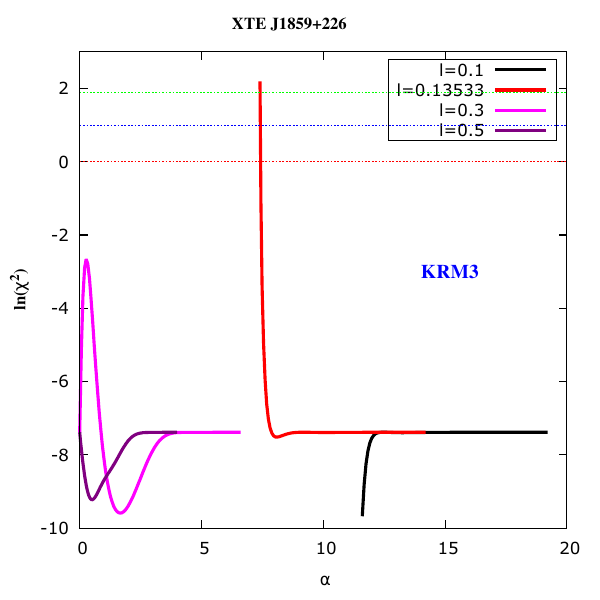}\label{g42}}
		\caption{(a) Variation of $\ln \chi^{2}$ with $(\alpha,l)$ for XTE J1859+226 using the model KRM3, 
        (b) Variation of $\ln \chi^{2}$ with the hair parameter $\alpha$ for $l = 0.1,0.13533,0.3,$ and $0.5$. The red, blue, and green horizontal dotted lines represent the $1\sigma$, $2\sigma$, and $3\sigma$ confidence limits, respectively. For  $l=e^{-2}$, $\alpha\simeq \alpha_{\rm crit}$ are excluded outside the $3\sigma$ confidence level. For all other values of $l$, the data cannot constrain the $l,\alpha$ parameter space.}
        \label{g4}
\end{figure}


\item{\bf{FRM1, KRM2, NADO1 and NADO2}}

\begin{figure}[hbt!]
	\centering
	\subfloat[]{
		\includegraphics[width=0.48\linewidth]{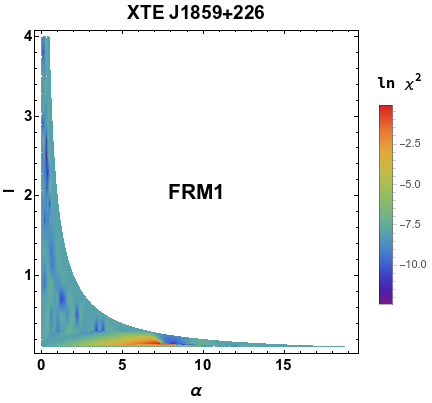}\label{g6}
	}
	\hfill
	\subfloat[]{
		\includegraphics[width=0.48\linewidth]{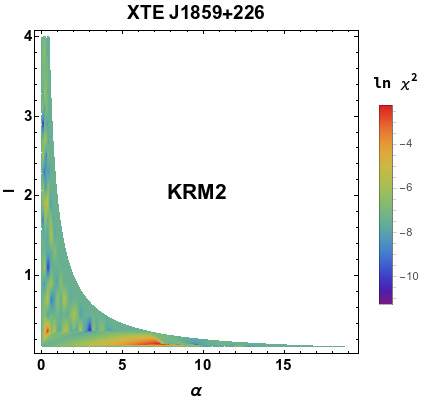}\label{g7}
	}

	\vspace{0.3cm}

	\subfloat[]{
		\includegraphics[width=0.48\linewidth]{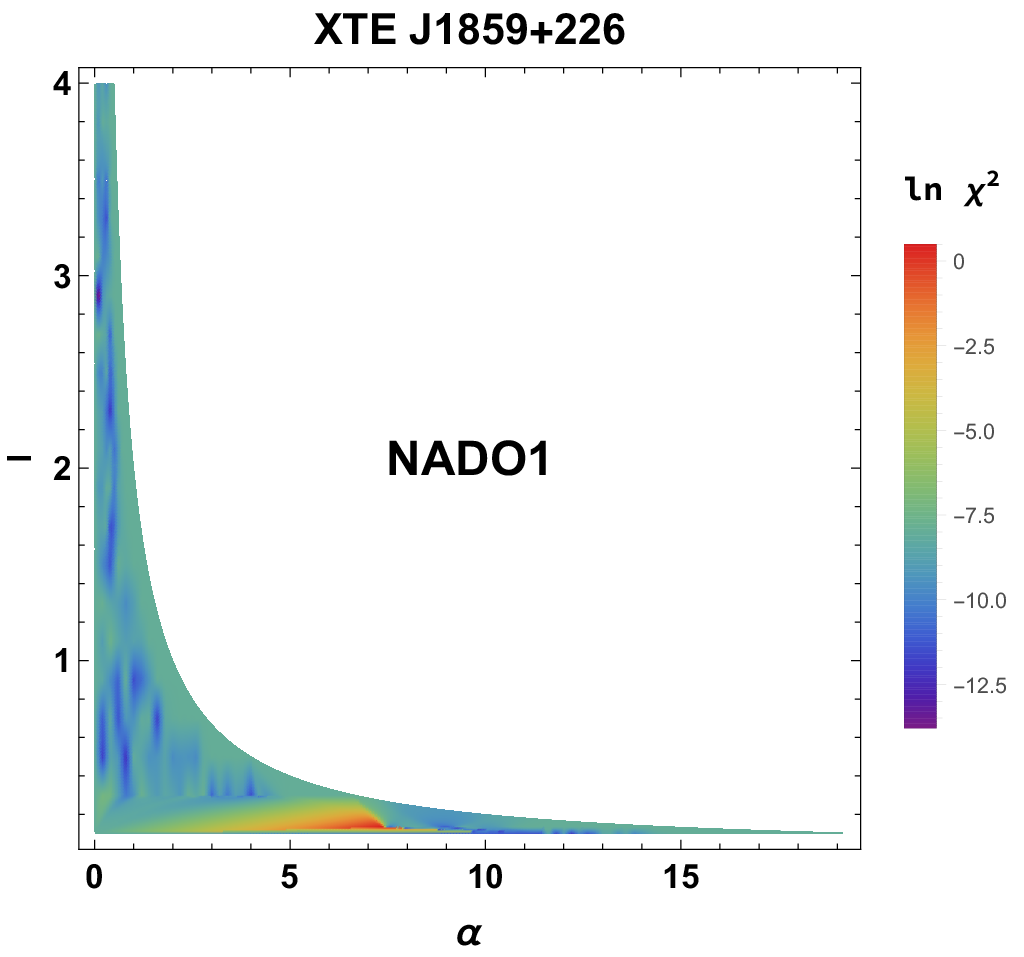}\label{g8}
	}
	\hfill
	\subfloat[]{
		\includegraphics[width=0.48\linewidth]{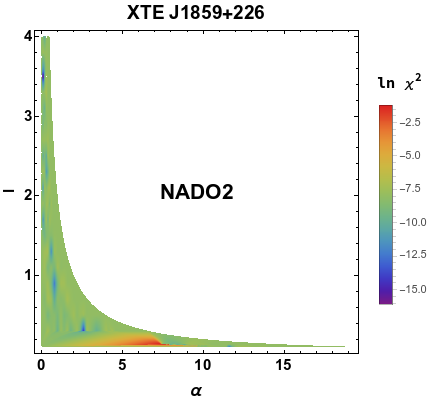}\label{g9}
	}

	\caption{Density plot showing the variation of $\ln \chi^{2}$ with the $(\alpha,l)$ parameter space for the  black hole XTE J1859+226 using the (a) Forced Resonance model 1 (FRM1), (b) Keplerian Resonance Model 2 (KRM2),  (c) Non Axisymmetric Disc Oscillation Model 1 (NADO1), and (d) Non Axisymmetric Disc Oscillation Model 2 (NADO2). It is evident that $\chi^2$ remains very small and shows only weak variation across the parameter space, leading to a broad nearly degenerate allowed region for all models. Consequently, the data includes the entire $l,\alpha$ parameter space and fails to distinguish between the Kerr and the hairy Kerr scenario based on these models.
    }
	\label{g10}
\end{figure}


\newpage
\item{\bf{Warped Disc Oscillation Model  (WDOM)}}
\begin{figure}[hbt!]
		\centering
		\subfloat[]
        {\includegraphics[width=0.48\linewidth]{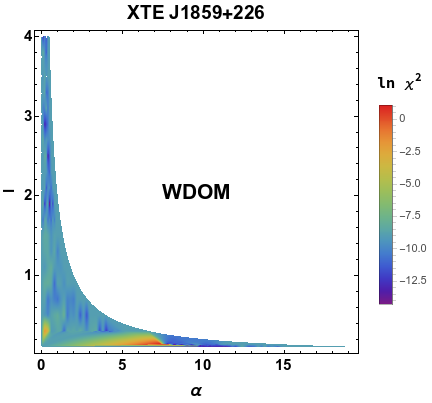}\label{g51}}
		\hfill
		\subfloat[]
        {\includegraphics[width=0.48\linewidth]{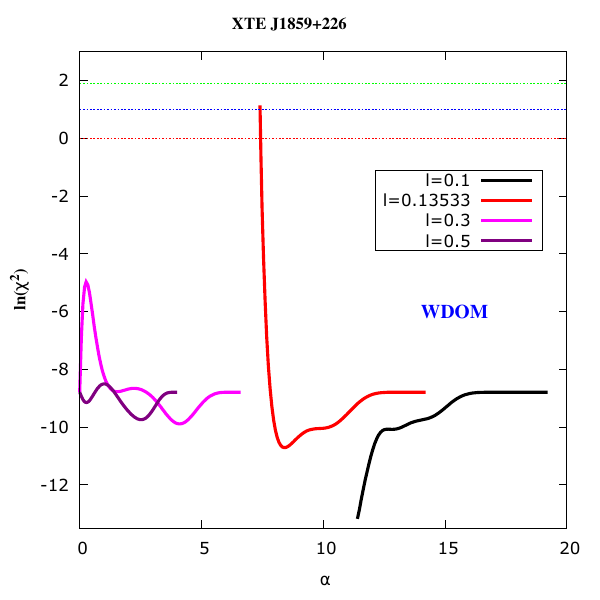}\label{g52}}
		\caption{(a) Density plot showing the variation of $\ln \chi^{2}$ with the $(\alpha,l)$ parameter space for the  black hole XTE J1859+226 using the Warped Disc Oscillation Model  (WDOM),  (b) Variation of $\ln \chi^{2}$ with the hair parameter $\alpha$ for $l = 0.1,0.13533,0.3,$ and $0.5$. The red, blue, and green horizontal dotted lines represent the $1\sigma$, $2\sigma$, and $3\sigma$ confidence limits, respectively. For  $l=e^{-2}$, $\alpha\simeq\alpha_{\rm crit}$ are excluded outside the $3\sigma$ confidence level. For all other values of $l$, the data cannot constrain the $l,\alpha$ parameter space.
        }
        \label{g5}
\end{figure}


\end{enumerate}

\section{Parameter constraints from MCMC analysis}
\label{MCMC}
In the previous section, the model parameters were obtained using a direct grid-search technique. In this method, we systematically varied the hairy parameters $l$,  $\alpha$, the black hole spin $a$, the mass $M$, and the emission radius $r_{\rm em}$ over their allowed ranges and evaluated the corresponding $\chi^2$ at each grid point. This has been discussed in detail in \autoref{Sec4}.  
Since the reliability of this procedure depends on the fineness of the parameter grid, it is important to verify that the true global minimum is not missed. To check the robustness of the grid-search results, we additionally perform a Bayesian parameter estimation using the Markov Chain Monte Carlo (MCMC) method.

The posterior probability distribution for the parameter vector
\[
\theta \equiv \{l,\alpha,a,M,r_{\rm em}\}
\]
is expressed as
\begin{align}
P(\theta|D)=\frac{P(D|\theta)\,P(\theta)}{P(D)},
\end{align}
where $P(\theta)$ denotes the prior distribution of the parameters, $P(D|\theta)$ represents the likelihood function, and $P(D)$ is the Bayesian evidence, which may be taken as unity since we are considering a fixed data set \cite{Verde:2009tu}.

We assume uniform priors for the parameters $l$, $\alpha$, $a$, and $r_{\rm em}$ within their physically allowed ranges. For the black hole mass $M$, Gaussian priors are adopted for GRO J1655-40 and XTE J1550-564 and uniform priors are adopted for H 1743+322, and Sgr A* using the independent observational constraints listed in \autoref{t1}.
In the present work, we consider only the twin high-frequency quasi-periodic oscillations (HFQPOs) as no resonance model take LFQPOs into account. Accordingly, the likelihood function is constructed solely from the upper and lower HFQPO frequencies . The total log-likelihood is therefore written as

\begin{figure}[H]

\centering

\begin{minipage}[b]{0.49\textwidth}
\centering
\includegraphics[width=\textwidth]{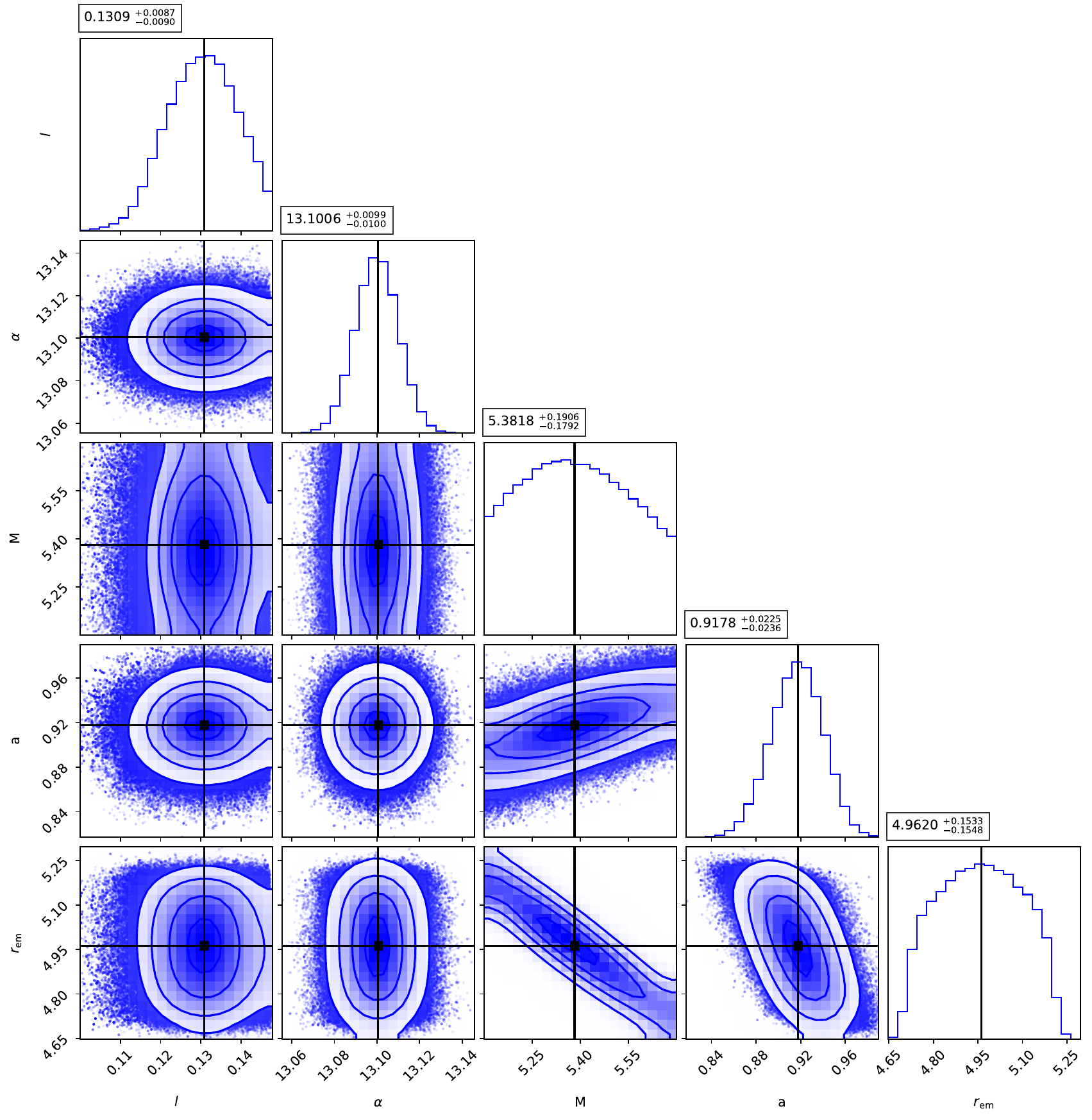}
\caption*{(a) GRO J1655-40 black hole (PRM model)}
\end{minipage}
\hfill
\begin{minipage}[b]{0.49\textwidth}
\centering
\includegraphics[width=\textwidth]{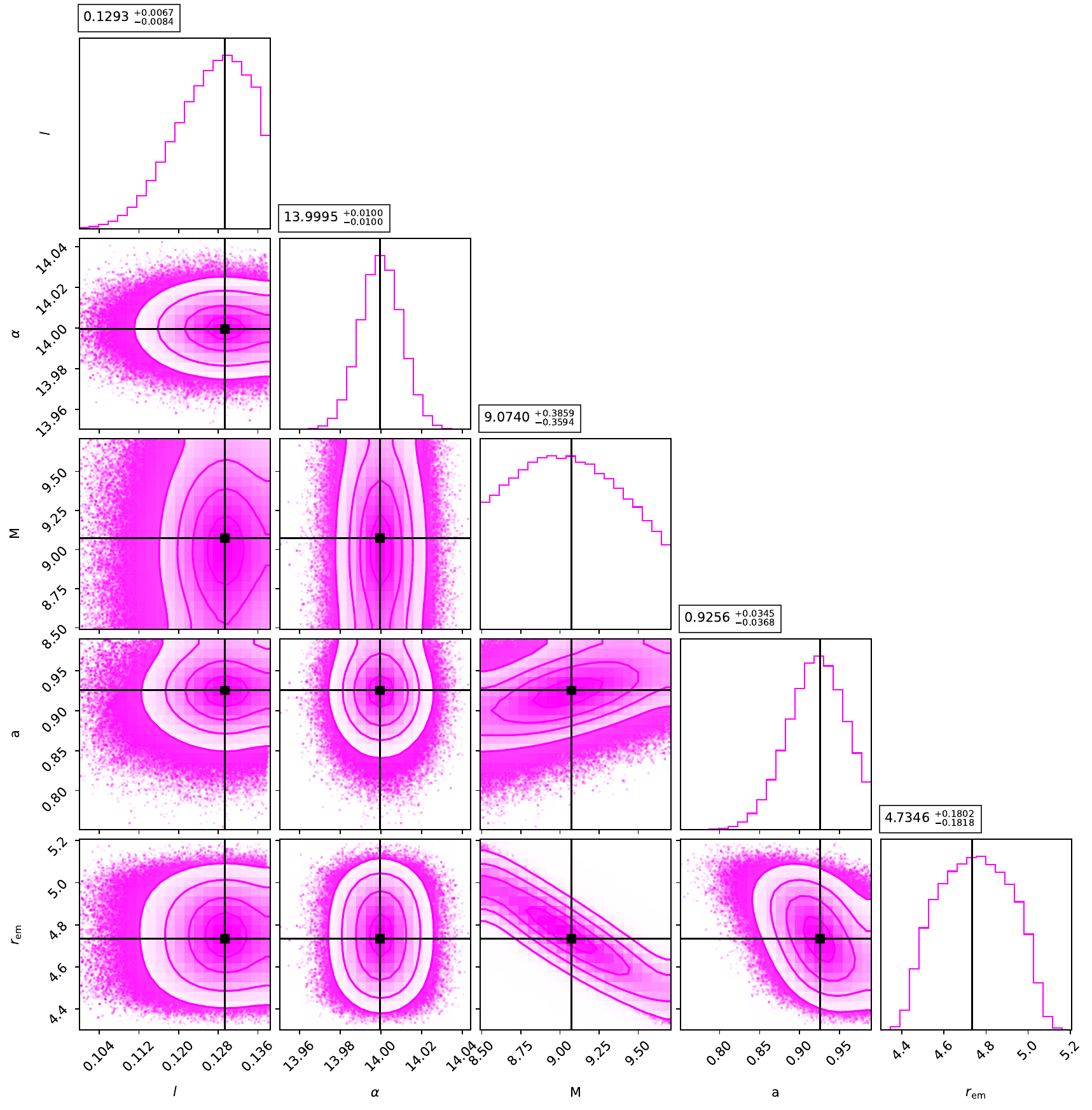}
\caption*{(b) XTE J1550-564  black hole (PRM model)}
\end{minipage}

\vspace{0.3cm}

\begin{minipage}[b]{0.49\textwidth}
\centering
\includegraphics[width=\textwidth]{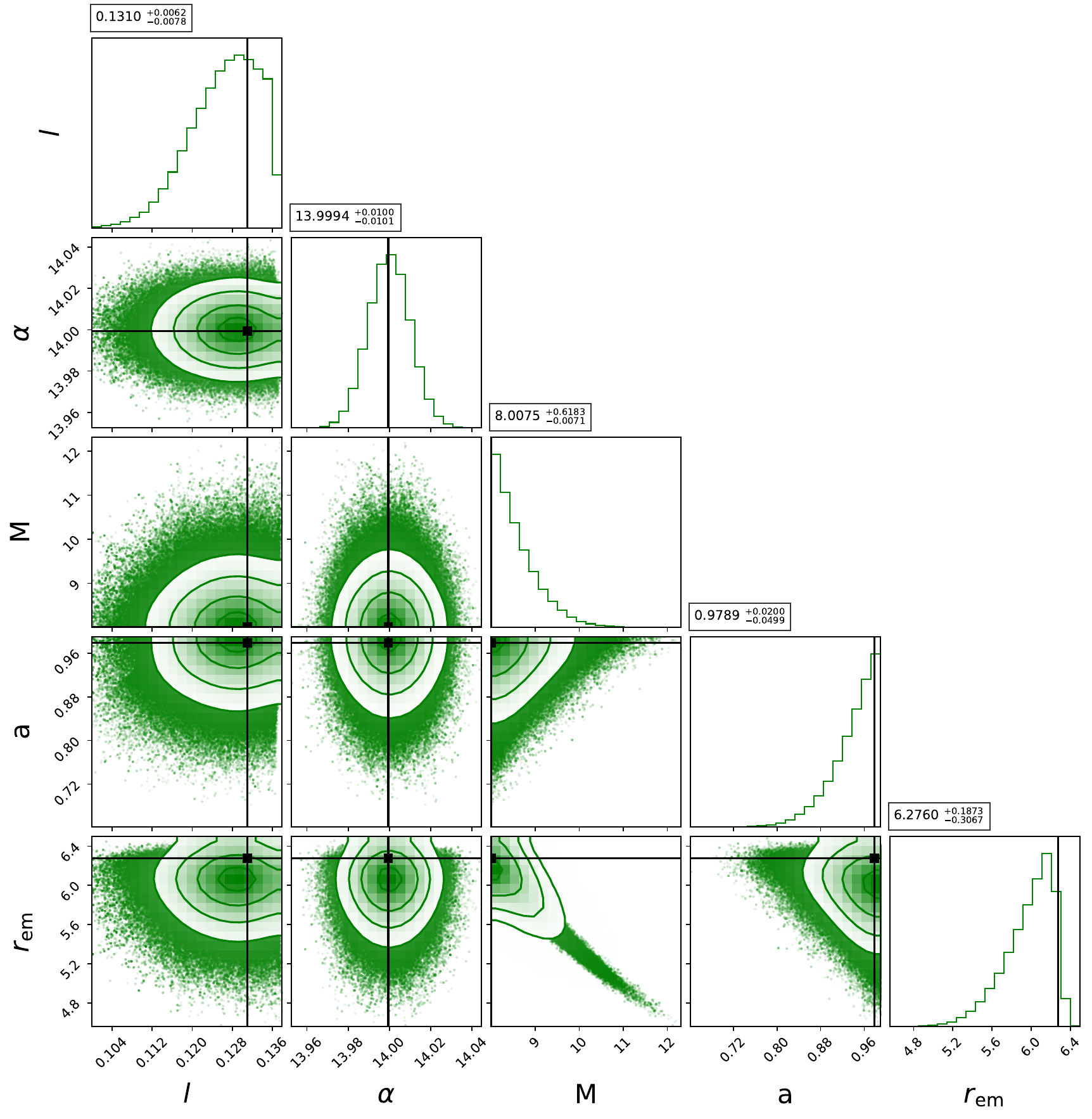}
\caption*{(c) H 1743+322 black hole (KRM1 model)}
\end{minipage}
\hfill
\begin{minipage}[b]{0.49\textwidth}
\centering
\includegraphics[width=\textwidth]{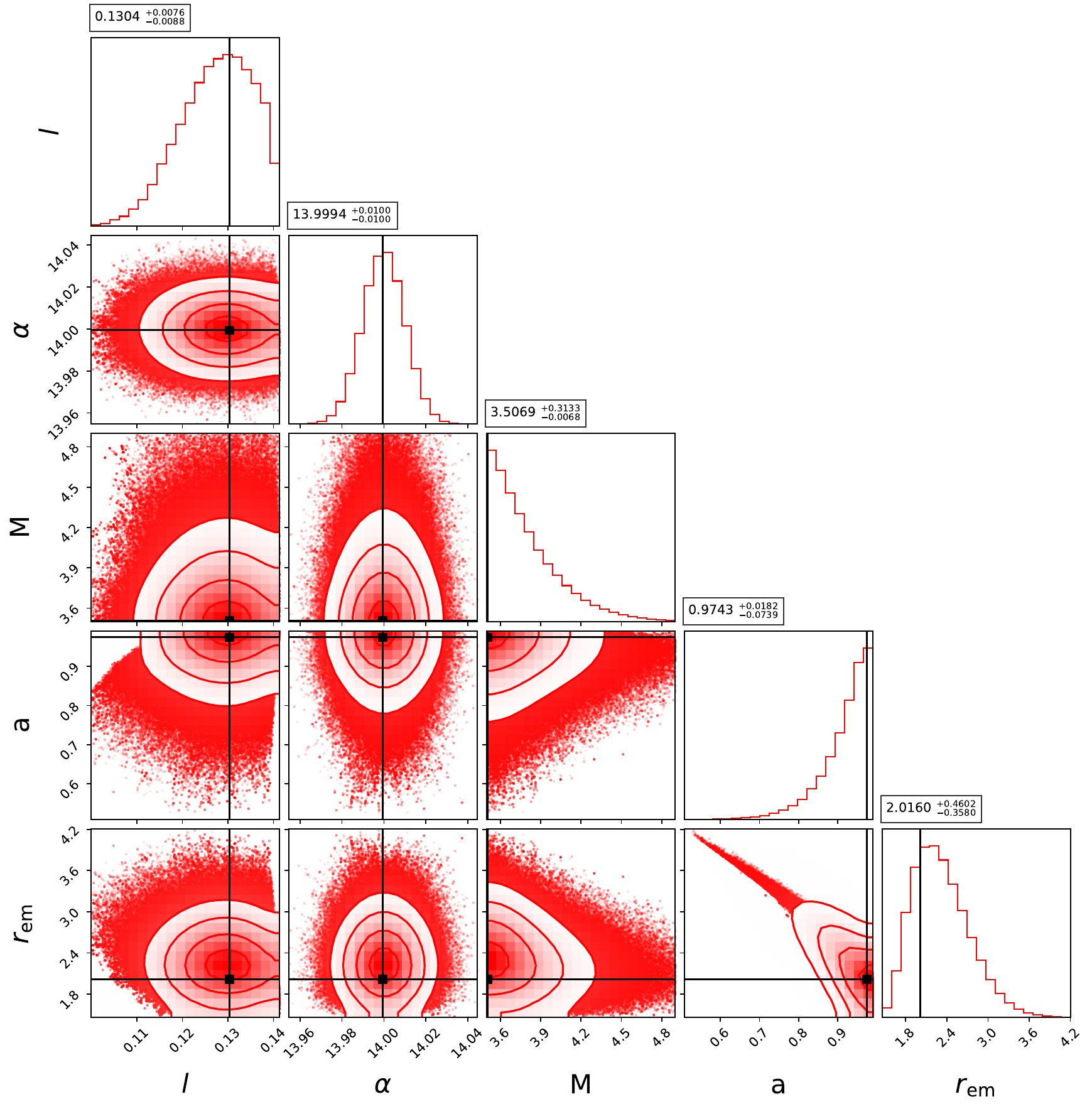}
\caption*{(d) Sgr A* black hole (FRM1 model)}
\end{minipage} 

\caption{Corner plots depicting the best-fit model parameters using the MCMC method}
\label{Fig3}
\end{figure}

\begin{align}
\log \mathcal{L} = \log \mathcal{L}_{U_1} + \log \mathcal{L}_{U_2},
\label{4}
\end{align}
where
\begin{eqnarray}
& & \log \mathcal{L}_{U_1} = -\frac{1}{2}\frac{\left[f_{{\rm up1}}-
f_{1}(l,\alpha,a,M,r_{\rm em})
\right]^2}{\sigma_{{\rm up1}}^{\,2}
},
\label{5} \\
& & \log \mathcal{L}_{U_2} = -\frac{1}{2} \frac{ \left[ f_{{\rm up2}}-
f_{2}(l,\alpha,a,M,r_{\rm em})
\right]^2}{\sigma_{{\rm up2},i}^{\,2}
}.
\label{6}
\end{eqnarray}
Here, $f_{{\rm up1}}$ and $f_{{\rm up2}}$ denote the observed upper and lower HFQPO frequencies, while $\sigma_{{\rm up1}}$ and $\sigma_{{\rm up2}}$ represent their corresponding observational uncertainties. We have done MCMC for those black holes and corresponding models which are able to constrain a good amount of parameter space in our previous grid search approach. As for GRS 1915+105 and XTE J1859+226, most models could not significantly constrain the parameter space, hence we do not present the MCMC results for these two black holes. For the rest of the four black holes listed in \autoref{t1}, some representative corner plots are shown in \autoref{Fig3}. We report that the best-fit parameters estimated using the MCMC approach are in agreement with the grid-search method discussed previously.

\subsection{Robustness of our results}
We further examine the sensitivity of our results to various factors, including the grid resolution and the choice of parameter ranges and priors. We make the following observations:


\begin{itemize}
\item \textit{Grid resolution and range of $l$:} We have carefully explored the theoretically allowed parameter space using a sufficiently fine grid. Specifically, we consider $0.1 \leq l \leq 4$, with a grid spacing of $\Delta l=0.2$. The choice $l=0$ is excluded, as it violates the strong energy condition (SEC). We have verified that our conclusions are not sensitive to the choice of the upper bound $l=4$. As evident from \autoref{a1}--\autoref{g5}, increasing $l$ does not introduce any qualitatively new features in the $\chi^2$ profiles. Instead, the deviation from the Kerr case decreases with increasing $l$, accompanied by a smooth decrease in the corresponding $\chi^2$ values. Moreover, well before $l=4$, the deviations from the observational constraints fall below the $1\sigma$ level for all sources. Thus, extending the parameter space to larger values of $l$ does not provide additional physically relevant information.

For each fixed value of $l$, the parameter $\alpha$ is varied over its physically allowed range, $\alpha_{\rm crit}\leq\alpha\leq\alpha_{\rm max}=2/l$, as determined by the SECs and the requirement of asymptotic flatness. For $l>4$, $\alpha_{\rm max}<0.5$, resulting in a highly restricted range of $\alpha$. We therefore restrict our analysis to $0.1\lesssim l\lesssim4$. This choice also allows us to consistently account for the substantially larger values of $\alpha$ permitted at smaller $l$; for example, $\alpha_{\rm max}\sim20$ for $l\sim0.1$.

\item \textit{Allowed $(l,\alpha)$ parameter space:} As demonstrated in \autoref{s2} and Paper I \cite{Dasgupta:2025blbe}, the parameter space can be naturally divided into three distinct regimes: (i) $l<e^{-2}$, (ii) $l=e^{-2}$, and (iii) $l>e^{-2}$. This distinction is important because, for a given $l$, the SECs do not permit the entire nominal range $0\lesssim\alpha\lesssim2/l$. In particular, only when $\alpha_{\rm crit}<0$ is the full range $0\lesssim\alpha\lesssim2/l$ consistent with the SECs. Furthermore, the SECs impose an additional restriction on the allowed spin for a given $(l,\alpha)$. Specifically, when $\alpha_{\rm crit}\lesssim\alpha\lesssim\alpha_{\rm coin}$, the maximum allowed spin is smaller than $a_{\rm max}$, the spin corresponding to an extremal black hole. These restrictions are fully incorporated when defining the parameter ranges used in our grid search.
The resolution in $\alpha$ is chosen adaptively. We use $\Delta\alpha=0.2$ for $0.1\leq l<1.7$ and $\Delta\alpha=0.1$ for $l\geq1.7$. This ensures adequate sampling of the parameter space, particularly in regions where the SECs significantly reduce the allowed range of $\alpha$.

\item \textit{Spin resolution:} For each $(l,\alpha)$, we calculate $\chi^2$ over the allowed spin range, $-a_{\rm max}\lesssim a\lesssim a_{\rm max}$ or $-a_{\rm SEC}\lesssim a\lesssim a_{\rm SEC}$, as appropriate, with an initial step size of $\Delta a=0.1$. As the spin approaches its maximum allowed value, $a_{\rm max}$ or $a_{\rm SEC}$, we further reduce the step size. This is necessary because the radius of the marginally stable circular orbit, $r_{\rm ms}$, decreases more rapidly as the spin approaches its maximum value, which can lead to rapid variations in the model-dependent QPO frequencies.

\item \textit{Resolution in $M$ and $r_{\rm em}$:} In addition to $(l,\alpha,a)$, the model-dependent QPO frequencies are sensitive to the black-hole mass $M$ and the characteristic radius $r_{\rm em}$. For stellar-mass black holes, we adopt $\Delta M=0.01~M_\odot$, whereas for the supermassive black hole Sgr A*, we use $\Delta M=10^{-5}~M_\odot$. The parameter $r_{\rm em}$ is varied over the range $r_{\rm ms}\lesssim r_{\rm em}\lesssim r_{\rm ms}+20R_g$, where $R_g=GM/c^2$ is the gravitational radius. We use $\Delta r_{\rm em}=0.1R_g$ for $r_{\rm em}>r_{\rm ms}+5R_g$ and a finer resolution of $\Delta r_{\rm em}=10^{-3}R_g$ for $r_{\rm ms}\lesssim r_{\rm em}\lesssim r_{\rm ms}+5R_g$. These choices ensure sufficiently fine sampling of the parameter space and minimize the possibility of missing the global minimum of $\chi^2$.

For most of the black holes considered here, the $\chi^2$ landscape does not exhibit a sharply localized minimum. Instead, $\chi^2_{\rm min}\sim0$ is obtained over a relatively broad range of parameter values. This behavior is also evident from the $\chi^2$ versus $l,\alpha$ plots presented in the previous section.

\item \textit{Choice of MCMC priors:} For the MCMC analysis, we adopt prior ranges that are consistent with the physically allowed parameter domains discussed above. We find that the best-fit parameter values remain stable under moderate variations of these prior ranges. While changing the priors can affect the detailed shape of the posterior distributions, it does not lead to appreciable changes in the inferred best-fit parameters. We therefore conclude that moderate changes in the prior ranges, as well as further refinement of the grid resolution, do not qualitatively alter our results.

\end{itemize}

Before concluding this section, we also note that Liu et al. \cite{Liu:2023ggze} previously performed a related analysis and obtained constraints on the hairy Kerr parameters using the Relativistic Precession Model (RPM). Our analysis differs from \cite{Liu:2023ggze} in the following respects:

\begin{itemize}

\item We have considered all the available resonance models in this work and the kinematic models in Paper I \cite{Dasgupta:2025blbe} while \cite{Liu:2023ggze} considers only the RPM. 
This allows us to assess the most suitable HFQPO model for each source which we report in the next section.

\item Liu et al.~\cite{Liu:2023ggze} adopts the priors $\alpha l\in[0,2)$ and $\alpha\in[0,5]$. In contrast, we impose the theoretical constraints from the SEC and asymptotic flatness. In particular, $l=0$ is excluded by the SEC, and we explore $0.1\lesssim l\lesssim4$, with the allowed range of $\alpha$ determined for each $l$. Since $\alpha l\lesssim2$, smaller $l$ permits substantially larger $\alpha$; for example, $\alpha_{\rm max}\sim20$ for $l\sim0.1$. Thus, our analysis covers regions of the physically allowed parameter space that are excluded by the priors adopted in Liu et al. ~\cite{Liu:2023ggze}.

\item We systematically incorporate the SEC constraints on both $\alpha$ and the black hole spin. As shown in \autoref{s2}, the allowed $(l,\alpha)$ parameter space comprises three distinct regimes, $l<e^{-2}$, $l=e^{-2}$, and $l>e^{-2}$. These restrictions are explicitly incorporated in our grid-search and MCMC analyses.

\item The restricted prior $\alpha\leq5$ in Liu et al. ~\cite{Liu:2023ggze} excludes, for example, the pronounced minimum we find around $l\sim0.1$ and $\alpha\sim12$ for GRO~J1655--40 within the RPM (Paper I \cite{Dasgupta:2025blbe}). Consequently, the parameter constraints obtained by us and Liu et al. \cite{Liu:2023ggze} are expected to differ.
\end{itemize}

\section{Model Comparison}
\label{sec6}
In the previous section, we considered nine well-established resonance models to constrain the hairy parameters from the six black holes listed in \autoref{t1}. Previously, in Paper I \cite{Dasgupta:2025blbe}, we considered the two available kinematic models (Relativistic Precession Model (RPM) and Tidal Disruption Model (TDM)) for constraining the hairy parameters. Now, it is essential to compare these models and analyze which of these eleven models can best fit the observational data for each black hole. In order to compare these models, we apply the Akaike Information Criterion ($AIC$) to get the best model corresponding to each black hole. $AIC$ is represented as,
\begin{align}
    AIC=-2\ln\mathcal L_{\rm max}+2p\,,
    \label{AIC}
\end{align}
where $\mathcal L_{\rm max}$ denotes the maximum likelihood and $p$ is the number of parameters involved in a particular model. The model that best fits the observational data will have the least $AIC$ value. 
It suggests that we can choose the best model only based on maximizing the likelihood. However, if this is achieved by increasing the number of model parameters, then the second term in \autoref{AIC} penalizes the model. Since in the present case all models have the same number of parameters ($l$, $\alpha$, $a$, $M$, $r_{\rm em}$), a comparison of $AIC$ amounts to a comparison of $-2\ln \mathcal L_{\rm max}$ which in turn is equivalent to $\chi^2_{\rm min}$. Hence, the model which has the lowest $\chi^2$ value will be treated as the best model for any particular black hole. 
Now, determining $AIC$ for a single model is not meaningful. This method becomes significant when we consider more than one model and compare their relative agreement with the observational data. We denote the $AIC$ of the best model as $AIC_{\rm min}$. Now, how much other models are significant relative to the best model can be determined by $\Delta_i=AIC_i-AIC_{\rm min}$, where $AIC_i$ denotes the $AIC$ of the i\textsuperscript{th} model. The criteria for model selection is described below \cite{Yang:2024mro},
     \begin{enumerate}
    \item $0 < \Delta_i \leq 2$: The $i^{\text{th}}$ model and the best model can be treated on equal footing.
    
    \item $2 < \Delta_i \leq 10$: The $i^{\text{th}}$ model is less significant than the best model.
    
    \item $\Delta_i > 10$: The $i^{\text{th}}$ model is unsupported by the observational data.
\end{enumerate}
We now compare the performance of the nine resonance models and also the two kinematic models (discussed in Paper I \cite{Dasgupta:2025blbe}) in explaining the QPO data for the six black hole sources, using AIC and the consistency of the inferred spins (based on the QPO based study) with independent spin estimates obtained from non-QPO observations. Since the physical origin of the HFQPOs is not yet conclusively established, it is not possible to identify apriori which model provides the most appropriate description for a given source. We therefore consider all the eleven models and assess their compatibility with the observations based on their statistical performance and inferred model parameters.

It is important to note that the existing independent spin estimates for these sources were obtained under the assumption that the black holes are described by the Kerr metric. Therefore, when comparing the spins inferred from the different models with these estimates, we consider the corresponding $a_{min}$ values in the $\alpha\simeq0$ (or $\alpha l\simeq 2$) limit, for which the spacetime approaches the Kerr case. This provides a consistent basis for comparison with the previously reported Kerr-based spin estimates. 

\begin{table}[htpb]
\centering
\setlength{\tabcolsep}{5pt}               
\renewcommand{\arraystretch}{1.6}         
\footnotesize
\begin{adjustbox}{max width=\textwidth}
\begin{tabular}{|l|l|l|l|l|l|l|l|}
\hline
\multicolumn{8}{|c|}{\textbf{GRO J1655-40}} \\ \hline
\multirow{2}{*}{\textbf{Previous constrains}}     & \multicolumn{3}{c|}{\textbf{Spin}} & \multicolumn{4}{c|}{\textbf{Mass $(M_{\odot})$} }\\  \cline{2-8} 
& {$\rm a\sim 0.65-0.75 $ \cite{shafee2005estimating}}  & {$\rm a\sim 0.94-0.98$ \cite{Miller:2009cw}} & {$ \rm a=0.29\pm 0.003  $ \cite{Motta:2013wga}}   &  \multicolumn{4}{|c|}{\textbf{$\rm 5.4\pm 0.3$} \cite{Beer:2001cg} } \\

   \hline \hline                 
\multirow{1}{*}{\textbf{Model}} & \multicolumn{7}{c|}{\textbf{}} \\ \cline{2-8} 
& \textbf{ $\mathbf{Best-fit (l,\alpha )}$} & \textbf{ $\mathbf{\chi^2_{\rm min}}$} & AIC &\textbf{Whether Kerr allowed within 1$\sigma$} & \textbf{Whether Kerr allowed within 3$\sigma$} & \textbf{$a_{\rm min}$ ~($\alpha\sim 0$)} & \textbf{$M_{\rm min}$ ~($\alpha\sim 0$)} \\\hline
PRM   & -   & 0 & 10 & Yes & Yes & $\sim 0.9$ &  5.17\\ \hline
FRM1   & -  & 0   & 10  & Yes & Yes & $\sim 0.4$ & 5.65\\  \hline
FRM2   & -   & 0 & 10 & Yes & Yes & $\sim 0.1$ &  5.32\\ \hline
KRM1   & -   & 2.3 & 12.3 & Yes & Yes & $\sim 0.999$ &  5.1\\ \hline
KRM2   & -   & 0 & 10 & Yes & Yes & $\sim 0.3$ &  5.26\\ \hline
KRM3   & -   & 0 & 10 & Yes & Yes & $\sim 0.2$ &  5.58\\ \hline
WDOM   & -  & 0   & 10  & Yes & Yes & $\sim 0.1$ & 5.33\\  \hline
NADO1   & -  & 0   & 10  & Yes & Yes & $\sim 0.5$ & 5.63\\  \hline
NADO2   & -  & 0   & 10  & Yes & Yes & $\sim 0.2$ & 5.13\\  \hline
\hline
RPM \cite{Dasgupta:2025blbe}  & $l \sim 0.1$, $\alpha \sim 12$    &  0.0146 & 10 & No & No & $\sim 0.3$ &  5.12\\ \hline
TDM \cite{Dasgupta:2025blbe}  & -  & 0   & 10  & Yes & Yes & $\sim 0.1$ & 5.32\\ 
\hline
\end{tabular}
\end{adjustbox}
\caption{Table summarizing the best-fit hairy parameters for GRO J1655-40 and comparing the spin derived from QPO related studies with previous independent estimates.
}
\label{gro}
\end{table}

In \autoref{gro} we summarize the best-fit hairy parameters for GRO J1655-40 and compare the spin derived from each of the QPO models studied here with the previous independent estimates. For completeness, we also report the best-fit model parameters based on the RPM and the TDM studied in Paper I \cite{Dasgupta:2025blbe}. From the table it is clear that for most models $\chi^2_{\rm min}\sim 0$ and hence $AIC\sim 10$, except KRM1. For KRM1, $\chi^2_{\rm min}\sim 2.31$, that gives $AIC=12.31$. Hence, based on the above discussion, this model seems to be less supported by the observational data. However, if we compare the spin predicted by this model ($a_{\rm min}\sim 0.999$) assuming the Kerr geometry, then it exhibits good agreement with the spin estimated independently by the Fe-line method \cite{Miller:2009cw}. Such agreement is also somewhat exhibited by the PRM model ($a_{\rm min}\sim 0.9$, \autoref{gro}). Interestingly, previous spin estimates predict the spin of GRO J1655-40 to be $a\sim 065-0.75$ from the Continuum-Fitting method \cite{shafee2005estimating}, $a\sim 0.94-0.98$ from the Fe-line method \cite{Miller:2009cw} and $a=0.29\pm0.003$ based on the QPO data assuming the RPM model \cite{Motta:2013wga}-and these are not mutually consistent. While the Fe-line method predicts high spin for GRO J1655-40, the spin predicted by the Continuum-Fitting method is moderate. Interestingly, the spin predicted by most of the QPO models in this work and Paper I are inconsistent with at least one of the previous estimates. In this regard, PRM and KRM1 score better as discussed above. But KRM1 has a higher $AIC$ compared to PRM, hence one may presume that PRM may be the most suitable model for this source. 

Except the RPM, all the remaining models cannot distinguish between the Kerr and the hairy Kerr scenario. 
From \autoref{gro} we find that the RPM provides the strongest constraint on the hairy parameters, strongly favoring $l\sim 0.1, \alpha \sim 12$ and ruling out the Kerr scenario outside 3$\sigma$. Moreover, the spin predicted by this model assuming the Kerr scenario is $a_{\rm min}\sim 0.3$ which agrees with the estimate reported in \cite{Motta:2013wga}. This, however, is expected since Motta et al. \cite{Motta:2013wga} derived the spin assuming the RPM model and the Kerr geometry. This spin however is inconsistent with the spin estimates obtained from the other independent methods \cite{shafee2005estimating, Miller:2009cw} assuming the Kerr geometry. Taken together, these results suggest that PRM (and possibly KRM1) warrant further consideration as possible models for describing the HFQPOs of GRO J1655-40.

The discrepancy between the spin estimates obtained from the Continuum-Fitting and the Fe-line methods \cite{shafee2005estimating,Miller:2009cw} is also noteworthy. Since both analyses assumed the Kerr metric, this discrepancy raises the possibility that additional black-hole parameters could play a role in reconciling the different observational inferences. This possibility is further strengthened when we note that the present data cannot rule out the hairy Kerr scenario.
A consistent determination of the additional hair parameter(s) and the spin parameter from independent spectral and timing observations would, however, require more precise X-ray data. Future high-quality spectral and timing observations of GRO J1655-40 will therefore be important for testing this possibility.


For the BH XTE J1550-564 all models are equally favored based on $AIC$ (\autoref{xte1}). KRM1 produces marginally higher $AIC \sim 10.5$ while the other ten models yield $AIC=10$. Since $\Delta_i<2$ for KRM1, we can take KRM1 in equal footing with the other ten models, based on $AIC$. Previous independent spin estimates for this source reveal that 
$a\sim 0.34$ from the Continuum-Fitting method, with an uncertainty range of $-0.11<a<0.71$, $a=0.55^{+0.15}_{-0.22}$ from the Fe-line analysis \cite{Steiner:2010bt}, and $0.3\lesssim a\lesssim 0.6$ from the jet-power analysis \cite{Banerjee:2020ubc}. 
In this case the earlier independent spin estimates are not so discrepant, although the error bar in the spin determined by the Continuum-Fitting method is quite high.
Assuming a Kerr background, the QPO models which best agree with at least one of the previous estimates are NADO1, FRM1, KRM2 and RPM, followed by FRM2, WDOM and NADO2 (\autoref{xte1}). Hence, comparison with independent spin measurements indicate that one of these models may be a suitable description of the HFQPO data of XTE J1550-564.
Notably, none of the previous estimates agree with the spin of this source predicted based on PRM and KRM1 (\autoref{xte1}), hence these models do not seem to be suitable for XTE J1550-564. Interestingly, none of these models 
can rule out the hairy Kerr scenario compared to the Kerr case.


\begin{table}[htpb]
\centering
\setlength{\tabcolsep}{5pt}               
\renewcommand{\arraystretch}{1.6}         
\footnotesize
\begin{adjustbox}{max width=\textwidth}
\begin{tabular}{|l|l|l|l|l|l|l|l|}
\hline
\multicolumn{8}{|c|}{\textbf{XTE J1550-564}} \\ \hline
\multirow{2}{*}{\textbf{Previous constrains}}     & \multicolumn{3}{c|}{\textbf{Spin}} & \multicolumn{4}{c|}{\textbf{Mass $(M_{\odot})$} }\\  \cline{2-8} 
& {$-0.11<a<0.71 $ \cite{Steiner:2010bt}}  & {$a=0.55^{+0.15}_{-0.22}$ \cite{Steiner:2010bt}} & $0.3\lesssim a \lesssim 0.6$ \cite{Banerjee:2020ubc}  &  \multicolumn{4}{|c|}{\textbf{$\rm 9.1\pm 0.61$} \cite{Orosz:2011ki} } \\

   \hline \hline                 
\multirow{2}{*}{\textbf{Model}} & \multicolumn{7}{c|}{\textbf{}} \\ \cline{2-8} 
& \textbf{ $\mathbf{Best-fit (l,\alpha )}$} & \textbf{ $\mathbf{\chi^2_{\rm min}}$} & AIC &\textbf{Whether Kerr allowed within 1$\sigma$} & \textbf{Whether Kerr allowed within 3$\sigma$} & \textbf{$a_{\rm min}$ ~($\alpha\sim 0$)} & \textbf{$M_{\rm min}$~($\alpha\sim 0$)}\\ \hline
PRM   & -   & 0 & 10 & Yes & Yes & $\sim 0.9$ &  8.58\\ \hline
FRM1   & -  & 0   & 10  & Yes & Yes & $\sim 0.4$ & 8.94\\  \hline
FRM2   & -   & 0 & 10 & Yes & Yes & $\sim 0.3$ &  9.65\\ \hline
KRM1   & -   & 0.5 & 10.5 & Yes & Yes & $\sim 0.999$ &  8.49\\ \hline
KRM2   & -   & 0 & 10 & Yes & Yes & $\sim 0.4$ &  9.29\\ \hline
KRM3   & -   & 0 & 10 & Yes & Yes & $\sim 0.2$ &  8.91\\ \hline
WDOM   & -  & 0   & 10  & Yes & Yes & $\sim 0.3$ & 9.64\\  \hline
NADO1   & -  & 0   & 10  & Yes & Yes & $\sim 0.5$ & 8.9\\  \hline
NADO2   & -  & 0   & 10  & Yes & Yes & $\sim 0.3$ & 9.1\\  \hline
\hline
RPM \cite{Dasgupta:2025blbe}  & -    & 0 & 10 & Yes & Yes & $\sim 0.4$ &   9.29\\ \hline
TDM \cite{Dasgupta:2025blbe}  & -  & 0   & 10  & Yes & Yes & $\sim 0.2$ &  8.91\\ 
\hline
\end{tabular}
\end{adjustbox}
\caption{Table summarizing the best-fit hairy parameters for XTE J1550-564  and comparing the spin derived from QPO related studies with previous independent estimates.}
\label{xte1}
\end{table}

For GRS 1915+105, \autoref{grs} reveals that all the QPO models yield $\chi^2_{\rm \min}\sim 0$, corresponding to $AIC\sim 10$. 
Thus, based on the $AIC$, the different models receive essentially comparable statistical support from the HFQPO data. Assuming a Kerr background, the spin estimates for GRS 1915+105 obtained from different independent methods exhibit considerable variation. For instance, Fe-line analyses have yielded $a\sim 0.6-0.98$ \cite{Blum:2009ez}, while Continuum-Fitting analyses have reported both intermediate spins around $a\sim 0.7$ \cite{Middleton:2006kj} and near-maximal spin, $a\sim 0.98$ \cite{McClintock:2006xd}. More recent analyses incorporating updated measurements of the mass and inclination of GRS 1915+105 constrain the spin to approximately $a\sim 0.4-0.98$ \cite{Mills:2021dxs}. Estimates based on jet-power observations similarly indicate a relatively high spin, $a\sim 0.6-0.9$ \cite{Banerjee:2020ubc}.
In the hairy Kerr model, with $\alpha\sim 0$, 
we note that that FRM2, KRM3, WDOM and TDM predict retrograde spin while FRM1, KRM2, NADO1, NADO2 and RPM predict static to very low BH spins.
Therefore the spin estimations based on these models are not in agreement with any of the previous spin estimates. In this regard, PRM and KRM1 seem more suitable for this source (\autoref{grs}). However, they do not provide a clear distinction between the Kerr and hairy Kerr scenarios (\autoref{c1}-\autoref{c2}).
This when coupled with the substantial spread among the previously reported spin estimates highlight the importance of reassessing these measurements within a more general gravitational framework.
In particular, a joint analysis of the Continuum-Fitting and Fe-line observations allowing for deviations from the Kerr geometry could help determine whether part of the discrepancy among the existing spin estimates can be attributed to the assumption of a Kerr background. Such an analysis, together with more precise spectral and timing observations, would be necessary to assess this possibility.


\begin{table}[htpb]
\centering
\setlength{\tabcolsep}{5pt}               
\renewcommand{\arraystretch}{1.6}         
\footnotesize
\begin{adjustbox}{max width=\textwidth}
\begin{tabular}{|l|l|l|l|l|l|l|l|}
\hline
\multicolumn{8}{|c|}{\textbf{GRS 1915+105}} \\ \hline
\multirow{2}{*}{\textbf{Previous constrains}}     & \multicolumn{5}{c|}{\textbf{Spin}} & \multicolumn{2}{c|}{\textbf{Mass $(M_{\odot})$} }\\  \cline{2-8} 
& {$ \rm a\sim 0.98$ \cite{McClintock:2006xd}} &  {$\rm a\sim 0.7$ \cite{Middleton:2006kj}} & {$ \rm a\sim 0.6-0.98  $ \cite{Blum:2009ez}}   &  $a \sim 0.4-0.98$ \cite{Mills:2021dxs} & $a \sim 0.6-0.9$ \cite{Banerjee:2020ubc} & \multicolumn{2}{|c|}{\textbf{$\rm 12.4^{+2.0}_{-1.8}$} \cite{Reid:2014ywa} } \\

   \hline \hline                 
\multirow{2}{*}{\textbf{Model}} & \multicolumn{7}{c|}{\textbf{}} \\ \cline{2-8} 
& \textbf{ $\mathbf{Best-fit (l,\alpha )}$} & \textbf{ $\mathbf{\chi^2_{\rm min}}$} & AIC &\textbf{Whether Kerr allowed within 1$\sigma$} & \textbf{Whether Kerr allowed within 3$\sigma$} & \textbf{$a_{\rm min}$ ~($\alpha\sim 0$)} & \textbf{$M_{\rm min}$~($\alpha\sim 0$)}\\ \hline
PRM   & -   & 0 & 10 & Yes & Yes & $\sim 0.8$ &  11.47\\ \hline
FRM1   & -  & 0   & 10  & Yes & Yes & $\sim 0.0$ & 11.0\\  \hline
FRM2   & -   & 0 & 10 & Yes & Yes & $\sim -0.1$ &  12.12\\ \hline
KRM1   & -   & 0. & 10 & Yes & Yes & $\sim 0.95$ &  11.77\\ \hline
KRM2   & -   & 0 & 10 & Yes & Yes & $\sim 0.0$ &  10.89\\ \hline
KRM3   & -   & 0 & 10 & Yes & Yes & $\sim -0.1$ &  11.91\\ \hline
WDOM   & -  & 0   & 10  & Yes & Yes & $\sim -0.3$ & 10.77\\  \hline
NADO1   & -  & 0   & 10  & Yes & Yes & $\sim 0.0$ & 11.0\\  \hline
NADO2   & -  & 0   & 10  & Yes & Yes & $\sim 0.1$ & 7.7\\  \hline
\hline
RPM \cite{Dasgupta:2025blbe}  & -   & 0 & 10 & Yes & Yes & $\sim 0.0$  & 11.00\\ \hline
TDM \cite{Dasgupta:2025blbe}  & -  & 0   & 10  & Yes & Yes & $\sim -0.1$ & 12.11\\ 
\hline
\end{tabular}
\end{adjustbox}
\caption{Table summarizing the best-fit hairy parameters for GRS 1915+105 and comparing the spin derived from QPO related studies with previous independent estimates.
}
\label{grs}
\end{table}


For H1743-322, a comparison of \autoref{d1}-\autoref{d2} show that all the models yield $\chi^2_{\rm \min}\sim 0$, while $\chi^2$ remains largely insensitive to variations in $l$ and $\alpha$. Thus, the HFQPO data do not provide a meaningful distinction between the Kerr and the hairy Kerr scenarios. 
For comparison, the Continuum-Fitting method gives a spin estimate of $a=0.2\pm0.3$ at the 
68\% confidence level, with an upper bound of $a<0.92$ at the $99.7\%$ confidence level \cite{Steiner:2011kd}. An independent estimate based on the observed jet power constrains the spin to $0.25\lesssim a\lesssim0.5$ \cite{Banerjee:2020ubc}, which is broadly consistent with the Continuum-Fitting result.
Based on the present analysis, the spin inferred from the model (assuming $\alpha\simeq 0$)  NADO2
exhibit the best agreement with the existing estimates \cite{Steiner:2011kd,Banerjee:2020ubc} while NADO1, WDOM, TDM, KRM2, KRM3, FRM2  are allowed within the error bars (\autoref{h}). In particular, RPM, KRM1, PRM and FRM1 do not seem to be a suitable description for the HFQPO data of this source.

\begin{table}[htpb]
\centering
\setlength{\tabcolsep}{5pt}               
\renewcommand{\arraystretch}{1.6}         
\footnotesize
\begin{adjustbox}{max width=\textwidth}
\begin{tabular}{|l|l|l|l|l|l|l|l|}
\hline
\multicolumn{8}{|c|}{\textbf{H1743-322}} \\ \hline
\multirow{2}{*}{\textbf{Previous constrains}}     & \multicolumn{2}{c|}{\textbf{Spin}} & \multicolumn{5}{c|}{\textbf{Mass $(M_{\odot})$} }\\  \cline{2-8} 
& {$ a=0.2\pm0.3 $ \cite{Steiner:2011kd}} &  {$0.25\lesssim a\lesssim0.5$ \cite{Banerjee:2020ubc}}   &  \multicolumn{5}{|c|}{\textbf{$8-14.07$} 
\cite{Pei:2016kka,Bhattacharjee:2017rbl,Petri:2008jc} } \\

   \hline \hline                 
\multirow{2}{*}{\textbf{Model}} & \multicolumn{7}{c|}{\textbf{}} \\ \cline{2-8} 
& \textbf{ $\mathbf{Best-fit (l,\alpha )}$} & \textbf{ $\mathbf{\chi^2_{\rm min}}\sim$} & AIC &\textbf{Whether Kerr allowed within 1$\sigma$} & \textbf{Whether Kerr allowed within 3$\sigma$} & \textbf{$a_{\rm min}$ ~($\alpha\sim 0$)} & \textbf{$M_{\rm min}$~($\alpha\sim 0$)} \\ \hline
PRM   & -   & 0 & 10 & Yes & Yes & $\sim 0.9$ &  9\\ \hline
FRM1   & -  & 0   & 10  & Yes & Yes & $\sim 0.7$ & 14.06\\  \hline
FRM2   & -   & 0 & 10 & Yes & Yes & $\sim -0.1$ &  8.7\\ \hline
KRM1   & -   & 0 & 10 & Yes & Yes & $\sim 0.999$ &  8.14\\ \hline
KRM2   & -   & 0 & 10 & Yes & Yes & $\sim 0.5$ &  11.62\\ \hline
KRM3   & -   & 0 & 10 & Yes & Yes & $\sim 0.0$ &  8.81\\ \hline
WDOM   & -  & 0   & 10  & Yes & Yes & $\sim 0.3$ & 11.68\\  \hline
NADO1   & -  & 0   & 10  & Yes & Yes & $\sim 0.4$ & 9.73\\  \hline
NADO2   & -  & 0   & 10  & Yes & Yes & $\sim 0.2$ & 9.46\\  \hline
\hline
RPM \cite{Dasgupta:2025blbe}  & -   & 0 & 10 & Yes & Yes & $\sim 0.6$ &  13.59\\ \hline
TDM \cite{Dasgupta:2025blbe}  & -  & 0   & 10  & Yes & Yes &  $\sim 0.1$ & 11.63\\ 
\hline
\end{tabular}
\end{adjustbox}
\caption{Table summarizing the best-fit hairy parameters for H1743-322 and comparing the spin derived from QPO related studies with previous independent estimates.}
\label{h}
\end{table}


\begin{table}[htpb]
\centering
\setlength{\tabcolsep}{5pt}               
\renewcommand{\arraystretch}{1.6}         
\footnotesize
\begin{adjustbox}{max width=\textwidth}
\begin{tabular}{|l|l|l|l|l|l|l|l|}
\hline
\multicolumn{8}{|c|}{\textbf{Sgr A*}} \\ \hline
\multirow{2}{*}{\textbf{Previous constrains}}     & \multicolumn{6}{c|}{\textbf{Spin}} & \multicolumn{1}{c|}{\textbf{Mass $(M_{\odot})$} }\\  \cline{2-8} 
& { $\rm a= 0.9\pm 0.06$ \cite{Daly:2023axh}}& 
{$ \rm a\sim 0.9$ \cite{Moscibrodzka:2009gw}} &  {$\rm a\sim 0.52$ \cite{Genzel:2003as}} & {$\rm a\sim 0.22$ \cite{Belanger:2006gm}} & {$ a \lesssim 0.1 $ \cite{Fragione:2020khu}}   & $\rm 0.4 \lesssim a \lesssim 1$ \cite{Meyer:2006fd} &
\multicolumn{1}{|c|}{\textbf{$(3.5-4.9)*10^6$} \cite{Ghez:2008ms,Gillessen:2008qv} } \\
    
   \hline \hline                 
\multirow{2}{*}{\textbf{Model}} & \multicolumn{7}{c|}{\textbf{}} \\ \cline{2-8} 
& \textbf{ $\mathbf{Best-fit (l,\alpha )}$} & \textbf{ $\mathbf{\chi^2_{\rm min}}\sim$} & AIC &\textbf{Whether Kerr allowed within 1$\sigma$} & \textbf{Whether Kerr allowed within 3$\sigma$} & \textbf{$a_{\rm min}$ ~($\alpha\sim 0$)} & \textbf{$M_{\rm min}$ ~($\alpha\sim 0$)}\\ \hline
PRM   & -   & 21.29 & 31.29 & Yes & Yes & $\sim 0.999$ &  $3.5\times10^6$\\ \hline
FRM1   & -  & 0.27   & 10.27  & Yes & Yes & $\sim 0.95$ & $3.5\times10^6$\\  \hline
FRM2   & -   & 0 & 10 & Yes & Yes & $\sim 0.999$ &  $4\times10^6$\\ \hline
KRM1   & -   & 24.86 & 34.86 & Yes & Yes & $\sim 0.999$ &  $3.5\times10^6$\\ \hline
KRM2   & -   & 0 & 10 & Yes & Yes & $\sim 0.9$ &  $3.9\times10^6$\\ \hline
KRM3   & -   & 0 & 10 & Yes & Yes & $\sim 0.99$ &  $4.2\times10^6$\\ \hline
WDOM   & -  & 0   & 10  & Yes & Yes & $\sim 0.999$ & $3.6\times10^6$\\  \hline
NADO1   & -  & 3.21   & 13.21  & Yes & Yes & $\sim 0.999$ & $3.5\times10^6$\\  \hline
NADO2   & -  & 0   & 10  & Yes & Yes & $\sim 0.8$ & $4.0\times10^6$\\  \hline
\hline
RPM \cite{Dasgupta:2025blbe}   & -   & 0 & 10 & Yes & Yes & $\sim 0.999$ & $4.4\times 10^6$ \\ \hline
TDM \cite{Dasgupta:2025blbe}  & -  & 0   & 10  & Yes & Yes & $\sim 0.999$ & $3.5\times 10^6$\\ 
\hline

\end{tabular}
\end{adjustbox}
\caption{Table summarizing the best-fit hairy parameters for Sgr A* and comparing the spin derived from QPO related studies with previous independent estimates.
}
\label{sgr}
\end{table}
From \autoref{f1}-\autoref{f9} we note that for the source Sgr A*, $\chi^2_{\rm min}\sim 0$ and $AIC\sim 10$ for the models FRM2, KRM2, KRM3, WDOM, NADO2, RPM and TDM. Although FRM1 has $AIC>10$, its $\Delta_i<2$ (\autoref{sgr}) and hence receive essentially comparable statistical support as the aforesaid models. 
NADO1 has $\Delta_i>2$ and hence has less support compared to FRM2, KRM2, KRM3, WDOM, NADO2, RPM, TDM and FRM1.
Models like PRM and KRM1 have significantly higher $AIC$ (\autoref{sgr}) with $\Delta_i>10$
and hence are unsupported by the HFQPO data of Sgr A*. 

Previous estimates of the spin of Sgr A* exhibit considerable variation. Some analyses favor a nearly vanishing spin, $a\lesssim0.1$ \cite{Fragione:2020khu}, while others indicate intermediate-to-low spins, with $a\sim0.52$, $a\sim0.5$, and $a\sim0.22$ \cite{Genzel:2003as,Shcherbakov:2010ki,Belanger:2006gm}. In contrast, several studies have reported high or near-extremal spins, including $a=0.9\pm0.06$, $a\sim0.9$, and $0.4\lesssim a\lesssim1$ \cite{Daly:2023axh,Moscibrodzka:2009gw,Meyer:2006fd}. It is important to note that these spin estimates were obtained under the assumption that Sgr A* is described by the Kerr geometry.
In the present analysis, we instead consider a hairy Kerr geometry. For $\alpha\sim0$, all models predict high to maximal spin for Sgr A* (\autoref{sgr}) in agreement with at least some existing independent spin estimates \cite{Daly:2023axh,Meyer:2006fd}. However, the best agreement is exhibited by the models FRM1 and KRM2 (\autoref{sgr}). These models are also preferred based on the $AIC$.

The substantial spread among the existing spin estimates, however, highlights the potential importance of the assumed background geometry in determining the spin of Sgr A*. A joint reanalysis of the existing observational constraints within a non-Kerr framework, such as the hairy Kerr geometry considered here, could therefore be useful in assessing whether the discrepancy among the previous spin measurements can be reduced when additional geometric parameters are allowed.


For XTE J1859+226, all the models yield $\chi^2_{\rm \min}\sim0$ with $AIC\sim 10$ and are therefore equally favored based on the $AIC$. The previous spin estimates of this source yield very different results.
QPO-based analyses using the RPM estimate a relatively low spin of $a=0.149\pm0.005$ \cite{Motta:2022rku}, which is consistent with the spin obtained in our RPM-based analysis (\autoref{xte2}). In contrast, independent spin measurements favor a near-maximal value of $a=0.987\pm0.003$ \cite{Mall:2023nab}. 
The TDM model under the Kerr assumption predicts $a_{\rm \min}\sim0.9$ (\autoref{xte2}), which is more consistent with the latter estimate \cite{Mall:2023nab}. The spin predicted by the remaining models yield low to intermediate values and hence are not consistent with \cite{Mall:2023nab}. Thus, the TDM seems to provide a potentially more consistent interpretation of the HFQPO data when compared with the existing independent spin measurements.
One may note that under the RPM, the sharp $\chi^2$ minimum and the MCMC analysis favor a hairy Kerr black hole with $l\sim0.1$ and $\alpha\sim12$ (\autoref{xte2} Paper I, \cite{Dasgupta:2025blbe}), excluding the Kerr case beyond $3\sigma$. However, this conclusion is model dependent because the correct QPO model for the source is not yet established.


\begin{table}[htpb]
\centering
\setlength{\tabcolsep}{5pt}               
\renewcommand{\arraystretch}{1.6}         
\footnotesize
\begin{adjustbox}{max width=\textwidth}
\begin{tabular}{|l|l|l|l|l|l|l|l|}
\hline
\multicolumn{8}{|c|}{\textbf{XTE J1859+226}} \\ \hline
\multirow{2}{*}{\textbf{Previous constrains}}     & \multicolumn{2}{c|}{\textbf{Spin}} & \multicolumn{5}{c|}{\textbf{Mass $(M_{\odot})$} }\\  \cline{2-8} 
& {$ 0.149\pm0.005 $ \cite{Motta:2022rku}} &  {$0.987\pm0.003$ \cite{Mall:2023nab}}   &  \multicolumn{5}{|c|}{\textbf{$\rm 7.85\pm 0.46$} \cite{motta2022black} } \\

   \hline \hline                 
\multirow{2}{*}{\textbf{Model}} & \multicolumn{7}{c|}{\textbf{}} \\ \cline{2-8} 
& \textbf{ $\mathbf{Best-fit (l,\alpha )}$} & \textbf{ $\mathbf{\chi^2_{\rm min}}\sim$} & AIC &\textbf{Whether Kerr allowed within 1$\sigma$} & \textbf{Whether Kerr allowed within 3$\sigma$} & \textbf{$a_{\rm min}$ ~($\alpha\sim 0$)} & \textbf{$M_{\rm min}$ ~($\alpha\sim 0$)}\\ \hline
PRM   & -   & 0 & 10 & Yes & Yes & $\sim 0.5$ &  8.13\\ \hline
FRM1   & -  & 0   & 10  & Yes & Yes & $\sim 0.2$ & 8.09\\  \hline
FRM2   & -   & 0 & 10 & Yes & Yes & $\sim 0.7$ &  8.02\\ \hline
KRM1   & -   & 0 & 10 & Yes & Yes & $\sim 0.5$ &  7.91\\ \hline
KRM2   & -   & 0 & 10 & Yes & Yes & $\sim -0.2$ &  7.42\\ \hline
KRM3   & -   & 0 & 10 & Yes & Yes & $\sim -0.3$ &  7.55\\ \hline
WDOM   & -  & 0   & 10  & Yes & Yes & $\sim 0.2$ & 7.7\\  \hline
NADO1   & -  & 0   & 10  & Yes & Yes & $\sim 0.1$ & 7.4\\  \hline
NADO2   & -  & 0   & 10  & Yes & Yes & $\sim 0.1$ & 7.7\\  \hline
\hline
RPM \cite{Dasgupta:2025blbe}  &  $l \sim 0.1$, $\alpha \sim 12$   & 0.016 & 10 & No & No & $\sim 0.2$  & 7.39\\ \hline
TDM \cite{Dasgupta:2025blbe}  & -  & 0   & 10  & Yes & Yes & $\sim 0.9$ & 7.49\\ 
\hline
\end{tabular}
\end{adjustbox}
\caption{Table summarizing the best-fit hairy parameters for XTE J1859+226  and comparing the spin derived from QPO related studies with previous independent estimates.
}
\label{xte2}
\end{table}

\section{Conclusion}
\label{sec7}
In this work we examine whether hairy Kerr black holes, characterized by the deviation parameters $\alpha$ and $l$, can explain the observed HFQPOs in black-hole sources. These solutions arise in modified gravity, effective field theories, or Einstein gravity with additional self-interacting fields. The Kerr limit is recovered for $\alpha\to0$ or $l\alpha\to2$, while the hairy Kerr solution satisfies the Einstein equations with a conserved energy-momentum tensor obeying the strong energy conditions. In Paper I \cite{Dasgupta:2025blbe} we studied the horizon structure of the hairy Kerr BHs which enabled us with an understanding of the physically allowed $l,\alpha$ parameter space. Thereafter we examined how the fundamental frequencies depend on the hairy parameters $(\alpha,l)$ and the BH spin $a$ relative to the Kerr case. Since the HFQPO models are determined by these fundamental frequencies, this analysis enabled us to assess the viability of the hairy Kerr solution and constrain its parameter space using the observed HFQPO data. The kinematic models (e.g., the Relativistic Precession Model (RPM) and the Tidal Disruption Model (TDM)) were tested with the HFQPO data of six BH sources (which exhibit HFQPOs in their power spectrum) in Paper I. In this paper we test the viability of the nine resonance models in explaining the HFQPO data of the six BH sources. 
Our analysis addresses two questions for each source: whether the HFQPO data favor the Kerr or the hairy Kerr spacetime, and the most suitable HFQPO model for each source. Since the physical origin of the QPOs remains uncertain, we consider multiple models rather than assuming a unique mechanism. Our conclusions are based on all the eleven QPO models, i.e., two kinematic models (considered in Paper I) and the nine resonance models considered here.

Our analysis based on the eleven QPO models reveal that in general for a given BH source, the present HFQPO data lack sufficient precision to conclusively distinguish between the Kerr and the hairy Kerr scenarios. 
For certain sources, the HFQPO data allows the entire $l,\alpha$ parameter space for some models (e.g. FRM1, FRM2, KRM2, KRM3, WDOM, NADO1 and NADO2 for GRS 1915+105, \autoref{c10}) or disfavor the intermediate $\alpha$ range for a given $l$ (e.g. Sgr A* with model FRM1, \autoref{f2}). This is particularly true for lower $l$ values (e.g. Sgr A* with model FRM2, \autoref{f6}). The only exception in this regard are the sources GRO J1655-40 and XTE J1859+226 which exhibit a QPO triplet in their power spectra. Relatively tight constraints on the model parameters are obtained only for these two sources within the framework of RPM (\autoref{gro} and \autoref{xte2}), the only model which explains the low frequency QPO (LFQPO) apart from the two HFQPOs. This has been discussed in detail in Paper I \cite{Dasgupta:2025blbe}. Thus, our results are sensitive to the specific identification of the observed QPOs, particularly to whether a complete QPO triplet can be reliably identified. The simultaneous identification of three QPO frequencies provides an additional independent observational constraint and substantially enhances the parameter-constraining power of the model.
For sources where only QPO doublets are available, the reduced number of observational constraints generally leads to broader allowed regions in the parameter space and correspondingly weaker parameter constraints. This demonstrates that the robustness of the inferred spacetime parameters depends not only on the theoretical model but also on the number and identification of the observed QPO frequencies. It therefore underscores the importance of reliable QPO identification and motivates future observations capable of providing more complete and precise QPO frequency measurements.

For each source, we compare the $AIC$ values of the considered HFQPO models to assess their relative observational support. When $\Delta_i>2$, the model with the lower $AIC$ is regarded as having stronger observational support.
We further distinguish between the eleven models by comparing the spin inferred from our analysis with previous independent estimates, while noting that such estimates often show significant discrepancies and generally assume a Kerr spacetime. To ensure a meaningful comparison, we therefore consider the spin $a_{\rm \min}$ corresponding to the $\alpha\simeq0$ limit. Tables \autoref{gro}–\autoref{xte2} summarize the spins obtained in our analysis, previous independent estimates, $\chi^2_{\rm \min}$, and the $AIC$ values corresponding to each model, for the six BH sources. Since $\chi^2_{\rm \min}\sim0$ is obtained over a broad parameter range for most sources, we do not quote specific best-fit hairy parameters in the tables.
We assume that at least one of the previously reported spin estimates for each source is reliable. A model whose predicted spin is consistent with at least one of these estimates is regarded as viable for that source. Since the previous spin measurements can differ substantially, multiple models may remain viable when the available data do not allow them to be distinguished. Conversely, a model whose predicted spin is inconsistent with all previous estimates is considered unsuitable for that source. Thus, for a given source, the most suitable models are those whose predicted spin remains consistent with at least one of the previous independent estimates and whose $AIC\sim 10$ corresponds to $\chi^2_{\rm min}\sim0$. Based on this argument, we summarize the most favored and disfavored models for each of the six BH sources in \autoref{summary}. For details, one is referred to \autoref{sec6}.

\begin{table}[H]
\centering
\begin{tabular}{|l|l|l|}
\hline
\textbf{Black Hole} & \textbf{Model favored} & \textbf{Model disfavored}\\ \hline
GRO J1655-40  & PRM and possibly KRM1 & FRM1, FRM2, KRM2, KRM3, WDOM, \\
 & & NADO1, NADO2, RPM, TDM \\  
  &  &  \\ \hline
XTE J1550-564 & NADO1, FRM1, KRM2, RPM (most favored) & PRM, KRM1     \\ 
 & followed by FRM2, WDOM, NADO2 & \\  
  &  &  \\ \hline

\hline
GRS 1915+105  & PRM, KRM1  & FRM2, KRM3, WDOM, TDM, FRM1,\\
&  &  KRM2, NADO1, NADO2, RPM      \\ 
 &  &  \\ 
\hline
H1743-322     & NADO2 (most favored), followed by NADO1 & RPM, KRM1, PRM, FRM1 \\
& WDOM, TDM, KRM2, KRM3, FRM2 &  \\
 &  &  \\ \hline
Sgr A*        & FRM1, KRM2 (most favored) followed by  & PRM, KRM1, followed by NADO1     \\
&FRM2, KRM3, WDOM, NADO2, & \\
& RPM, TDM & \\
\hline
XTE J1859+226 & TDM   & PRM, FRM1, FRM2, KRM1, KRM2 \\
&  & KRM3, WDOM, NADO1, NADO2  \\
 &  &  \\
\hline
\end{tabular}
\caption{Favored and disfavored models for each of the BH sources. }
\label{summary}
\end{table}

Before concluding, we would like to mention that the mutual inconsistency among previous spin estimates may itself warrant a reassessment of the Kerr paradigm. Combined with the limited ability of current HFQPO data to distinguish between Kerr and hairy Kerr spacetimes, and the occasional preference for alternative Kerr scenarios when QPO triplets are available, this motivates revisiting existing spin estimates obtained from the Continuum-Fitting and the Fe-line spectroscopy within non-Kerr backgrounds.
Our conclusions are necessarily cautious, as the current HFQPO data often lack sufficient precision to distinguish between the Kerr and hairy Kerr scenarios. These limitations arise primarily from the small number of BH sources exhibiting HFQPOs and the limited precision of existing observations, rather than from the methodology adopted here. Consequently, our results should be regarded as preliminary constraints on the allowed parameter space rather than statistically definitive conclusions. Future high-throughput X-ray missions such as the LOFT could provide substantially more precise QPO measurements, enabling tighter constraints on BH spin and possible deviations from the Kerr spacetime.

\section*{Acknowledgments}
Research of I.B. is funded by the Start-Up
Research Grant from SERB, DST, Government of India
(Reg. No. SRG/2021/000418). The work of S. M. is supported by the core research
grant from the Science and Engineering Research Board,
a statutory body under the Department of Science and
Technology, Government of India, under Grant
Agreement No. CRG/2023/007670.\\

\bibliography{Reee,mybib,QPO,Re,new-ref,newref,revision2}
\bibliographystyle{unsrt}

\end{document}